\documentclass[aps,twocolumn,superscriptaddress,groupedaddress]{revtex4}

\usepackage{graphics}
\usepackage[normalem]{ulem}
\usepackage{amssymb}
\usepackage{color}
\usepackage{graphicx}
\usepackage{amsthm}
\usepackage{paralist}
\usepackage{verbatim}
\usepackage{chemarr}
\usepackage{braket}
\usepackage{tikzit}
\usetikzlibrary{circuits.ee.IEC}
\usetikzlibrary{patterns}
\usepackage[caption=false]{subfig}

\newtheorem{lemma}{Lemma}

\newcommand{\spc}[1]{\mathcal{#1}}

\def\>{\rangle}
\def\<{\langle}

\newcommand{\st}[1]{\mathbf{#1}}

\newcommand{\map}[1]{\mathcal{#1}}
\newcommand{\Tr}{\operatorname{Tr}}

\newcommand{\St}{{\mathsf{St}}}

\newcommand{\Chan}{{\mathsf{Chan}}}

\newtheorem{theo}{Theorem}

\newtheorem{prop}{Proposition}

\newtheorem{defi}{Definition}

\newcommand{\Proof}{{\bf Proof. \,}}

\begin{document}

\title{Quantum Shannon theory with superpositions of trajectories}

\author{Giulio Chiribella}
\affiliation{Department of Computer Science, The University of Hong Kong, Pokfulam Road, Hong Kong}
\affiliation{Department of Computer Science, University of Oxford, Wolfson Building, Parks Road, Oxford, United Kingdom}
\author{Hl\'er Kristj\'ansson}
\affiliation{Department of Computer Science, University of Oxford, Wolfson Building, Parks Road, Oxford, United Kingdom}






\begin{abstract}
Shannon's   theory  of  information  was  built  on  the  assumption  that  the  information  carriers were
classical systems.   
 Its quantum counterpart, quantum
Shannon theory, explores the new possibilities  arising when the information carriers are quantum systems.   Traditionally, quantum Shannon theory has  focussed on scenarios where  the internal  state of the information carriers is quantum, while their trajectory   is classical. 
Here we propose a second level of quantisation where both the information and its  propagation in spacetime is treated quantum mechanically.   The framework is illustrated with a number of examples, showcasing  some of the counterintuitive phenomena taking place when information travels simultaneously through multiple transmission lines. 
\end{abstract}

\maketitle

\tableofcontents




\section{Introduction}
    When Claude Elwood Shannon laid the foundations of information theory \cite{shannon1948mathematical},     he modelled the transmission of data according to  the laws of classical physics. Specifically, he assumed  that the information carriers  had perfectly distinguishable internal states and  travelled along well-defined trajectories in spacetime.     Shannon's model worked extremely well for all practical purposes (even {\em too} well, according to Shannon himself \cite{shannon1956bandwagon}). 
      However,  at the very bottom Nature is described by the laws of quantum physics, which radically differ from the classical laws assumed by Shannon. When information is encoded into quantum systems,  some of Shannon's most fundamental  conclusions  no longer hold, giving rise  to  new  opportunities, 
       such as the opportunity to communicate securely  without pre-established keys  \cite{BennettCh1984,ekert1991quantum}.   

The extension of Shannon's  theory to the quantum domain,  called {\em Quantum Shannon theory}, is now a  highly developed  research area \cite{wilde2013quantum}.        However, there is a sense in which the transition from classical to quantum  is still incomplete.
  Traditionally, quantum Shannon theory has explored scenarios where  a number of parties exchange quantum messages,  {\em i.e.}\ messages encoded into quantum states.  
     While the  messages are allowed to be quantum,  their trajectory in spacetime is assumed to be classical. 
     However,  quantum particles can also propagate  simultaneously among multiple trajectories, as illustrated by the iconic double slit experiment \cite{feynman1964sands,wheeler2014quantum}.   
     
     The ability to propagate  along multiple paths allows quantum particles to experience   coherent superpositions  of alternative evolutions \cite{Aharonov1990,oi2003interference}, or to experience a set of evolutions  in a superposition of alternative  orders   \cite{chiribella2013quantum,Chiribella2012}.    When  a particle travels along alternative paths,  the interference of noisy processes taking place   on different paths   can result in 
        cleaner communication channels overall  \cite{gisin2005error,abbott2018communication}.      Similarly, when a particle experiences  noisy processes in a superposition of orders,  the interference between alternative orderings  can  boost  the capacity to communicate classical and quantum bits \cite{ebler2018enhanced,salek2018quantum,chiribella2018indefinite}.       The communication  advantages of  superposing alternative channel configurations  have been recently demonstrated experimentally, both for superpositions of  independent channels  \cite{lamoureux2005experimental} and of orders in time \cite{goswami2018communicating,guo2018experimental}.

The above examples  indicate, both theoretically and experimentally,  the potential of extending quantum Shannon theory to a broader framework where not only the content of the messages, but also their trajectory   can be quantum.  This extension can be regarded as a second level of quantisation of Shannon's information theory:  the first level of quantisation was to quantise   the internal degrees of freedom of the information carriers, while the second level is to quantise the external degrees of freedom, thus allowing for  the coherent propagation of messages along multiple trajectories.  
  Such quantisation, however, poses a few challenges.

 The first challenge is to formulate a clear-cut separation between the role of the internal degrees of freedom (in which information is encoded) and the external degrees of freedom (along which  information propagates).      This   separation  seems hard to enforce, due to the possibility of information flow from the internal to the external degrees of freedom, via the mechanism known as the phase kickback  \cite{cleve1998quantum}.      If the sender  exploits  phase kickback to encode information into the path,  then the path itself becomes part of the message,  and the communication protocol can be described within  the framework of standard Shannon theory, just with an enlarged quantum system playing the role of  the message-carrying degree of freedom.   In contrast,  a genuine extension of  quantum Shannon theory  should assign the message and the path two qualitatively distinct  information-theoretic roles.

   The second challenge is to formulate a model of communication where the superposition of transmission lines can be operationally built from the devices available to the communicating parties. A sender and a receiver who have access to multiple communication devices  should be able to combine them into a new communication device, corresponding to a quantum superposition of the original devices.    However, the existing definition of  superposition of channels     \cite{Aharonov1990,oi2003interference,abbott2018communication}  depends not only on the channels themselves, but also on the way in which the channels are realised through interactions with the environment. For a sender and receiver who only know the input-output description of their devices, it is not possible to know in advance which superposition will arise when the information carrier is sent through them along multiple  paths. In a Shannon-theoretic context, it is important to pinpoint what exactly are the basic resources available to the communicating parties, and show how these resources determine the transmission of information from the sender to the receiver.

This paper provides a framework  that meets  the above desiderata, laying  the foundation for an extended quantum Shannon theory where the information carriers propagate in a coherent superposition of trajectories.   A  key ingredient of the framework  is  an abstract notion of vacuum, describing the situation where no input is provided to the communication devices. The communication devices  are described by quantum channels 
capable of acting on the quantum system used as the message,  on the vacuum, and generally on coherent superpositions of the vacuum and the message.    The propagation of the message along a 
superposition of trajectories is  realised by coherently controlling which devices receive the message and which devices receive the vacuum as their input.   Such controlled routing provides an operational  way to build superpositions of channels from the local devices available to the sender and the receiver,  without the need to specify the interaction with  the environment.

Our communication model prohibits  all encoding operations that could be used to encode  messages (in part or in full) in the path of the information carrier.   This condition is satisfied when the information carrier is prepared  in a superposition of paths and sent directly to the communication devices: in this case, the encoding operation is just to initialise the path in a fixed state, independent of the message. 
 In general communication networks,  where the information has to pass through a sequence of noisy channels, our model also rules out intermediate repeater operations that create correlations between the message and the path.   In this way, it guarantees that the path is not used as an additional information carrier.  Global operations on the  message and path  are allowed only in the decoding stage, after the message has reached the receiver.   

The main resources used in our model are the vacuum-extended communication channels, the number of paths that are coherently superposed, and the total number of transmitted particles (where the term {\em particle} is used broadly to denote  any quantum system with an external degree of freedom determining its trajectory in spacetime).  For a fixed communication device and for a fixed number  of paths $N$, one can compute the number of classical or quantum bits that can be reliably transmitted  in the limit of asymptotically many particles. This construction defines a sequence of channel capacities, one capacity for every  value of $N$.  The sequence of capacities is  monotonically increasing with $N$, and the base case $N=1$ corresponds to the standard channel capacities studied in quantum Shannon theory \cite{wilde2013quantum}.  Higher values of $N$ correspond to communication protocols with  increasing levels of delocalisation of the paths.  In addition, one can also consider more general configurations where the paths of the transmitted particles are correlated.  In general, the communication model proposed in this paper opens up  the study of a range of new quantum channel capacities and the search for new techniques for quantifying them.

We illustrate our communication model in a series of examples. First, we consider the scenario where the communication channels on alternative paths are independent, showing a number of  interesting phenomena arising when the number of superposed paths $N$ becomes large. For example, we show that a qubit erasure channel, which cannot transfer any information when the path is fixed, can become a perfect classical bit channel when many paths are superposed.  Similarly, a qubit dephasing channel can become a perfect qubit channel in the large $N$ limit.    Then, we extend our analysis to scenarios exhibiting correlations.  An example of this situation arises when the local environments encountered on different paths are correlated  due to previous interactions between them.  Another example arises when a particle visits a given set of regions in a superposition of multiple orders \cite{chiribella2013quantum}. In this setting, the  correlations arising from the memory in the environment can give rise to a noiseless transmission of quantum information through noisy channels, a phenomenon that cannot take place with the superposition of independent channels \cite{chiribella2018indefinite}.

 The remainder of the paper is organised as follows.       In Section \ref{sec:framework} we provide the theoretical foundation of our communication model. We provide a general  definition of superposition of channels, which includes the superposition of independent channels, as well as superpositions of correlated channels.   We then provide an operational recipe to construct a superposition of channels from the communication devices available to the communicating parties.   In Section \ref{sec:alternativepaths} we formulate a communication model where information can propagate  through multiple independent channels.  We also provide several examples of communication protocols admitted by our model, including seemingly counterintuitive effects such as the possibility of classical communication through a superposition of pure erasure channels, or the possibility of quantum communication through a superposition of entanglement-breaking channels.  The model is extended in Section \ref{sec:alternativeorders} to scenarios where the channels on different paths are correlated, including correlations in space and correlations in time. 
  Finally, conclusions are drawn in Section \ref{sec:conclusions}.

 \section{Framework}\label{sec:framework}
  Here we provide the basic framework upon which our communication model is built.  
\subsection{Systems and sectors} 
Quantum Shannon theory describes communication in terms of abstract quantum systems, representing the degrees of freedom used to carry information.    An abstract quantum system  $A$ is associated to a Hilbert space $\spc H_A$. Its state space is the set  $  \St (A)$  containing all density operators on system $\spc H_A$, {\em i.e.}~all linear operators  $\rho\in L (\spc H_A)$, satisfying the conditions $\Tr[\rho]=1$ and $\<\psi|\rho|\psi\> \ge 0$ for every $|\psi\>\in  \spc H_A$.

 A process transforming  an input system $A$ into an output system $B$ is described by a {\em quantum channel} \cite{holevo1972mathematical}, namely a  linear, completely positive, trace-preserving map from   $\St (A)$ to  $\St (B)$. 
  The action of a quantum channel on an input state can be  written in the Kraus representation  $\map C  (\rho)  =  \sum_{i=1}^r \,  C_i  \rho  C_i^\dag$, where $\{C_i\}_{i=1}^r$ is a set of operators from $\spc H_A$ to $\spc H_B$, called Kraus operators and satisfying the normalisation condition $\sum_{i=1}^r  \,  C_i^\dag C_i   =  I_A$, $I_A$ being the identity  on $\spc H_A$.  
  The Kraus representation is non-unique, and the number  $r$ can be made arbitrarily large, {\em e.g.}~by appending null Kraus operators, or by replacing a Kraus operator $C_i$ with two Kraus operators $\sqrt p \,  C_i$ and $\sqrt{1-p} \,  C_i$. 

  The set of all channels from system $A$ to system $B$  will be denoted as $\Chan (A, B)$.   When $A  = B$, we use the shorthand  $\Chan (A): = \Chan (A,B)$. All throughout the paper, we use calligraphic fonts for channels (such as $\map C$) and standard italic for the corresponding Kraus operators (such as $C_i$).

In reality, an abstract quantum system $A$ is only the effective description of a subset of degrees of freedom that are accessible to the experimenter in a certain region of spacetime  \cite{viola2001constructing,zanardi2004quantum,chiribella2018agents}.   For example,  
 a polarisation qubit is identified by the two orthogonal states $| 1\>_{  \st k,   H } \otimes | 0\>_{   \st k,   V } $ and  $| 0\>_{   \st k,   H } \otimes | 1\>_{   \st k,   V }$, corresponding to a single photon of wavevector $\st k$ in the mode of horizontal polarisation and  a single photon  in the   mode of vertical polarisation.      The ``polarisation qubit'' description holds as long as the state of the electromagnetic field is constrained within the subspace spanned by these two  vectors.  

In general, the Hilbert space $\spc H_A $  of an abstract quantum system $A$ is a subspace of a larger   Hilbert space $\spc H_{S}$ describing all the degrees of freedom that in principle  could be accessed. The states of system $A$  are the density operators  $\rho$ satisfying the constraint 
\begin{align} \Tr[ \rho  P_A ]=1 \, ,
\end{align} where $P_A$ is the projector onto $\spc H_A$. When a system corresponds to a subspace of a larger Hilbert space, we call it a {\em sector}.   
 
The evolution of the larger system is  described by a channel $\widetilde {\map C}  \in \Chan (S)$.
       Such a channel defines an effective evolution of the sector $A$     only if  it  maps the sector  $A$ into itself, that is, if it satisfies  the {\em No Leakage Condition}  
 \begin{align}\label{noleak}  \Tr  \left[  P_A  \,    \widetilde {\map C}  (\rho)  \right]  =1 \,   \qquad \forall \rho \in  {\sf St}  (A) \, ,
 \end{align} 
 meaning that if we set up an experiment to test whether the system is in the sector $A$  after the action of the process $\map C$, the test will always respond positively provided that the initial state was in the sector $A$.    
In turn, the No Leakage Condition holds if and only if  the Kraus  operators of $\widetilde {\map C}$   satisfy the relation 
\begin{align}\label{noleakkraus}
	P_A  \widetilde C_i  P_A  =    \widetilde C_i  P_A \qquad \forall  i\in  \{1,\dots, r\} \, 
\end{align}   
(see Lemma 1 in Appendix A).
When equation \eqref{noleakkraus} is satisfied, one can define an effective channel $\map C\in \Chan (A)$  with  Kraus operators  $  C_i   :  =   P_A  \widetilde C_i P_A $.   In this case, we say that $\map C$ is the {\em restriction} of $\widetilde {\map C}$ to sector $A$ and that $\widetilde {\map C}$ is an {\em extension} of $\map C$.    


\subsection{Superposition of channels} 

Two sectors  $A$ and $B$ are {\em orthogonal} if the corresponding Hilbert spaces are orthogonal subspaces of the larger  Hilbert space $\spc H_S$.    Given two quantum systems $A$ and $B$, one can build  a new system $S  =  A\oplus B$, in which $A$ and $B$ are orthogonal sectors. Mathematically, this is done by taking the direct sum Hilbert space  $\spc H_A \oplus \spc H_B$.  
Physically, $A\oplus B$ represents a quantum system that can be in sector $A$, or in sector $B$, or in a coherent superposition of the two.

We are now ready to provide a general definition of superposition of channels: 
 \begin{defi}\label{def:superposition}  
  A  superposition of   two channels $\map A  \in  \Chan (A)$ and $\map B  \in  \Chan (B)$ is any channel  $\map S \in \Chan (A\oplus B)$ such that  {\em (i)}  $\map S$  satisfies  the No Leakage Condition with respect to  $A$ and $B$,  and  {\em (ii)}    the restrictions of $\map S$ to sectors $A$ and $B$ are $\map A$ and $\map B$, respectively.  
\end{defi}
The channel $\map S$ describes a process that can take an input  in the sector $A$,  in the sector  $B$, or   in a  coherent superposition of these two sectors.

One example of superposition is the superposition of  two unitary channels   $\map U  =  U  \cdot U^\dag$ and $\map V  =  V\cdot   V^\dag$ in terms of  the unitary channel $\map S  =  S\cdot S^\dag$ defined by  $S    =   U  \oplus  V$. 
Another example of a superposition is the non-unitary channel  $\map S    =  S_1\cdot S_1^\dag  +  S_2 \cdot S_2^\dag$, with $S_1   =   U \oplus 0_B$ and $S_2  =  0_A  \oplus  V$, where $0_A$ and $0_B$ are the null  operators on $\spc H_A$ and $\spc H_B$, respectively.     One way to realise the second example is to perform a non-demolition measurement that distinguishes between the sectors $A$ and $B$ while preserving coherence within each sector. Then, channel $\map S$ can be implemented by  performing  either channel $\map U$ or channel $\map V$ depending on the measurement outcome. This second type of superposition is {\em incoherent}, in the sense that it collapses every superposition state in $\St (A\oplus B)$ into a classical mixture of states of $A$ and states of $B$.

For noisy channels $\map A$ and $\map B$,  one can pick two Kraus representations $\{A_i\}_{i=1}^r$ and $\{B_i\}_{i=1}^r$ with the same number  of Kraus operators,  and define a superposition channel   $\map S$ with   Kraus operators
  \begin{align}\label{generalsuperposition}
  S_i  : =  A_i  \oplus  B_i   \qquad i\in  \{1,\dots, r\}\ .
  \end{align}
 Note that if two Kraus representations  $\{A_i\}_{i=1}^{r_A}$ and $\{B_i\}_{i=1}^{r_B}$    have different numbers of operators, one can always extend them to Kraus representations with the same number of operators, {\em e.g.} by appending null operators. 
   
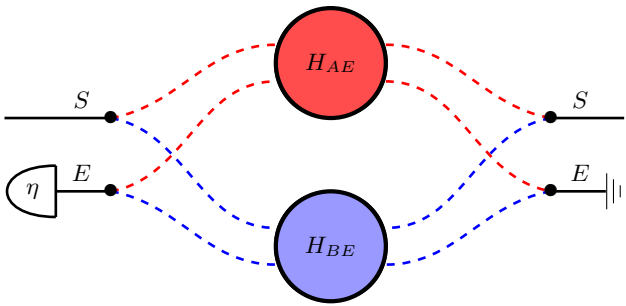
\begin{figure}
	\centering
		\begin{tikzpicture}[scale=1.95, circuit ee IEC]
	\begin{pgfonlayer}{nodelayer}
		\node [style=none] (0) at (0, 2) {};
		\node [style=none] (1) at (0, 0.5) {};
		\node [style=none] (2) at (0, -0.5) {};
		\node [style=none] (3) at (0, -2) {};
		\node [style=none, font ={\large}] (4) at (3, 0.5) {$\bullet$};
		\node [style=none, font={\large}] (5) at (-3, 0.5) {$\bullet$};
		\node [style=none] (6) at (-0.75, 1.5) {};
		\node [style=none] (7) at (0.75, 1.5) {};
		\node [style=none] (8) at (-0.75, -1) {};
		\node [style=none] (9) at (-0.75, -1.5) {};
		\node [style=none] (10) at (0.75, -1) {};
		\node [style=none] (11) at (-0.75, -1.5) {};
		\node [style=none] (12) at (-0.75, 1) {};
		\node [style=none, font={\large}] (13) at (-3, -0.5) {$\bullet$};
		\node [style=none, font={\large}] (14) at (3, -0.5) {$\bullet$};
		\node [style=none] (15) at (0.75, 1) {};
		\node [style=none] (16) at (0.75, -1.5) {};
		\node [style=none] (17) at (0, 1.25) {$H_{AE}$};
		\node [style=none] (18) at (0, -1.25) {$H_{BE}$};
		\node [style=none] (19) at (4, 0.5) {};
		\node [style=none] (20) at (3.75, -0.5) {};
		\node [style=none] (21) at (-4.45, 0.5) {};
		\node [style=none] (22) at (-3.75, -0.5) {};
		
		\node [style=none] (35) at (-3.75, -0.15) {};
		\node [style=none] (36) at (-3.75, -0.85) {};
		
		\node [style=none, xshift=-0.2em] (24) at (-4, -0.5) {$\eta$};

		\node [style=none] (27) at (-3.4, 0.75) {$S$};
		\node [style=none] (28) at (-3.4, -0.25) {$E$};
		\node [style=none] (29) at (3.4, 0.75) {$S$};
		\node [style=none] (30) at (3.4, -0.25) {$E$};
		\node [ground=none, style=none, xshift=-0.45em] (32) at (4, -0.5) {};
	\end{pgfonlayer}
	\begin{pgfonlayer}{edgelayer}
		\draw [bend left=90, looseness=1.75] (0.center) to (1.center);
		\draw [bend left=90, looseness=1.75] (1.center) to (0.center);
		\draw [bend left=90, looseness=1.75] (2.center) to (3.center);
		\draw [bend left=90, looseness=1.75] (3.center) to (2.center);
		\draw [red, style=dashed, line width=1pt, in=180, out=10, looseness=1.00] (5.center) to (6.center);
		\draw [blue, style=dashed, line width=1pt, in=180, out=-10, looseness=1.00] (5.center) to (8.center);
		\draw [red, style=dashed, line width=1pt, in=180, out=10, looseness=1.00] (13.center) to (12.center);
		\draw [blue, style=dashed, line width=1pt, in=180, out=-10, looseness=1.00] (13.center) to (11.center);
		\draw [red, style=dashed, line width=1pt, in=170, out=0, looseness=1.00] (7.center) to (4.center);
		\draw [blue, style=dashed, line width=1pt, in=0, out=190, looseness=1.00] (4.center) to (10.center);
		\draw [red, style=dashed, line width=1pt, in=170, out=0, looseness=1.00] (15.center) to (14.center);
		\draw [blue, style=dashed, line width=1pt, in=0, out=190, looseness=1.00] (14.center) to (16.center);
		\draw [line width=1pt] (21.center) to (5.center);
		\draw [line width=1pt] (22.center) to (13.center);
		\draw [line width=1pt] (4.center) to (19.center);
		\draw [line width=1pt] (14.center) to (20.center);
		\draw [fill=red!70!white, ultra thick] (0,1.25) circle [radius=0.75];
		\draw [fill=blue!40!white, ultra thick] (0,-1.25) circle [radius=0.75];
		
		\draw [line width=1pt] (35.center) to (36.center);
		\draw[line width=1pt] [in=180, out=180, looseness=3.25] (35.center) to (36.center);
		
	\end{pgfonlayer}
\end{tikzpicture}
		\caption{\label{fig:superposition} {\em Hamiltonian realisation of an arbitrary superposition of channels.}    The system $S$ and the environment $E$ are jointly routed to one of  two regions, $R_A$ and $R_B$, depending on whether the system is in sector $A$ or sector $B$.    In region $R_A$ ($R_B$) the system and the environment interact through a  Hamiltonian $H_{AE}$ ($H_{BE}$).  After the interaction, the two paths of   both the system and the environment are recombined, and finally the environment is discarded.  In general, this realisation requires the ability to control the environment.  }
\end{figure}

One may wonder whether there exist other ways to superpose two channels.  The answer is negative: 
\begin{theo}\label{theo:allsup}
The following are equivalent: 
\begin{enumerate}
\item Channel $\map S$ is a superposition of channels $\map A$ and $\map B$.
 \item  The Kraus operators of $\map S$ are of the form $S_i  =  A_i  \oplus  B_i  $  for {\em some} Kraus representations  $\{A_i\}$ and $\{B_i\}$ of channels $\map A$ and $\map B$, respectively.
\item There exists an environment $E$, a pure state $|\eta\>  \in  \spc H_E$,   two Hamiltonians $H_{AE}$ and $H_{BE}$, with supports in the orthogonal subspaces $\spc H_A\otimes \spc H_E$ and $\spc H_B \otimes \spc H_E$, respectively, and an interaction time $T$, such that $\map A (\rho)   =    \Tr_E   \big[    U_{AE}     (  \rho \otimes \eta)    U_{AE}^\dag \big]$ with $U_{AE}   =    \exp  \big[  -i   H_{AE} T/\hbar  \big]$, and  $\map B (\rho)   =    \Tr_E   \big[    U_{BE}     (  \rho \otimes \eta)    U_{BE}^\dag \big]$ with $U_{BE}   =    \exp  \big[  -i   H_{BE} T/\hbar  \big]$, and
$\map S (\rho)   =    \Tr_E   \big[    U     (  \rho \otimes \eta)    U^\dag \big]$ with $U   =    \exp  \big[  -i   (  H_{AE} \oplus H_{BE} ) T/\hbar  \big] $,    having used the notation $\eta := \ket{\eta}\bra{\eta}$. 
\end{enumerate}
\end{theo}
Theorem \ref{theo:allsup}, proven in Appendix A,   characterises all the possible  superpositions of  two given quantum channels  $\map A$ and $\map B$.    Condition 3 provides a physical realisation, illustrated in Figure \ref{fig:superposition}:  a general way to realise a  superposition of channels is to jointly route the system and the environment to two distinct regions, $R_A$ and $R_B$, depending on whether the system is in the sector $A$ or in the sector $B$.    In the two  regions, the system and the environment  interact either through the Hamiltonian $H_{AE}$ or through the Hamiltonian $H_{BE}$.  After the interaction, the two alternative paths are recombined, and the environment is discarded.

Theorem \ref{theo:allsup} pinpoints a few issues with the notion of ``superposition of channels''.    First,  the superposition of two channels $\map A$ and $\map B$ is not determined by the channels $\map A$ and $\map B$ alone: different choices of Kraus representations, $\{A_i\}_{i=1}^r$ and $\{B_i\}_{i=1}^r$ generally give rise to different superpositions.  
 Physically, the dependence on the choice of Kraus representation can be understood in terms of the unitary realisation of channels $\map A$ and $\map B$ \cite{oi2003interference,abbott2018communication}: in general, the superposition of two channels depends not only on the channels themselves, but also on the way in which the channels are realised through an interaction with the surrounding environment.

A further issue is that, even if complete access to the environment is granted,  the superposition of two unitary gates cannot be implemented in a circuit if the two unitaries are unknown \cite{soeda2013limitations,Araujo2014quantum,chiribella2016optimal,thompson2018quantum}.     In other words, it is impossible to generate the coherent superposition  $U\oplus V$  of two  arbitrary unitaries $U$ and $V$   by inserting the corresponding devices into a quantum circuit with two open slots. 
 The impossibility to build the superposition of two gates from the gates themselves is reflected in the fact that the superposition is not a {\em quantum supermap} \cite{chiribella2008transforming,chiribella2009theoretical,chiribella2013quantum,chiribella2016optimal}, {\em i.e.}\ is not a physically admissible transformation  of quantum channels.        This prevents a resource-theoretic formulation where the communicating parties are given a set of communication resources and a set of allowed operations to manipulate them \cite{ebler2018enhanced}.

Finally, Theorem \ref{theo:allsup} shows one physical realisation of the superposition of channels, in which the system {\em and the environment} are jointly routed to two different regions, where they experience two different interactions.  Routing the environment  is problematic in a theory of communication,  especially in cryptographic scenarios where the environment is under the control of an adversary.

In the following  we will address the above issues by upgrading the physical description of the communication devices: instead of describing them as quantum channels acting on the information carrier  alone, we will describe them as quantum channels acting  on the information carrier,  on the vacuum, or on coherent superpositions of the information carrier and the vacuum.  

\subsection{The vacuum extension of a quantum channel}  

In information theory, each use of a communication  channel is counted as a resource. However, the physical apparatus used to communicate does not come  into existence in the moment when it is used to transmit a signal.  For example, an optical fibre is in place also when no photon is sent through it. 
   When the fibre is not used, we can model its input as being the vacuum state.  Abstracting from this example, we assume that the system  used to communicate is a sector $A$ of a larger system $S$,  containing another sector, called the {\em vacuum sector}  ${\rm Vac}$, and orthogonal to $A$.  Orthogonality of  $A$ and $\rm Vac$   means that one can perfectly distinguish between situations where a signal is sent and situations where it is not.     
   
   Our abstract notion of the vacuum is directly inspired by the vacuum in quantum optics, which is orthogonal to the polarisation states of single photons, that are used as information carriers   in many quantum communication protocols.    Orthogonality of the message with the vacuum may not be exactly satisfied in some scenarios, {\em e.g.}~when single-photon states are replaced by weak coherent states, or in certain scenarios of quantum field theory, where the vacuum may not be exactly orthogonal to the states the sender can generate locally in order to transmit a message. In  these scenarios, it is understood that our exact orthogonality condition should be relaxed to an approximate orthogonality condition.  In this paper, however, we will stick to to the exact case, which allows for a considerably simpler presentation. 
       
   A communication device is modelled as a quantum channel on the direct sum system $\widetilde A   :  = A\oplus \rm Vac$.    In this picture, the input of the device can be interpreted as an abstract {\em mode}, which can be either in the one-particle sector $A$ or in the vacuum sector $\rm Vac$.   This description  is consistent with the standard framework of  quantum optics, where the action of  physical devices like  beam splitters or  phase shifters are represented by  quantum channels from a set of input modes to a set of output modes. 
    
      \begin{defi}  
     Channel $\widetilde {\map C}  \in  \Chan  (\widetilde A )$ is a {\em vacuum extension}  of channel $\map C  \in  \Chan  (A)$ if {\em (i)} $\widetilde {\map C}$ satisfies the No Leakage Condition with respect to sectors $A$ and $\rm Vac$, and {\em (ii)} the restriction of $\widetilde {\map C}$ to sector $A$ is  $\map C$. 
     \end{defi}
The vacuum extension of channel $\map C$ is a superposition of $\map C$  with some other channel $\map C_{\rm Vac}$, representing the action of the communication device on the vacuum sector.

 For simplicity, in the following we  will assume that the vacuum sector is one-dimensional, meaning that there is a unique vacuum state $|\rm vac\>$, up to global phases.   In this case, Theorem \ref{theo:allsup} implies that the Kraus operators of the vacuum extension $\widetilde {\map C}$  are of the form
   \begin{align}\label{vacext}
   \widetilde C_i  =  C_i  \oplus  \gamma_i \,  |\rm vac \>\<\rm vac  |   \, ,
   \end{align} 
  where    $\{C_i\}_{i=1}^r$ is a Kraus representation of $\map C$, and $\{\gamma_i\}_{i=1}^r$ are complex amplitudes satisfying the normalisation condition 
  \begin{align}
\sum_i \, |   \gamma_i|^2   =  1  \, .
  \end{align}   Hereafter, we will call  $\{\gamma_i\}_{i=1}^r$  the  {\em vacuum amplitudes of $\widetilde{\map C}$}.    The case of vacuum sectors of arbitrary dimension is discussed in Appendix B. In the main body of the paper we will always assume that the vacuum sector is one-dimensional.

  Note that the vacuum  extension   is  highly non-unique: it depends on the choice of Kraus representation {\em and} on the choice of vacuum amplitudes.     For example, the vacuum extension of a unitary channel $\map U$ could be  a unitary channel $\widetilde{\map U}$ with  $\widetilde U  =   U\oplus  |{\rm vac} \>\<{\rm vac } |$, or also a non-unitary channel $\map C (\cdot) =  C_1\cdot C_1^\dag +  C_2  \cdot C_2^\dag$, with   $C_1  =  U  \oplus 0_{\rm Vac}$ and $C_2   =  0_A  \oplus  |{\rm vac}\>\<{\rm vac}|$. In the second case,  coherence with the vacuum is not preserved, and channel $\map C$ transforms every superposition in $A\oplus \rm Vac$ into a classical mixture of a state of $A$ and the vacuum.  
  
   For a non-unitary channel $\map C$ with Kraus operators $\{C_i\}_{i=1}^r$ one can define many vacuum extensions, {\em e.g.} by defining the  Kraus operators $  \widetilde C_i  :  =   C_i \oplus   |{\rm vac }\>\<{\rm vac } |/\sqrt{r}$, or the Kraus operators $\widetilde C_i  :  =  C_i  \oplus  0_{\rm Vac}$, for $i\in  \{1,\dots, r\}$, and $\widetilde C_{r+1}  =   0_A\oplus  |\rm vac \>\<\rm vac  |$.   Note that the second example does not preserve coherence between the sectors $A$ and $\rm Vac$.   In general, we say that a vacuum extension has {\em no coherence with the vacuum}  if  $\gamma_i   =  0$ whenever $C_i\not  =  0$.

Even though the vacuum extension is mathematically not unique, the choice of vacuum extension is uniquely determined by the physics of the device. 
     For example, consider a device that rotates the polarisation of a single- photon about the $z$-axis by an angle $\theta_k$, chosen at random with probability $p_k$.    Physically, the single-photon polarisation corresponds to the two-dimensional subspace spanned by the logical states $|0\>_L:  = |1\>_{\st k,H} \otimes |0\>_{\st k,V}$ and  $|1\>_L:  = |0\>_{\st k, H} \otimes |1\>_{\st k,V}$, where $\st k$ is the wavevector, and $H$ and $V$ label the modes with vertical and horizontal polarisation.  The polarisation rotation is generated by the Hamiltonian  $H  =  (a_{\st k,  H}^\dag  a_{\st k,  H}   -   a_{\st k,  V}^\dag  a_{\st k,  V})/2$  (in suitable units), which induces the unitary transformation  $U_\theta  =    \exp[ - i  \theta  H]$ on the modes.   When restricted to the one-photon subspace, the unitary rotation $U_\theta$ acts in the familiar way, as $R_\theta  =  e^{-i\theta/2} \,  |0\>\<0|_L   +  e^{i\theta/2}\,  |1\>\<1|_L$. When restricted to the vacuum  $|{\rm vac}\> :  =  |0\>_{\st k,  H}\otimes |0\>_{\st k, V}$, it acts trivially as $U_\theta |{\rm vac}\> =  |{\rm vac}\>$.  When acting on a coherent superposition, it acts as  the direct sum $\widetilde R_{\theta}  =   R_\theta   \oplus  |{\rm vac}\>\<{\rm vac}|$.     In this example, the communication channel is $\map C (\cdot)  =  \sum_k \,  p_k  \,  R_{\theta_k}  \cdot  R_{\theta_k}^\dag$ and its vacuum extension is $\widetilde {\map C} (\cdot)  =  \sum_k  \,  p_k \, \widetilde R_{\theta_k}  \cdot  \widetilde R_{\theta_k}^\dag$.    
 Explicitly, the original Kraus operators are $C_k  =       \sqrt{  p_k}  \,  R_k$  and the vacuum-extended Kraus operators are $\widetilde C_k   =  \sqrt {p_k}  \,  R_k  \oplus \sqrt {p_k }  \,   |{\rm vac}\>\<{\rm vac}|$.

 Physically,  the vacuum extension is the complete description of the communication resource available to the sender and receiver, and it can be determined experimentally by an input-output tomography of the communication device.  Its specification is part of the specification of the communication scenario.     It is important to stress that specifying the vacuum extension  does not mean specifying the full unitary realisation of the channel $\map C$.   Thanks to this fact, our communication model maintains a separation between the system and its environment, which potentially can be under the control of an adversary. 
    The relation between the vacuum extension and the unitary realisation is  discussed in Appendix C.

    The vacuum extensions of a given channel  form a convex set.     Its extreme points  correspond to vacuum extensions that are free from classical randomness.   In Appendix D we characterise the extreme vacuum extensions of a given channel, proving bounds on the number of linearly independent Kraus operators and on the structure of the vacuum amplitudes.   An interesting consequence of this characterisation is that the extreme vacuum extensions should have coherence with the vacuum: the vacuum amplitude $\gamma_i$ should be non-zero whenever the corresponding Kraus operator $C_i$ is non-zero.

 \subsection{The  superposition of two independent channels}

 We now provide an operational way to build the superposition of two channels from  their vacuum extensions. 
      The idea is that  physical systems always come in alternative to the vacuum.  Given two systems $A$ and $B$,    we construct the vacuum-extended  systems  $\widetilde A  :  =  A \oplus \rm Vac$  and $\widetilde B  :=   B\oplus \rm Vac$, and  we consider the composite system $\widetilde A\otimes \widetilde B$. 
Such a system contains a no-particle sector ${\rm Vac}\otimes {\rm Vac}$, a  one-particle sector  $\left(A\otimes {\rm Vac} \right)  \oplus \left( {\rm Vac} \otimes B\right)$, and a two-particle sector $A\otimes B$.   Since  the vacuum sector is one-dimensional, the one-particle sector  is isomorphic to the direct sum $ A \oplus B$.

Given two   vacuum extensions $\widetilde{\map A}$ and $\widetilde{\map B}$ of the channels $\map A$ and $\map B$, we can consider the product channel  $\widetilde{\map A} \otimes \widetilde{\map B}$ representing the independent action of $\widetilde{\map A}$ and $\widetilde{\map B}$.  Then,  we can define a superposition of channels   $\map A$ and $\map B$ as the restriction of the
 product channel $\widetilde {\map A} \otimes \widetilde {\map B}$ to the one-particle sector  $\left(A\otimes {\rm Vac}\right)  \oplus \left({\rm Vac}\otimes B\right)$.    More formally: 
 \begin{defi}\label{sup_def_simple}
The  superposition of channels $\map A$ and $\map B$ specified  by the vacuum extensions $\widetilde{\map A}$ and $\widetilde{\map B}$  is the channel 
\begin{align}\label{canS}
\map S_{  \widetilde {\map A}  , \widetilde {\map B}} :  =    \map V^\dag  \circ  \left(  \widetilde {\map A}  \otimes \widetilde {\map B}\right) \circ \map V \, ,
\end{align}
where $\map V$ and $\map V^\dag$ are the quantum channels  $\map V   (\cdot):=   V  (\cdot) V^\dag$, $\map V^\dag (\cdot) :=   V^\dag (\cdot) V$, and $V$ is the unitary operator $V:  \spc H_A  \oplus \spc H_B  \to \left(\spc H_A\otimes \spc H_{\rm Vac}\right)  \oplus \left(\spc H_{\rm Vac}\otimes \spc H_B\right) $, defined by the relation 
\begin{align}\label{V}  V  \Big(|\alpha\>  \oplus |\beta\>\Big)  : =   \Big(|\alpha\>\otimes |{\rm vac}\>\Big) \oplus  \Big( |{\rm vac}\>\otimes |\beta\>\Big) \, ,
\end{align}
for every $|\alpha\>$ in $\spc H_A$ and every $|\beta\> $ in $\spc H_B$. 
 \end{defi}  
Operationally,  the superposition $\map S_{\widetilde {\map A},  \widetilde {\map B}}$  is  built from  the physical devices described by the   vacuum extensions $\widetilde {\map A}$ and $\widetilde {\map B}$.  The two devices  are used in parallel, and their input is constrained to be  a superposition of    one particle travelling through $\widetilde{ \map  A}$  (with the vacuum in $\widetilde {\map B}$)   and  one particle travelling through $\widetilde{ \map  B}$ (with the vacuum in $\widetilde {\map A}$).  Mathematically, the transformation  from  the pair of channels $(\widetilde {\map A}, \widetilde {\map B})$ to the channel $\map S_{  \widetilde {\map A}  , \widetilde {\map B}}$ is a legitimate {\em quantum supermap} \cite{chiribella2008transforming,chiribella2009theoretical,chiribella2013quantum,chiribella2016optimal}.

The Kraus operators of the  superposition $\map S_{\widetilde {\map A},  \widetilde {\map B}}$  are    
\begin{align}\label{canonicalsup}
S_{ij}   =      A_i       \,  \beta_j  \oplus  \alpha_i \, B_j \, ,
\end{align} 
where $\alpha_i$ and $\beta_j$ are the vacuum amplitudes associated to channels $\map A$ and $\map B$, respectively.  Note that these Kraus operators may or may not have coherence between the sectors $A$ and $B$, depending on whether or not the vacuum extensions $\widetilde {\map A}$ and $\widetilde {\map B}$ have coherence with the vacuum.    For example, if the  vacuum extension of $\map A$ has no coherence with the vacuum  ($\alpha_i=  0$ whenever $A_i  \not  = 0$), then the superposition has no coherence between the sectors $A$ and $B$,  meaning that the Kraus operators are either of the form $S_{ij}  = A_i\, \beta_j  \oplus 0_B$, or of the form $S_{ij}  =   0_A \oplus \alpha_i  \,  B_j$.

\subsection{Superposition of multiple independent channels} 

   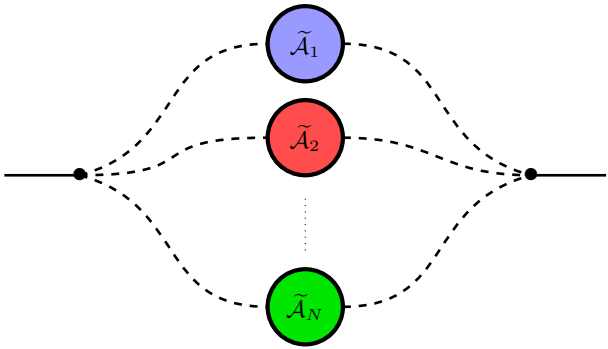
\begin{figure}
   	\centering
   	\begin{tikzpicture}[scale=2.0]
	\begin{pgfonlayer}{nodelayer}
		\node [style=none, font={\large}] (0) at (-3, 0) {$\bullet$};
		\node [style=none] (1) at (-4, 0) {};
		\node [style=none, font={\large}] (2) at (3, 0) {$\bullet$};
		\node [style=none] (3) at (4, 0) {};
		\node [style=none] (4) at (0, 1.75) {$\map{\widetilde{A}}_1$};
		\node [style=none] (5) at (0, 2.25) {};
		\node [style=none] (6) at (0, 1.25) {};
		\node [style=none] (7) at (0, -1.25) {};
		\node [style=none] (8) at (0, -2.25) {};
		\node [style=none] (9) at (0, 1) {};
		\node [style=none] (10) at (0, 0) {};
		\node [style=none] (11) at (0, 0.5) {$\map{\widetilde{A}}_2$};
		\node [style=none] (12) at (0, -1.75) {$\map{\widetilde{A}}_N$};
		\node [style=none] (13) at (-0.5, 1.75) {};
		\node [style=none] (14) at (-0.5, 0.5) {};
		\node [style=none] (15) at (-0.5, -1.75) {};
		\node [style=none] (16) at (2.5, -0.25) {};
		\node [style=none] (17) at (0.5, 1.75) {};
		\node [style=none] (18) at (0.5, 0.5) {};
		\node [style=none] (19) at (0.5, -1.75) {};
		\node [style=none] (20) at (0, -0.25) {};
		\node [style=none] (21) at (0, -1) {};
	\end{pgfonlayer}
	\begin{pgfonlayer}{edgelayer}
		\draw [bend left=90, looseness=1.75] (5.center) to (6.center);
		\draw [bend right=90, looseness=1.75] (5.center) to (6.center);
		\draw [bend left=90, looseness=1.75] (7.center) to (8.center);
		\draw [bend right=90, looseness=1.75] (7.center) to (8.center);
		\draw [bend left=90, looseness=1.75] (9.center) to (10.center);
		\draw [bend right=90, looseness=1.75] (9.center) to (10.center);
		\draw [line width=1pt] (1.center) to (0.center);
		\draw [line width=1pt] (2.center) to (3.center);
		\draw [style=dashed, line width=1pt, in=180, out=-15, looseness=1.25] (0.center) to (15.center);
		\draw [style=dashed, line width=1pt, in=180, out=0, looseness=1.75] (0.center) to (14.center);
		\draw [style=dashed, line width=1pt, in=180, out=15, looseness=1.25] (0.center) to (13.center);
		\draw [style=dashed, line width=1pt, in=-165, out=0, looseness=1.25] (19.center) to (2.center);
		\draw [style=dashed, line width=1pt, in=180, out=0, looseness=1.25] (18.center) to (2.center);
		\draw [style=dashed, line width=1pt, in=165, out=0, looseness=1.25] (17.center) to (2.center);
		\draw [style=dotted] (21.center) to (20.center);
		
		\draw [fill=red!70!white, ultra thick] (0,0.5) circle [radius=0.5];
		\draw [fill=blue!40!white, ultra thick] (0,1.75) circle [radius=0.5];
		\draw [fill=green!90!black, ultra thick] (0,-1.75) circle [radius=0.5];
		
	\end{pgfonlayer}
\end{tikzpicture}
	\caption{\label{fig:generalsup}  {\em Operational superposition of $N$ independent channels.} An input system $A  =  A_1\oplus A_2 \oplus \cdots \oplus A_N$    branches out according to its sectors, with the $j$-th branch  sent through the vacuum extension  $\widetilde{\map A}_j$. 
		The outputs are finally recombined to form the overall output of the superposition channel.}  
\end{figure}

The generalisation to superpositions of more than two channels is immediate, and is illustrated in Figure \ref{fig:generalsup}.  The  superposition of the channels $\map A^{(1)},  \dots  ,  \map A^{(N)}$  specified by the vacuum extensions  $\widetilde{\map A}^{(1)},  \dots  ,  \widetilde{\map A}^{(N)}$   is the channel with Kraus operators 
\begin{align}
S_{i_1  \cdots i_N}  =       \bigoplus_{j=1}^N  \, \alpha^{(1)}_{i_1}   \cdots  \alpha^{(j-1)}_{i_{j-1}}  A^{(j)}_{i_j} \alpha^{(j+1)}_{i_{j+1}}  \cdots    \alpha^{(N)}_{i_{N}}    \, ,            
 \end{align} 
  where   $\{A_{i_j}^{(j)}\}$ is a Kraus representation of the $j$-th channel, and  $\{\alpha^{(j)}_{i_j}\}$ are the corresponding vacuum amplitudes.  
  Physically, the above Kraus operators describe  a coherent superposition of scenarios where a  particle is sent to the $j$-th process, while the remaining $N-1$ processes act on  the vacuum.

\section{Communication through a superposition of independent channels}\label{sec:alternativepaths}     

In this section we formulate a communication model where information can be transmitted simultaneously through multiple independent channels.   The central idea is to maintain a separation of roles between the internal degrees of freedom, in which information is encoded, and the external degrees of freedom, which control the propagation of  information in spacetime. 
The general model is illustrated in a number of examples, highlighting some counterintuitive features of the superposition of channels.

\subsection{Communication model with  coherent control over independent transmission lines} 
Here we develop a  model of communication where the information carrier can  travel  along multiple alternative paths, experiencing an   independent noisy process on each path.    In the typical scenario, the paths are spatially separated, meaning that they visit  non-overlapping  regions. 

 \subsubsection{Single-particle communication}   
Let  us consider first  the simplest scenario of  communication from a single sender to a single receiver using a single particle. The particle has an internal degree of freedom  $M$  (the ``message-carrying system''), and an external degree of freedom $P$  (the ``path'').   
For simplicity, we assume that the particle can travel through two alternative devices, and therefore the path degree of freedom $P$ can be  effectively described as a qubit. 

 The action of the two devices on the internal degree of freedom $M$ is specified by two channels $\map A$ and $\map B$, with inputs $A$ and $B$, respectively, with $A\simeq B\simeq M$.  When the message is not sent through a device, the input of that device is the vacuum.      The  full description of the communication devices  is provided by the vacuum extensions $\widetilde {\map A}$ and $\widetilde  {\map B}$, acting on the extended systems $\widetilde A  =  A \oplus {\rm Vac}$ and $\widetilde B =  B \oplus {\rm Vac}$.     The channels $\widetilde {\map A}$ and $\widetilde  {\map B}$ are assumed to be  {\em independent}, meaning that the evolution of the composite system $\widetilde A\otimes \widetilde B$ is the product channel $\widetilde {\map A} \otimes \widetilde  {\map B}$.  

The transmission of a single particle in a superposition of paths is described by initialising the input of the channel $\widetilde {\map A} \otimes \widetilde  {\map B}$ in the one-particle sector  $\big( A \otimes {\rm Vac}\big) \oplus  \big({\rm Vac} \otimes B \big)$. Since the sectors $A$ and $B$ are both  isomorphic to $M$,  the one-particle subspace is isomorphic to $M\otimes P$, where  $P$ is the path qubit.  The isomorphism is implemented by the unitary gate 
$U$, defined as 
\begin{align}
\nonumber U   (  |\psi\>  \otimes |0\>)    &:=  |\psi\>  \otimes |{\rm vac}\>  \\
U   (  |\psi\>  \otimes |1\>)    &:=  |{\rm vac}\> \otimes  |\psi\>   \, , \label{QPtoSum}
\end{align} 
where the basis states $|0\>$ and $|1\>$ correspond to the  two alternative paths that the particle can take.
The isomorphism $U$ is crucial, in that it implements the change of description  from the ``particle picture'' with system  $M\otimes P$ to the ``mode picture''  with system $\big( A \otimes {\rm Vac}\big) \oplus  \big({\rm Vac} \otimes B \big)$.

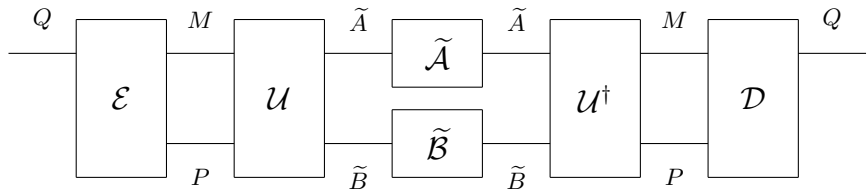
\begin{figure*}
	\centering
	\begin{tikzpicture}[scale=1.2]
	\begin{pgfonlayer}{nodelayer}
		\node [style=none] (0) at (-1, 0.25) {};
		\node [style=none] (1) at (-1, 1.75) {};
		\node [style=none] (2) at (1, 1.75) {};
		\node [style=none] (3) at (1, 0.25) {};
		\node [style=none] (4) at (-1, -0.25) {};
		\node [style=none] (5) at (-1, -1.75) {};
		\node [style=none] (6) at (1, -1.75) {};
		\node [style=none] (7) at (1, -0.25) {};
		\node [style=none] (8) at (2.5, 1.75) {};
		\node [style=none] (9) at (2.5, -1.75) {};
		\node [style=none] (10) at (4.5, 1.75) {};
		\node [style=none] (11) at (4.5, -1.75) {};
		\node [style=none] (12) at (6, -1.75) {};
		\node [style=none] (13) at (6, 1.75) {};
		\node [style=none] (14) at (8, 1.75) {};
		\node [style=none] (15) at (8, -1.75) {};
		\node [style=none] (16) at (-2.5, 1.75) {};
		\node [style=none] (17) at (-2.5, -1.75) {};
		\node [style=none] (18) at (-4.5, -1.75) {};
		\node [style=none] (19) at (-4.5, 1.75) {};
		\node [style=none] (20) at (-6, 1.75) {};
		\node [style=none] (21) at (-8, 1.75) {};
		\node [style=none] (22) at (-8, -1.75) {};
		\node [style=none] (23) at (-6, -1.75) {};
		\node [style=none] (24) at (-6, 1) {};
		\node [style=none] (25) at (-4.5, 1) {};
		\node [style=none] (26) at (-4.5, -1) {};
		\node [style=none] (27) at (-6, -1) {};
		\node [style=none] (28) at (-2.5, 1) {};
		\node [style=none] (29) at (-1, 1) {};
		\node [style=none] (30) at (-1, -1) {};
		\node [style=none] (31) at (-2.5, -1) {};
		\node [style=none] (32) at (1, 1) {};
		\node [style=none] (33) at (2.5, 1) {};
		\node [style=none] (34) at (1, -1) {};
		\node [style=none] (35) at (2.5, -1) {};
		\node [style=none] (36) at (4.5, 1) {};
		\node [style=none] (37) at (6, 1) {};
		\node [style=none] (38) at (6, -1) {};
		\node [style=none] (39) at (4.5, -1) {};
		\node [style=none] (40) at (-8, 1) {};
		\node [style=none] (41) at (-9.5, 1) {};
		\node [style=none] (42) at (8, 1) {};
		\node [style=none] (43) at (9.5, 1) {};
		\node [style=none] (44) at (-8.75, 1.75) {$Q$};
		\node [style=none] (45) at (-5.25, 1.75) {$M$};
		\node [style=none] (46) at (-1.75, 1.75) {$\widetilde{A}$};
		\node [style=none] (47) at (1.75, 1.75) {$\widetilde{A}$};
		\node [style=none] (48) at (5.25, 1.75) {$M$};
		\node [style=none] (49) at (8.75, 1.75) {$Q$};
		\node [style=none] (50) at (-5.25, -1.75) {$P$};
		\node [style=none] (51) at (-1.75, -1.75) {$\widetilde{B}$};
		\node [style=none] (52) at (1.75, -1.75) {$\widetilde{B}$};
		\node [style=none] (53) at (5.25, -1.75) {$P$};
		\node [style=none, font={\large}] (54) at (-7, 0) {$\map{E}$};
		\node [style=none, font={\large}] (55) at (-3.5, 0) {$\map{U}$};
		\node [style=none, font={\large}] (56) at (0, 1) {$\map{\widetilde{A}}$};
		\node [style=none, font={\large}] (57) at (0, -1) {$\map{\widetilde{B}}$};
		\node [style=none, font={\large}] (58) at (3.5, 0) {$\map{U^{\dagger}}$};
		\node [style=none, font={\large}] (59) at (7, 0) {$\map{D}$};
	\end{pgfonlayer}
	\begin{pgfonlayer}{edgelayer}
		\draw (21.center) to (20.center);
		\draw (20.center) to (23.center);
		\draw (23.center) to (22.center);
		\draw (22.center) to (21.center);
		\draw (19.center) to (16.center);
		\draw (16.center) to (17.center);
		\draw (17.center) to (18.center);
		\draw (18.center) to (19.center);
		\draw (1.center) to (2.center);
		\draw (2.center) to (3.center);
		\draw (3.center) to (0.center);
		\draw (0.center) to (1.center);
		\draw (4.center) to (7.center);
		\draw (7.center) to (6.center);
		\draw (6.center) to (5.center);
		\draw (5.center) to (4.center);
		\draw (8.center) to (10.center);
		\draw (10.center) to (11.center);
		\draw (11.center) to (9.center);
		\draw (9.center) to (8.center);
		\draw (13.center) to (14.center);
		\draw (14.center) to (15.center);
		\draw (15.center) to (12.center);
		\draw (12.center) to (13.center);
		\draw (43.center) to (42.center);
		\draw (37.center) to (36.center);
		\draw (38.center) to (39.center);
		\draw (35.center) to (34.center);
		\draw (33.center) to (32.center);
		\draw (29.center) to (28.center);
		\draw (30.center) to (31.center);
		\draw (26.center) to (27.center);
		\draw (25.center) to (24.center);
		\draw (40.center) to (41.center);
	\end{pgfonlayer}
\end{tikzpicture}
	\caption{\label{fig:model} {\em Communication via the transmission of a single particle through  two independent channels.} The sender uses the encoding channel $\map{E}$ to encode the state of a quantum system $Q$ into  the composite system $M \otimes P$, representing a quantum particle with internal degree of freedom $M$ and external degree of freedom  $P$.  The unitary channel $\map{U}$  maps the composite system $M\otimes P$ into the one-particle subspace of the composite system $\widetilde{A} \otimes \widetilde{B}$, on which the product channel $\widetilde{ \map  A} \otimes \widetilde{ \map  B}$ acts.   Finally, the unitary channel $\map{U}^\dagger$ converts the output back into the system $M\otimes P$ and the decoding channel $\map{D}$ outputs the decoded state of system $Q$. } 
\end{figure*}  

With these notions, we are ready to construct communication protocols like the one shown  in Figure \ref{fig:model}. 
  Initially, the sender has a message,   represented by a quantum state $\rho$ of some abstract quantum system $Q$.  Then, the sender encodes the message in the state of the particle, using an encoding channel 
\begin{align}\label{encoding}
\map E:  \St (Q)  \to \St  (
  M\otimes P) \, .
\end{align}  
In general, the encoding channel could encode information not only in the internal degree of freedom $M$, but also in the path $P$.   In that case, however, there would be no difference of roles between $M$ and $P$ as far as information theory is concerned: the composite system $M\otimes P$ would just become the new message $M'$.   In contrast, here we are interested in the scenario where the information is encoded  into the original message system  $M$, while the path system $P$ is used to route the message  in space. From this perspective,    it is natural to demand that the encoding channel $\map E$  satisfies the no-signalling condition 
\begin{align}\label{nosig}
\Tr_{M}  \big[\map E (\rho) \big]  =  \Tr_{M}  \big[\map E (\sigma) \big]  \qquad \forall \rho,\sigma \in\St(Q)\, . 
\end{align}  
In fact, we will demand an even stronger condition, which guarantees that the encoding does not create any correlations between the message and the path. Explicitly, we require  the encoding operation to have  the product form  
\begin{align}\label{productenc}
\map E    =   \map M   \otimes \omega  \, , 
\end{align}
where $\map M$ is a channel from $\St (Q)$ to $\St (M)$, and $\omega  \in  \St (P)$ is a fixed state of the path.  
This means that, for every initial state $\rho \in \St (Q)$, the encoded state has the product form $\map M (\rho)  \otimes \omega$, with the state of the path  completely uncorrelated with the  message.  Conditions (\ref{nosig}) and (\ref{productenc})   will be further discussed  in  Subsection \ref{subsect:nosig}.

After the message is encoded, it is sent through the channels  $\widetilde {\map A}$ and $\widetilde  {\map B}$.    The transmission  is  described by the channel 
\begin{align}\label{SU}
\map S_{\widetilde {\map A}, \widetilde {\map B}}    =   \map U^\dag (\widetilde {\map A} \otimes \widetilde{ \map  B})  \,\map U \,,
\end{align}
and has  Kraus operators  
\begin{align}\label{ctrlchan}
S_{ij}  = A_i  \, \beta_j    \otimes |0\>\<0|  +  \alpha_i \, B_j    \otimes |1\>\<1|  \, ,
\end{align} 
 where $\{\alpha_i\}$ and $\{\beta_j\}$ are the vacuum amplitudes of channels $\map A$ and $\map B$, respectively.  Note that in this stage the transmission can generate correlations between the message and the path, due to the interaction of the particle  with the communication devices.  
  
 Finally, the receiver will perform a decoding operation, described by a quantum channel $\map D:  \St  (M\otimes P)  \to \St(  Q)$.   Since in this stage the particle has already reached the receiver, we assume no constraints on the decoding operations.

\subsubsection{Extension {\rm vs} restriction}

The aim of our communication  model is to provide an extension of quantum Shannon theory where messages can propagate in a coherent superposition of trajectories.  In a genuine extension, the role of the path should be qualitatively different from the role of the message,  for otherwise the  ``extension'' would only consist of using a larger quantum system as the message.

Now, the separation of roles between internal and external degrees of freedom is not automatically guaranteed. For example, if no restriction is imposed, the sender could  send a bit to the receiver by encoding the value of the bit in the path. Explicitly,  the sender could encode the bit value $0$ into the state $|\psi_0\>  \otimes |0\>$ and the bit value $1$ into the state $|\psi_1\>  \otimes |1\>$, which remain orthogonal when the particle is sent through the communication channel [cf. Equations (\ref{SU}) and (\ref{ctrlchan})], no matter which communication channel is  used.

Perhaps counterintuitively, the {extension} of quantum Shannon theory to the scenario where messages propagate in a superposition requires   a {\em restriction} on the allowed encoding operations.  The contradiction is only apparent, because the extension consists in giving the message system $M$ an additional feature (the ability to propagate in a superposition of paths), which was not present in the standard model of quantum Shannon theory.   
As we will see in the following, this extension allows us to define a hierarchy of  quantum channel  capacities that includes the standard quantum channel capacities as its first level.

\subsubsection{No-signalling {\rm vs} signalling encodings}\label{subsect:nosig}

The condition that the encoding operation must not signal to the path rules out a number of  communication protocols based on  a
``quantum superposition of circuits'', which has some similarities with our model, but is  conceptually rather different.

\begin{figure}
	\centering
	\begin{tikzpicture}[circuit ee IEC, scale=1.5]
	\begin{pgfonlayer}{nodelayer}
		\node [style=none] (0) at (-4, 0) {};
		\node [style=none] (1) at (-5, 1.5) {};
		\node [style=none] (2) at (-3, 2) {};
		\node [style=none] (3) at (-3, -0.5) {};
		\node [style=none] (4) at (-1.5, 2) {};
		\node [style=none] (5) at (-1.5, -0.5) {};
		\node [style=none] (6) at (-3, 1.5) {};
		\node [style=none] (7) at (-3, 0) {};
		\node [style=none] (8) at (-1.5, 1.5) {};
		\node [style=none] (9) at (-1.5, 0) {};
		\node [style=none] (10) at (-0.5, 1.5) {};
		\node [style=none] (11) at (-0.5, 0) {};
		\node [style=none] (12) at (-0.5, 2) {};
		\node [style=none] (13) at (1, 2) {};
		\node [style=none] (14) at (-0.5, 1) {};
		\node [style=none] (15) at (1, 1) {};
		\node [style=none] (16) at (-0.5, 0.5) {};
		\node [style=none] (17) at (1, 0.5) {};
		\node [style=none] (18) at (-0.5, -0.5) {};
		\node [style=none] (19) at (1, -0.5) {};
		\node [style=none] (20) at (2, 0) {};
		\node [style=none] (21) at (2, 1.5) {};
		\node [style=none] (22) at (2, 2) {};
		\node [style=none] (23) at (3.5, 0) {};
		\node [style=none] (24) at (3.5, -0.5) {};
		\node [style=none] (25) at (3.5, 2) {};
		\node [style=none] (26) at (1, 1.5) {};
		\node [style=none] (27) at (3.5, 1.5) {};
		\node [style=none] (28) at (2, -0.5) {};
		\node [style=none] (29) at (1, 0) {};
		\node [style=none] (30) at (4.5, 0) {};
		\node [style=none, font={\Large}] (31) at (2.75, -1.5) {$\bullet$};
		\node [style=none] (32) at (-2.25, -0.5) {};
		\node [style=none] (33) at (2.75, -0.5) {};
		\node [style=none, font={\footnotesize}] (34) at (-2.25, 0.75) {$\map{SWAP}$};
		\node [style=none] (35) at (0.25, 1.5) {$\map{\widetilde{A}}$};
		\node [style=none] (36) at (-2.25, 0.75) {};
		\node [style=none, font={\footnotesize}] (37) at (2.75, 0.75) {$\map{SWAP}$};
		\node [style=none] (38) at (0.25, 0) {$\map{\widetilde{B}}$};
		\node [style=none] (39) at (0.25, 0) {};
		\node [style=none] (40) at (0.25, 0) {};
		\node [style=none] (41) at (0.25, 0) {};
		\node [style=none] (43) at (-4.5, 0) {\hspace{0.07cm}vac};
		\node [style=none, ground, xshift=-0.45em] (46) at (4.75, 0) {};
		\node [style=none] (47) at (4.75, -1.5) {};
		\node [style=none] (48) at (-5, -1.5) {};
		\node [style=none] (49) at (-3, -1.25) {};
		\node [style=none, font={\Large}] (50) at (-2.25, -1.5) {$\bullet$};
		\node [style=none] (51) at (4.75, 1.5) {};
		\node [style=none] (57) at (-4, 0.5) {};
		\node [style=none] (58) at (-4, -0.5) {};
		\node [style=none] (59) at (-4, -1) {};
	\end{pgfonlayer}
	\begin{pgfonlayer}{edgelayer}
		\draw (2.center) to (4.center);
		\draw (3.center) to (5.center);
		\draw (2.center) to (3.center);
		\draw (4.center) to (5.center);
		\draw (1.center) to (6.center);
		\draw (0.center) to (7.center);
		\draw (8.center) to (10.center);
		\draw (9.center) to (11.center);
		\draw (12.center) to (14.center);
		\draw (14.center) to (15.center);
		\draw (15.center) to (13.center);
		\draw (13.center) to (12.center);
		\draw (16.center) to (17.center);
		\draw (17.center) to (19.center);
		\draw (19.center) to (18.center);
		\draw (18.center) to (16.center);
		\draw (22.center) to (25.center);
		\draw (28.center) to (24.center);
		\draw (22.center) to (28.center);
		\draw (25.center) to (24.center);
		\draw [in=180, out=0] (26.center) to (21.center);
		\draw (29.center) to (20.center);
		\draw (23.center) to (30.center);
		\draw (33.center) to (31.center);
		\draw (48.center) to (47.center);
		\draw (32.center) to (50.center);
		\draw (27.center) to (51.center);
		\draw [bend left=270, looseness=3.50] (57.center) to (58.center);
		\draw (57.center) to (58.center);
	\end{pgfonlayer}
\end{tikzpicture}
	\caption{\label{fig:swap} {\em A quantum circuit realising the operational superposition of two channels using  controlled-{\tt SWAP} operations}.  The state of the input system  goes through either channel $\widetilde{\map A}$ or $\widetilde{\map B}$ depending on the state of the control system (bottom wire),  while  the other channel takes the vacuum as input. At the output, the system in the vacuum is discarded.} 
\end{figure}
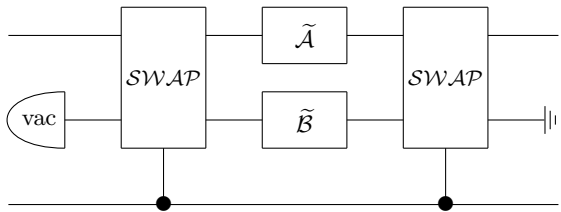

 To analyse these scenarios, it is convenient to reformulate the superposition of channels (\ref{ctrlchan}) as a controlled circuit,  where a control qubit determines  which  of the two channels $\widetilde {\map A}$ and $\widetilde {\map B}$ receives a particle as its input, and which one receives the vacuum, as illustrated  in  Figure \ref{fig:swap}.  One  choice of encoding is to initialise the control system  in a fixed  state $\omega$.  In this way, the encoding channel is 
\begin{align}
\map E  (\rho)   =  {\tt CSWAP}   \Big(   \rho  \otimes |{\rm vac}\>\<{\rm vac}|  \otimes \omega \Big)  \, {\tt CSWAP}  \, ,
\end{align}
where ${\tt CSWAP}  =  I\otimes I \otimes |0\>\<0|  +  {\tt SWAP} \otimes |1\>\<1|$ is the control-swap operator. 
Since the vacuum is orthogonal to all states used to encode the message, the encoding operation $\map E$ satisfies the no-signalling condition (\ref{nosig}):    no information flows from the  information-carrying system (represented by the two top wires in  Figure \ref{fig:swap}) and the control system (represented by the bottom wire).   

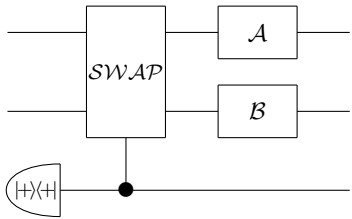
\begin{figure}
	\centering
		\begin{tikzpicture}[scale=1.4, circuit ee IEC]
	\begin{pgfonlayer}{nodelayer}
		\node [style=none] (0) at (3.25, 1) {};
		\node [style=none] (1) at (1.75, 2) {};
		\node [style=none] (2) at (-1.25, 0.5) {};
		\node [style=none] (3) at (-0.75, 1) {};
		\node [style=none] (4) at (0.75, 1.5) {};
		\node [style=none] (5) at (0.75, 3.5) {};
		\node [style=none] (6) at (2.5, 1.5) {$\map{B}$};
		\node [style=none] (7) at (-0.75, 3.5) {};
		\node [style=none] (8) at (0.75, 1) {};
		\node [style=none] (9) at (-0.75, 1.5) {};
		\node [style=none] (11) at (-1.25, -0.5) {};
		\node [style=none] (13) at (-2.25, 1.5) {};
		\node [style=none] (14) at (4.25, 1.5) {};
		\node [style=none] (15) at (-1.25, 0) {};
		\node [style=none] (16) at (4.25, 0) {};
		\node [style=none, font={\footnotesize}] (17) at (0, 2.25) {$\map{SWAP}$};
		\node [style=none] (18) at (1.75, 1.5) {};
		\node [style=none] (19) at (3.25, 1.5) {};
		\node [style=none, font={\scriptsize}] (20) at (-1.75, 0) {\hspace{0.06cm}$\ket{\!+\!}\!\!\bra{\!+\!}$};
		\node [style=none] (21) at (3.25, 2) {};
		\node [style=none, font={\Large}] (22) at (0, 0) {$\bullet$};
		\node [style=none] (23) at (3.25, 1.5) {};
		\node [style=none] (24) at (0, 1) {};
		\node [style=none] (25) at (1.75, 1) {};
		\node [style=none] (26) at (2.5, 3) {$\map{A}$};
		\node [style=none] (28) at (3.25, 3) {};
		\node [style=none] (29) at (3.25, 3) {};
		\node [style=none] (30) at (1.75, 3) {};
		\node [style=none] (31) at (3.25, 2.5) {};
		\node [style=none] (32) at (3.25, 3.5) {};
		\node [style=none] (33) at (1.75, 2.5) {};
		\node [style=none] (34) at (1.75, 3.5) {};
		\node [style=none] (35) at (4.25, 3) {};
		\node [style=none] (36) at (-2.25, 3) {};
		\node [style=none] (37) at (-0.75, 3) {};
		\node [style=none] (38) at (0.75, 3) {};
	\end{pgfonlayer}
	\begin{pgfonlayer}{edgelayer}
		\draw (7.center) to (5.center);
		\draw (5.center) to (8.center);
		\draw (8.center) to (3.center);
		\draw (3.center) to (7.center);
		\draw (24.center) to (22.center);
		\draw (13.center) to (9.center);
		\draw (15.center) to (16.center);
		\draw [bend right=90, looseness=3.50] (2.center) to (11.center);
		\draw (2.center) to (11.center);
		\draw (1.center) to (21.center);
		\draw (21.center) to (0.center);
		\draw (0.center) to (25.center);
		\draw (25.center) to (1.center);
		\draw (4.center) to (18.center);
		\draw (23.center) to (14.center);
		\draw (34.center) to (32.center);
		\draw (32.center) to (31.center);
		\draw (31.center) to (33.center);
		\draw (33.center) to (34.center);
		\draw (28.center) to (35.center);
		\draw (36.center) to (37.center);
		\draw (38.center) to (30.center);
	\end{pgfonlayer}
\end{tikzpicture}
\caption{\label{fig:forbidden1} 
{\em Bypassing noise via the {\tt SWAP} test.} A two-qubit message undergoes a {\tt SWAP}   controlled by the path.   Using the {\tt SWAP} test \cite{buhrman2001swap}, this encoding can transfer one bit from the message to the path, thus bypassing channels $\map A$ and $\map B$.  Such a signalling encoding is forbidden by our model.}
\end{figure}

Extrapolating from Figure \ref{fig:swap}, one could think of  a similarly-looking setup, where {\em two} particles are sent to the input ports of channels $\widetilde {\map A}$ and $\widetilde {\map B}$, as in Figure \ref{fig:forbidden1}.   In this case, using controlled-$\tt SWAP$ operations does lead to signalling, due to the possibility of performing a  {\tt SWAP} test \cite{buhrman2001swap}.   Explicitly,   suppose  that the two inputs are two  qubits, prepared either in a symmetric state $|\Phi^+\>   =  {\tt SWAP}  |\Phi^+\>$  or in an antisymmetric state $|\Phi^{-}\>  =   -   {\tt SWAP} |\Phi^-\>$.       Then,  the control swap operation transforms the input states $|\Phi^{\pm} \>  \otimes |+\>$ into the states    $|\Phi^{\pm}\>  \otimes |\pm\>$, transferring one bit of  information from the message to the control qubit.   This bit reaches the receiver {\em independently of channels $\map A$ and $\map B$}.  Since the control qubit is unaffected by noise, this type of encoding bypasses any noisy process occurring on the system.  Our model rules out such bypassing, allowing us to highlight non-trivial ways in which the superposition of communication devices can boost the communication from sender to receiver.

\begin{figure}
	\centering
	\begin{tikzpicture}[scale=1.5, circuit ee IEC]
	\begin{pgfonlayer}{nodelayer}
		\node [style=none, font={\Large}] (0) at (0, 0) {$\bullet$};
		\node [style=none] (1) at (0, 1) {};
		\node [style=none] (2) at (-0.5, 2) {};
		\node [style=none] (3) at (-0.5, 1) {};
		\node [style=none] (4) at (0.5, 1) {};
		\node [style=none] (5) at (0.5, 2) {};
		\node [style=none] (6) at (-0.5, 1.5) {};
		\node [style=none] (7) at (-2, 1.5) {};
		\node [style=none] (8) at (-1, 0) {};
		\node [style=none] (9) at (0.5, 1.5) {};
		\node [style=none] (10) at (3, 1.5) {};
		\node [style=none] (11) at (4, 0) {};
		\node [style=none] (12) at (-1, 0.5) {};
		\node [style=none] (13) at (-1, -0.5) {};
		\node [style=none] (14) at (3, 2) {};
		\node [style=none] (15) at (1.5, 2) {};
		\node [style=none] (16) at (3, 1) {};
		\node [style=none] (17) at (1.5, 1.5) {};
		\node [style=none] (18) at (1.5, 1) {};
		\node [style=none] (19) at (2.5, 1) {};
		\node [style=none] (20) at (3, 1.5) {};
		\node [style=none] (21) at (4, 1.5) {};
		\node [style=none] (22) at (0, 1.5) {$\map{X}$};
		\node [style=none] (23) at (2.25, 1.5) {$\map{A}$};
		\node [style=none, font={\scriptsize}] (24) at (-1.5, 0) {\hspace{0.06cm}$\ket{\!+\!}\!\!\bra{\!+\!}$};
	\end{pgfonlayer}
	\begin{pgfonlayer}{edgelayer}
		\draw (2.center) to (5.center);
		\draw (5.center) to (4.center);
		\draw (4.center) to (3.center);
		\draw (3.center) to (2.center);
		\draw (1.center) to (0.center);
		\draw (7.center) to (6.center);
		\draw (8.center) to (11.center);
		\draw [bend right=90, looseness=3.50] (12.center) to (13.center);
		\draw (12.center) to (13.center);
		\draw (15.center) to (14.center);
		\draw (14.center) to (16.center);
		\draw (16.center) to (18.center);
		\draw (18.center) to (15.center);
		\draw (9.center) to (17.center);
		\draw (20.center) to (21.center);
	\end{pgfonlayer}
\end{tikzpicture}
\caption{\label{fig:forbidden}  {\em Bypassing noise  via phase kickback.}  A single-qubit message undergoes a bit flip $\map{X}$ controlled by the path.  This encoding can transfer one bit  from the message to the path, exploiting the phase kickback property of the ${\tt CNOT}$ gate  \cite{cleve1998quantum}. 
The transfer of information to the path trivialises the communication protocol, making the communication channel $\map A$  irrelevant.  }
\end{figure}
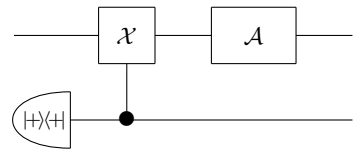

Another example of a  communication protocol excluded in our model is a protocol that uses  the  $\tt CNOT$ gate in the encoding stage.    Suppose that a $\tt NOT$ operation is applied to a target qubit, depending on the state of a control qubit, as in Figure \ref{fig:forbidden}.   The roles of the control and the target in the $\tt CNOT $ gate can be exchanged, as shown explicitly by the relation
  \begin{align}
 \nonumber  {\tt CNOT}     &=     I  \otimes |0\>\<0|  +   X  \otimes  |1\>\<1|  \\
  &    =  |+\>\<+| \otimes I +  |-\>\<-| \otimes Z  \, ,
  \end{align}
  expressing the phase kickback of the $\tt CNOT$ gate \cite{cleve1998quantum}. 
Hence, a $\tt CNOT$ applied to the states $|\pm  \>  \otimes |+\>$, will generate the states $|\pm  \> \otimes |\pm\>$, transferring one bit of information from the target to the control.    This is another example of signalling encoding that can be used to transfer classical information, independently of the noisy channel acting on the target qubit.   Also in this case, the ability to communicate does not reveal any interesting feature of the original channel,  and instead it is   an artefact of the signalling from the target to the control.  
  
\begin{figure}
	\centering
	\begin{tikzpicture}[scale=1.14]
	\begin{pgfonlayer}{nodelayer}
		\node [style=none] (0) at (-1, 1) {};
		\node [style=none] (1) at (1, 1) {};
		\node [style=none] (2) at (-1.75, -1) {$P$};
		\node [style=none] (3) at (1, 1.75) {};
		\node [style=none] (4) at (-1.75, 1.5) {$M$};
		\node [style=none] (5) at (-1, 0.25) {};
		\node [style=none] (6) at (1.75, 1.5) {$M$};
		\node [style=none] (7) at (-1, 1.75) {};
		\node [style=none, font={\large}] (8) at (0, 1) {$\map{A}$};
		\node [style=none] (9) at (1, 0.25) {};
		\node [style=none] (10) at (2.5, 1) {};
		\node [style=none] (11) at (2.5, -1.75) {};
		\node [style=none] (12) at (6, 1.75) {};
		\node [style=none] (13) at (2.5, 1.75) {};
		\node [style=none] (14) at (6, -1.75) {};
		\node [style=none, font={\large}] (15) at (4.25, 1) {$\map{D}$};
		\node [style=none] (16) at (2.5, -0.5) {};
		\node [style=none] (17) at (6, 1) {};
		\node [style=none] (18) at (6, -1) {};
		\node [style=none] (19) at (-6, 1) {};
		\node [style=none] (20) at (-6, -1.75) {};
		\node [style=none] (21) at (-2.5, 1.75) {};
		\node [style=none] (22) at (-6, 1.75) {};
		\node [style=none] (23) at (-2.5, -1.75) {};
		\node [style=none, font={\large}] (24) at (-4.25, 1) {$\map{E}$};
		\node [style=none] (25) at (-6, -1) {};
		\node [style=none] (26) at (-2.5, 1) {};
		\node [style=none] (27) at (-2.5, -0.5) {};
		\node [style=none] (28) at (-7.5, 1) {};
		\node [style=none] (29) at (7.5, 1) {};
		\node [style=none] (30) at (-6.75, 1.5) {$Q$};
		\node [style=none] (31) at (6.75, 1.5) {$Q$};
		\node [style=none] (32) at (-4.5, -0.5) {};
		\node [style=none] (33) at (-3, -0.5) {};
		\node [style=none] (34) at (4.5, -0.5) {};
		\node [style=none] (35) at (3, -0.5) {};
		\node [style=none] (36) at (-3.75, -0.5) {$\map{U}_i$};
		\node [style=none] (37) at (3.75, -0.5) {$\map{U}_i^{\dagger}$};
		\node [style=none] (38) at (-3.75, 0.25) {};
		\node [style=none] (39) at (6, 1) {};
		\node [style=none] (40) at (4.5, -0.5) {};
		\node [style=none] (41) at (2.5, 1) {};
		\node [style=none] (42) at (3.75, 0.25) {};
		\node [style=none] (43) at (2.5, 1) {};
		\node [style=none] (44) at (3.75, 0.25) {};
		\node [style=none] (45) at (2.5, 1) {};
		\node [style=none] (46) at (3.75, 0.25) {};
		\node [style=none] (47) at (-5.5, -0.25) {$P$};
		\node [style=none] (48) at (5.5, -0.25) {$P$};
	\end{pgfonlayer}
	\begin{pgfonlayer}{edgelayer}
		\draw [line width=1pt] (7.center) to (3.center);
		\draw [line width=1pt] (3.center) to (9.center);
		\draw [line width=1pt] (9.center) to (5.center);
		\draw [line width=1pt] (5.center) to (7.center);
		\draw [line width=1pt] (13.center) to (12.center);
		\draw [line width=1pt] (12.center) to (14.center);
		\draw [line width=1pt] (14.center) to (11.center);
		\draw [line width=1pt] (11.center) to (13.center);
		\draw [line width=1pt] (22.center) to (21.center);
		\draw [line width=1pt] (21.center) to (23.center);
		\draw [line width=1pt] (23.center) to (20.center);
		\draw [line width=1pt] (20.center) to (22.center);
		\draw [line width=1pt] (26.center) to (0.center);
		\draw [line width=1pt] (27.center) to (16.center);
		\draw [line width=1pt] (1.center) to (10.center);
		\draw [line width=1pt] (28.center) to (19.center);
		\draw [line width=1pt] (17.center) to (29.center);
		\draw [line width=1pt, bend left=90, looseness=1.75] (32.center) to (33.center);
		\draw [line width=1pt, bend right=90, looseness=1.75] (32.center) to (33.center);
		\draw [line width=1pt, bend left=90, looseness=1.75] (35.center) to (34.center);
		\draw [line width=1pt, bend right=90, looseness=1.75] (35.center) to (34.center);
		\draw [line width=1pt, in=165, out=-15, looseness=0.75] (19.center) to (32.center);
		\draw [line width=1pt] (33.center) to (27.center);
		\draw [line width=1pt, dotted, in=180, out=75, looseness=0.75] (38.center) to (26.center);
		\draw [line width=1pt, in=15, out=-165, looseness=0.75] (39.center) to (40.center);
		\draw [line width=1pt] (16.center) to (35.center);
		\draw [line width=1pt, dotted, in=0, out=105, looseness=0.75] (46.center) to (45.center);
	\end{pgfonlayer}
\end{tikzpicture}
	\caption{\label{fig:sneaky} {\em Bypassing a complete dephasing channel via a quantum one-time pad.} The sender encodes a qubit $Q$ into a four-dimensional message system $M$ and a two-dimensional  path system $P$, using a quantum version of the one-time pad  \cite{ambainis2000private}.  The state of the qubit $Q$ is first transferred to the path $P$, and then rotated by a random  Pauli  gate $U_i$, with index $i$ chosen uniformly at random in the set $\{0,1,2,3\}$.  The  index $i$  is written on the  state $|i\>$ of the message  $M$, as in  Equation  (\ref{signalandlock}). The message then goes through the completely dephasing channel $\map A$, which however does not affect the states of the basis $\{ |i\>\}_{i=0}^3$.  Thanks to this fact, the receiver can unlock the quantum information encoded in the path  by reading the value $i$ from $M$ and performing the correction operation $U_i^\dag$ on $P$.  } 
\end{figure}
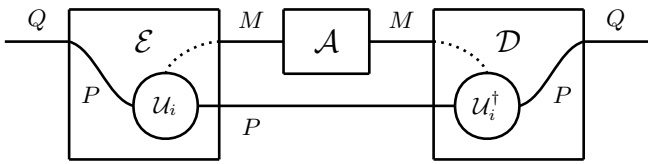  

The no-signalling condition (\ref{nosig}) prevents communication protocols that encode information  in the path, in such a way that this information can be retrieved even if the message is lost.  However, there exist protocols that encode information in the path, but hide it  in such a way that the information can only be retrieved if one has access to a ``key'', written in the message.  For example, consider the encoding channel  
\begin{align}\label{signalandlock}
\nonumber
\map E (\rho_Q)  &=  \frac {1}{4} \big[ |0\>\<0|_M  \otimes \rho_P  + |1\>\<1|_M \otimes (X\rho X)_P \\
 &+  |2\>\<2|_M  \otimes  (Y\rho Y)_P  +  |3\>\<3|_M \otimes (Z\rho Z)_P  \big] \, ,  
\end{align} 
 where the input system $Q$ is a qubit, the internal degree of freedom $M$ is a four-dimensional system, and the path $P$ is a qubit.     If system $M$ is discarded, one obtains the depolarising channel $\Tr_M  [\map E (\rho)]  =  (\rho  +  X\rho X  +  Y\rho Y+  Z\rho Z)/4  =  I/2$, and therefore the no-signalling condition (\ref{nosig}) is satisfied. Still, one may argue that information {\em has} been encoded in the path, although the message system is necessary to unlock it.    With this kind of encoding, the sender  could send the message through the completely dephasing channel $\map A (\rho)  =  \sum_{i=0}^3  |i\>\<i|  \rho  |i\>\<i|$, and the receiver would still be able to recover the quantum state $\rho$ without any error.  Also in this case, it appears that the path has been used to circumvent the  channel $\map A$, allowing the transmission of quantum information through the path degree of freedom. This protocol is illustrated in Figure \ref{fig:sneaky}.
  In our model we forbid this type of transmission  by demanding that the encoding operation does not create any correlations between $M$ and $P$.  The product encoding condition (\ref{productenc}) guarantees that the sender does not use the path to sneak information through the path,  even in an indirect way as in Equation  (\ref{signalandlock}).

\subsubsection{Communication capacities assisted by superposition of paths}  

\begin{figure*}
	\centering
    \input{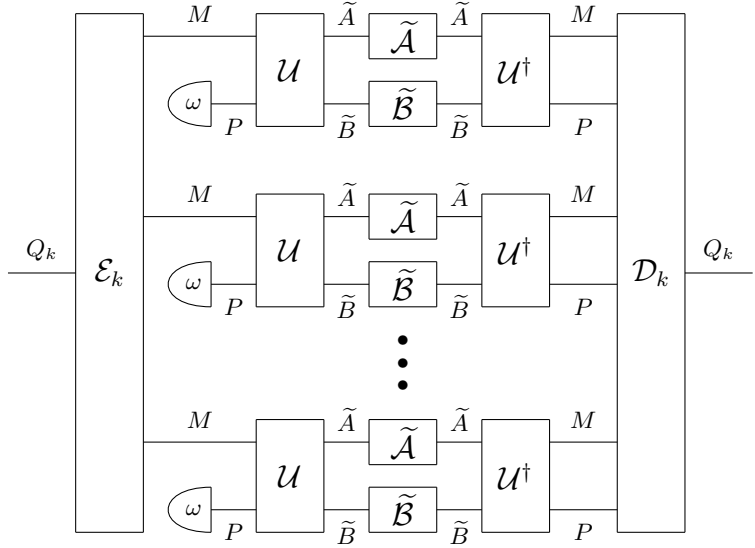}
    \caption{\label{fig:many_particles} {\em Communication of a message encoded in $k$ particles, with each particle travelling in a superposition of two paths. } The sender encodes the state of a quantum system $Q_k$ onto the internal state of $k$ particles, using a global encoding operation $\map E_k$. 
    Each particle is sent through one of the channels $\widetilde{\map  A}$ and $\widetilde {\map  B}$, with the choice of the channel controlled by the state $\omega$ of the particle's external degree of freedom.  Finally, the receiver applies a global decoding operation  $\map D_k$ on the output particles, returning a quantum state in system $Q_k$. 
    } 
\end{figure*}

In our communication model, the sender is only allowed to use product encodings, of the form $\map E (\rho)  = \map M(\rho)  \otimes \omega$, where $\omega$ is a fixed state of the path. For simplicity, we consider the case where the initial system $Q$ has the same dimension as the internal degree of freedom $M$, so that  $\map M$ can be chosen to be  the identity channel.  As we will see this simplification can be made without loss of generality, because the  framework of channel capacities already includes global encoding operations, in which map $\map M$ can be  incorporated  (cf. Figure  \ref{fig:many_particles}).  The evolution of the   internal degree of freedom is then described by  the {\em effective channel} $\map S_{\widetilde{\map A},  \widetilde {\map B},\omega}$  given by 
\begin{align}
\map S_{\widetilde{\map A},  \widetilde {\map B},\omega} (\rho)    :=  \map S_{\widetilde{\map A},  \widetilde {\map B}} (\rho \otimes \omega) \, ,
\end{align}
where $\map S_{\widetilde{\map A},  \widetilde {\map B}}$ is the superposition defined in Equation (\ref{SU}). 
One can then study various communication capacities of the effective channel, considering the asymptotic scenario when the channel is used $k$ times, with $k\to \infty$, as illustrated in Figure \ref{fig:many_particles}.    

 As in standard quantum Shannon theory, one can consider several  types of capacities, such as the classical  (quantum) capacity,  corresponding to the maximum number of bits (qubits) that can be reliably transmitted per use of the channel  $\map S_{\widetilde {\map A}, \widetilde {\map B}, \omega}$, in the limit of asymptotically many uses.

An interesting special case is $\widetilde{\map A} =\widetilde {\map B}$, meaning that the particle can travel through two identical transmission lines.   In this case, the (classical or quantum) capacity of the channel $\map S_{\widetilde {\map A}, \widetilde {\map A}, \omega}$ is a new type  of (classical or quantum) capacity of the channel $\widetilde {\map A}$. We call it the (classical or quantum) {\em two-path capacity}.    
  More generally,  the sender could send a single particle along one of $N$ identical transmission lines, and one could evaluate  the communication capacity of the resulting channel.  Once the path state $\omega$ has been optimised, the capacity is a non-decreasing function of $N$, and the base case $N=1$ corresponds to the usual channel capacity considered in quantum Shannon theory.    The {\em $N$-path capacity}, where $N$  is the number of paths that are coherent with each other, can be regarded as the amount of information transmitted  per particle in the asymptotic limit where $k\to \infty$ particles are sent through the $N$ transmission lines. In the next Subsection we will see examples where increasing $N$ leads to interesting capacity enhancements. 
  
In passing, we mention that the scenario of Figure \ref{fig:many_particles} lends itself to several generalisations. For example, instead of assuming that the path of each particle is in the same state $\omega$, one could allow different states $\omega_1\otimes\omega_2\otimes\cdots \otimes \omega_k$, or even generally correlated states $\omega_{12\dots k}$. Likewise, the number of paths available for each particle could be different from particle to particle. Finally, instead of taking the limit $k\to \infty$ for fixed $N$, one could consider different asymptotic regimes where both $k$ and $N$ tend to infinity together.

\subsection{Examples}

 The model defined in the previous Subsection   allows for new communication protocols that are not possible in the standard quantum  Shannon theory. Some examples that fit into this model have been recently presented in \cite{abbott2018communication}.    Here we illustrate a few new examples, some of which exhibit rather striking features in the limit of large numbers of paths $N$. 
 
 \subsubsection{Classical communication through pure erasure channels} 
  Suppose that a sender and a receiver have access to two communication channels, $\map A$ and $\map B$, each of which acts  on the message as a complete erasure   $\map E (\rho)   =  |\psi_0\>\<\psi_0|$, {\em i.e.}\  $\map A  = \map B  =\map E$.     Clearly, no information can be sent to the receiver using a conventional communication protocol where  the two channels $\map A$ and $\map B$ are  in a definite configuration.  Now, suppose that the communication devices used in the protocol  can take the vacuum as the input, and are described by a vacuum extension  $\widetilde {\map E}$ with Kraus operators  $\widetilde E_i  =  |\psi_0\>  \<i|  \oplus  \alpha_i  \,  |{\rm vac}\>\<{\rm vac}|$, for $i  \in  \{1,\dots, d\}$ and for some amplitudes $\{\alpha_i\}$. 
     Then, the sender can transmit  the message in a superposition of travelling through $\map A$ and travelling through $\map B$, initialising the path  in the state $|+\>  =  (|0\> +|1\>)/\sqrt{2}$.     
The output state,  computed according to   (\ref{ctrlchan}), is 
\begin{align}\label{erasureonthepath}
 \map S_{\widetilde {\map E}, \widetilde{\map E}} (\rho\otimes |+\>\<+|)    &=     |\psi_0\>\<\psi_0|  \otimes  \left( p    |+\>\<+|     + (1-p)   \frac{I}2 \right)  \\   \nonumber p   & =  \<\alpha|  \rho |\alpha\>, \quad |\alpha\> = \sum_{i} \alpha_{i} | i \>,      
\end{align} 
  Since the output state depends on the input, the receiver will be able to decode some of the information in the original message.  Precisely, the overall channel from the sender's input $\rho$ to the receiver's output is a measure and re-prepare channel, equivalent to the orthogonal measurement with projectors $\{   |\alpha\>\<\alpha|   ,  I-  |\alpha\>\<\alpha|\}$ 
 followed by a re-preparation of the states $|+\>\<+|$ or $I/2$, depending on the outcome.  In turn, this channel is equivalent to a classical binary asymmetric channel, with $0$  mapped deterministically to $0$, and $1$ mapped to a uniform mixture of $0$ and $1$.   The capacity of this channel, also known as the $Z$ channel, is $\log_2  (5/4)\approx .32$  \cite{richardson2008modern} and can be achieved using polar codes \cite{honda2013polar}.   In the quantum setting,  the sender has only to encode   0 in the  state $|\alpha\>$ and 1 in an orthogonal state $|\alpha_\perp\>$, and then use the optimal classical  code. 
 
Note that the possibility to communicate with the vacuum-extended erasure channel $\widetilde {\map E}$ depends essentially on the fact that such a channel preserves coherence between the message and the vacuum.  If we had chosen a  vacuum extension without coherence, such as the extension with Kraus operators $\widetilde E_i'    :=  |\psi\>\<i  |  \oplus 0_{\rm Vac}$, $i\in  \{1,\dots, d\}$  and $\widetilde E'_{d+1}:  =  0_{A} \oplus |{\rm vac}\>\<{\rm vac} |$, the overall channel would be equivalent to a measurement on the path followed by an erasure channel on the message.   The output state, computed according to  Equation (\ref{ctrlchan}),   would have been 
\begin{align}
\map S_{\widetilde{\map E}' , \widetilde {\map E}'}  (\rho  \otimes |+\>\<+|)   =  |\psi_0\>\<\psi_0|  \otimes \frac I 2 \, ,  
\end{align}
which is independent of the input state $\rho$, and therefore prevents any kind of communication. 
  In this  and the following examples, the resource that enables communication is the coherence in the initial state of the path, and the availability of communication devices that preserve such coherence.  The advantage of having a device that can act coherently on the vacuum is similar in spirit to the advantages of counterfactual quantum computation \cite{mitchison2001counterfactual}, cryptography \cite{noh2009counterfactual}, and communication \cite{salih2013protocol,vaidman2014comment}. Another related effect is two-way classical communication using one-particle states\cite{del2018two}. 
   
\subsubsection{Quantum communication through entanglement-breaking channels}

In the previous example quantum communication is not possible, because the overall channel  $\map S_{\widetilde {\map E}, \widetilde{\map E}}$   in Equation (\ref{erasureonthepath}) is entanglement-breaking  \cite{ruskai2003entanglement}, and all such channels have zero quantum capacity \cite{holevo2001evaluating}. 
Examples where the superposition of channels enables quantum communication do exist, however. 
Before showing an explicit example, it is useful to get some general insight into the superposition of two identical channels.  Suppose that a message propagates in superposition through  two transmission lines, each described by the  vacuum extension  $\widetilde {\map A}$ of  channel $\map A$.  Assuming that the  path is initialised in the state $|+\>$, a message encoded into the input state $\rho$  is transformed into the output state
\begin{align} \label{cleaner}
\nonumber
 \map S_{\widetilde {\map A}, \widetilde{\map A}}  (\rho  \otimes |+\>\<+|)  &=    \frac{ \map A  (\rho)  + F  \rho  F^\dag   }2  \otimes |+\>\<+|  \\
&+  \frac{  \map A(\rho)  -  F\rho F^\dag}2 \otimes |-\>\<-|  \, ,
\end{align}  
where $F  :=  \sum_i  \overline{\alpha}_i \,  A_i$  depends on the specific vacuum extension describing the communication devices.     We will call $F$ the \emph{vacuum interference} operator.

 By measuring in the Fourier basis $\{|+\>  ,  |-\>\}$, the receiver can separate the two quantum operations  $\map Q_{\pm}   =  ( \map A  \pm  F\cdot F^\dag)/2$.    The outcome $+$ heralds a constructive interference among the noisy processes along the two paths, while the outcome $-$ heralds a destructive interference.  This observation  is the working principle  of the error filtration technique of Gisin {\em et al} \cite{gisin2005error}, which allows probabilistically reducing the noise by selecting events where one of the two operations  $\map Q_\pm$ is less noisy than the original channel.  
 
Equation (\ref{cleaner})  offers several important insights.   First, it shows that the optimal decoding strategy consists in measuring the path and conditionally operating on the message. 
Second, it shows that it is sometimes possible to obtain a noiseless  probabilistic transmission of the quantum state, thanks to the destructive interference term $\map A  - F\cdot F^\dag$.  For example, consider the  complete decoherence   channel $\map D  (\rho)  =  |0\>\<0|\rho|0\>\<0|  +  |1\>\<1|  \rho |1\>\<1|$, which {\em per se} cannot transmit any quantum information.      For   a vacuum extension   with Kraus operators  $\{|i\>\<i|  \oplus  \frac 1 {\sqrt 2}  \,  |{\rm vac}\>\<{\rm vac}| \}_{i=0}^1$,   the vacuum interference operator is $F  =  I/\sqrt 2$, and the destructive interference term is proportional to the unitary gate $Z  =  |0\>\<0| -  |1\>\<1|$, which can be undone by the receiver.     The probability of  destructive interference is $1/4$, meaning that the superposition of channels allows us to transmit a single qubit  25\%  of the times. In the remaining cases, one has constructive interference,  and the conditional evolution of the message amounts to the 
the channel $ \frac{2}{3}( \map A  (\rho)  + F  \rho  F^\dag ) =  \frac{2}{3} ( |0\>\<0|\rho|0\>\<0|  +  |1\>\<1|  \rho |1\>\<1| + \rho /2 )  = \frac 23 \rho  +  \frac 13 Z\rho Z$, 
whose quantum capacity is $1-  h(1/3)\approx .08$ with $h(x)  =  -x \log_2 x  -  (1-x)  \log(1-x)$ being the binary entropy \cite{wilde2013quantum}.   

It is worth noting that  the superposition of independent noisy channels never leads to a  noiseless communication channel.  Indeed,  in order for the superposition channel  (\ref{cleaner}) to be perfectly correctable, both maps $\map A  \pm  F\rho F^\dag$ must be proportional to unitary channels.  However,  the map    $\map A  + F\cdot F^\dag$ is proportional to a unitary gate if and only if the original channel $\map A$ was a unitary gate itself.   In fact, the same result holds for the superposition of two different channels $\map A$ and $\map B$,  and more generally, of $N$ independent channels: superpositions of independent noisy channels never lead to a noiseless channel, as long as $N$ is a finite number   \cite{chiribella2018indefinite}.

\subsubsection{Perfect communication through asymptotically many paths} It is interesting to see what happens when a single particle is sent through  $N$  identical and independent  transmission lines, each described by the vacuum extension $\widetilde {\map A}$ of some channel $\map A$. 
  Initialising the path in the maximally coherent state $|e_0\>   =  \sum_{k=0}^{N-1}  |k\>/\sqrt 
  N$ we obtain the output state  
\begin{align}\label{muchcleaner}
\nonumber
 \map C (\rho  \otimes |e_0\>\<e_0|)  &=    \frac{ \map A  (\rho)  + (N-1)\, F  \rho  F^\dag   }N  \otimes |e_0\>\<e_0| \\
  &+  \frac{  \map A(\rho)  -  F\rho F^\dag}N \otimes \big(I  - |e_0\>\<e_0|\big)  \, ,
\end{align}  
with $F  :=  \sum_i  \overline{\alpha}_i \,  A_i$.  
Again, the state of the  path is diagonal  in the Fourier basis, and one has the possibility  of constructive and destructive interference.   In the large $N$ limit, the channel tends to become a  mixture of the two quantum operations   $F  \cdot F^\dag$ and $\map A  -  F \cdot F^\dag$.   This limiting behaviour leads to striking results:  
\begin{enumerate}
\item For the pure erasure channel   $\map A   (\rho)   =   \, |\psi_0\> \<  0|  \rho  |0\>  \<\psi_0 |+  |\psi_0\> \<  1|  \rho  |1\>  \<\psi_0 |  \,$, perfect classical communication of one bit is  achieved in the limit $N\to \infty$ if one has access to the vacuum extension with Kraus operators $\widetilde A_0  =  |\psi_0\>  \<0|  \oplus  \alpha_0  \,  |{\rm vac}\>\<{\rm vac}|$ and $\widetilde A_1  =  |\psi_0\>  \<1|  \oplus  \alpha_1  \,  |{\rm vac}\>\<{\rm vac}|$. In this case,  one has has $F =  |\psi_0\>\<\alpha|$ and Equation  (\ref{muchcleaner}) yields $\map C (\rho  \otimes |e_0\>\<e_0|) \to   \<\alpha| \rho  |\alpha\>  \,  |\psi_0\>\<\psi_0|  \otimes  |e_0\>\<e_0|     +  \<\alpha_\perp|  \rho |\alpha_\perp\> \, |\psi_0\>\<\psi_0|  \otimes  \omega_\perp $ with $\<\alpha_\perp|\alpha\>  = 0$ and $\omega_\perp   :=  (  I-  |e_0\>\<e_0|)/(N-1)$.    This channel is equivalent to a measurement on the basis $\{|\alpha\>  ,  |\alpha_\perp\>\}$, followed by preparation of one of the orthogonal states $|e_0\>\<e_0|$ and $\omega_\perp$, depending on the outcome. Since these  states are orthogonal, this channel acts as a perfect channel for communicating classical bits. 
   
  \item   For the complete dephasing   channel  $\map A  (\rho)   =  |0\>\<0|  \rho  |0\>\<0| +  |1\>\<1| \rho  |1\>\<1|$, perfect quantum communication is achieved  for $N\to \infty$ if one has access to the vacuum extension with Kraus operators $|0\>\<0|   \oplus  |{\rm vac}\>\<{\rm vac}|/\sqrt 2$ and $|1\>\<1|    \oplus  |{\rm vac}\>\<{\rm vac}|/\sqrt 2$.  Indeed,  one has $ F=  I/\sqrt 2$ and $\map A   (\rho)-   F\rho F^\dag  =   Z\rho Z/2$.  Hence, both quantum operations $F\cdot F^\dag$ and $\map A -  F \cdot F^\dag$ are proportional to unitary channels.  By measuring the path, the receiver can find out which unitary channel acted and correct it.   
    
\item For the complete depolarising channel $\map A (\rho)  =    (\rho  +  X\rho  X  +  Y\rho Y +  Z\rho Z)/4$, noiseless quantum communication with probability 25\% becomes possible in the $N\to \infty$ limit if one has access to the vacuum extension       with  Kraus operators  $\widetilde A_0  =  ( I  \oplus  1)/2$,  $\widetilde  A_1  =  ( X   \oplus i)/2$, $\widetilde  A_2  =  ( Y   \oplus i)/2$, and $\widetilde  A_3  =  ( Z   \oplus i)/2$.  In this case, the vacuum interference term  is proportional to a unitary gate, as one has $F  = \left(     \cos \theta  \, I  -  i \sin \theta S \right)/2$, with $\cos \theta = 1/2$ and $S   =  (  X+Y +Z)/\sqrt 3$.   Hence, when the measurement on the path heralds the quantum operation $F\cdot F^\dag$, the noiseless transmission of a qubit occurs. 
\end{enumerate}

\section{Communication  through a superposition of correlated channels}\label{sec:alternativeorders}   

Here we extend our communication model to scenarios where the channels on alternative paths are correlated.  We consider first correlations in space, and then correlations in time.

\subsection{Spatially correlated channels}

When a particle travels through a given region, its internal degree of freedom  $M$ interacts with degrees of freedom in that region, which play the role of a local environment  $E$.  Without loss of generality, the interaction can be modelled as a unitary channel $\map V$, acting jointly on $M$ and $E$.    When  no particle is sent in the region,  we assume that the state of the environment remains unchanged.    This can be modelled by defining an extended system $\widetilde A  =  A  \oplus \rm Vac$, with $A  \simeq M$, and by  extending the unitary channel $\map V $ to a unitary  channel $\widetilde {\map V}$ acting on the extended system $\widetilde A\otimes E$, in such a way that 
\begin{align}\label{locvacext} 
\widetilde{\map V}  \Big(  |{\rm vac}\>\<{\rm vac}|  \otimes   \eta  \Big )  =    |{\rm vac}  \>\<{\rm vac}| \otimes   \eta \qquad \forall \eta  \in  \St (E)\, .
\end{align}
 We call $\widetilde {\map V}$ a {\em local vacuum extension} of channel $\map V$, emphasising that it specifies the action of the channel $\map V$ when system $A$ is replaced by  the vacuum, while the environment is in a non-vacuum  sector $E$.  In the terminology of  Definition \ref{def:superposition}, the local vacuum extension $\widetilde {\map V}$ is a coherent superposition of the unitary channel $\map V$ acting on the sector $A\otimes E$ with the identity channel acting on the sector ${\rm Vac} \otimes E$. 
 
 An example of a local vacuum extension is the unitary gate generated by the Hamiltonian $H  =  a_{\st k,  V}^\dag a_{\st k,  V}  \otimes Z$ (in suitable units), representing an interaction between a vertically polarised mode with wavevector $\st k$ and a two-level atom with Pauli operator $Z$. Here, the system $A$ is the polarisation of a single photon, and the environment $E$ is the two-level atom.    The unitary operator $U  =  \exp[i\pi H/2]$ acts as the  identity on the sector ${\rm Vac} \otimes E$ defined by the vacuum state $|{\rm vac}\>  =  |0\>_{\st k, H} \otimes |0\>_{\st k,  V}$.   On the sector $A\otimes E$, the operator $U$ acts as the entangling  gate  $W =  |0\>\<0|\otimes   I  +   \, |1\>\<1| \otimes   i  Z $.     This means that, in general, the interaction with the environment will lead to an irreversible evolution of the polarisation degree of freedom.

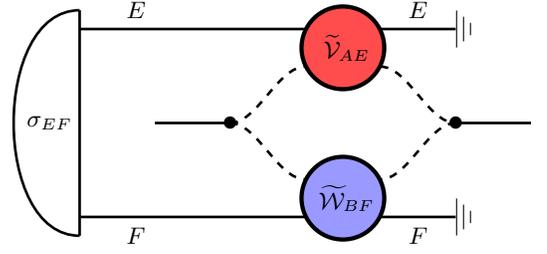
\begin{figure}
	\centering
	\begin{tikzpicture}[scale=2.0, circuit ee IEC]
	\begin{pgfonlayer}{nodelayer}
		\node [style=none, font={\large}] (0) at (0, 0) {$\bullet$};
		\node [style=none, font={\large}] (1) at (3, 0) {$\bullet$};
		\node [style=none] (2) at (-1, 0) {};
		\node [style=none] (3) at (4, 0) {};
		\node [style=none] (4) at (2, 0.75) {};
		\node [style=none] (5) at (2, 1.25) {};
		\node [style=none] (6) at (1, 1.25) {};
		\node [style=none] (7) at (1, 0.75) {};
		\node [style=none] (8) at (1, -0.75) {};
		\node [style=none] (9) at (2, -0.75) {};
		\node [style=none] (10) at (2, -1.25) {};
		\node [style=none] (11) at (1, -1.25) {};
		\node [style=none] (12) at (3, 1.25) {};
		\node [style=none] (13) at (3, -1.25) {};
		\node [style=none] (14) at (-2, 1.5) {};
		\node [style=none] (15) at (-2, -1.5) {};
		\node [style=none] (16) at (-2, -1.25) {};
		\node [style=none] (17) at (-2, 1.25) {};
		\node [style=none] (18) at (-2.5, 0) {\hspace{0.2cm}$\sigma_{EF}$};
		\node [style=none] (19) at (1.5, 1) {\hspace{0.1cm}$\map{\widetilde{V}}_{AE}$};
		\node [style=none] (20) at (1.5, -1) {\hspace{0.1cm}$\map{\widetilde{W}}_{BF}$};
		\node [style=none, ground=none, xshift=-0.45em] (21) at (3.25, 1.25) {};
		\node [style=none, ground=none, xshift=-0.45em] (22) at (3.25, -1.25) {};
		\node [style=none] (23) at (-1.25, 1.5) {$E$};
		\node [style=none] (24) at (-1.25, -1.5) {$F$};
		\node [style=none] (25) at (2.5, 1.5) {$E$};
		\node [style=none] (26) at (2.5, -1.5) {$F$};
	\end{pgfonlayer}
	\begin{pgfonlayer}{edgelayer}
		\draw [fill=red!70!white, ultra thick] (1.5,1) circle [radius=0.55];
	    \draw [fill=blue!40!white, ultra thick] (1.5,-1) circle [radius=0.55];
	
		\draw [line width=1pt] (2.center) to (0.center);
		\draw [line width=1pt] (1.center) to (3.center);
		\draw [line width=1pt](5.center) to (12.center);
		\draw [line width=1pt](10.center) to (13.center);
		\draw [line width=1pt](17.center) to (6.center);
		\draw [line width=1pt](16.center) to (11.center);
		\draw [line width=1pt,in=180, out=180, looseness=1.00] (15.center) to (14.center);
		\draw [line width=1pt](14.center) to (15.center);
		\draw [style=dashed, line width=1pt, in=180, out=0, looseness=0.75] (0.center) to (7.center);
		\draw [style=dashed, line width=1pt,in=180, out=0, looseness=0.75] (0.center) to (8.center);
		\draw [style=dashed, line width=1pt, in=180, out=0, looseness=0.75] (4.center) to (1.center);
		\draw [style=dashed, line width=1pt, in=0, out=180, looseness=0.75] (1.center) to (9.center);
	\end{pgfonlayer}

\end{tikzpicture}
	\caption{\label{fig:correlatedenvironments}{\em Quantum particle travelling through two alternative paths, interacting  with two correlated environments $E$ and $F$.}  The two environments are in a correlated state $\sigma_{EF}$ as the result of a previous interaction.   Each environment interacts locally with the particle via a unitary channel  ($\map{\widetilde{V}}_{AE}$ and $\map{\widetilde{W}}_{BF}$ in the picture).    After the interactions, the environments are discarded and the paths of the particle are recombined.}
\end{figure}

Now, suppose that a particle can be sent through two alternative paths, as in Figure \ref{fig:correlatedenvironments}. Along the two paths, the particle can interact with two environments, $E$ and $F$, which may have previously interacted with each other. As a result of the  interaction, their state $\sigma_{EF}$  may exhibit correlations.  We denote by  $A$ and $B$ the input systems on the two paths ($A\simeq B\simeq M$)  and by $\widetilde{\map V}_{AE}$ and $\widetilde{\map W}_{BF}$ the unitary channels describing the interaction with the environments $E$ and $F$, respectively.  
 Then, the evolution of the particle is described by the effective channel 
\begin{align}\label{correlatedC}
\map C  (\rho  \otimes \omega) \! &  = \map U^\dag \! \left( \!  \Tr_{EF} \! \left\{  (\widetilde {\map V}_{AE}\otimes \widetilde {\map W}_{BF})  [  \map U  (\rho \otimes \omega)   \otimes \sigma_{EF}  ]\right\} \! \right) \! ,
\end{align} 
where $\map U$ is the unitary channel  that implements the isomorphism between the ``particle picture'' $M\otimes P$ and  the ``mode picture''  $(A\otimes {\rm Vac})  \oplus ({\rm Vac} \otimes B)$  [cf.\ Eq.\ (\ref{QPtoSum})].  
Notice that the original superposition of independent channels can be recovered by letting each
channel interact with uncorrelated environments.

 The generalisation to $N\ge 2$ paths is immediate: the local environments on the $N$ paths can be generally in an $N$-partite correlated state, and the interaction between a particle on a path and the corresponding environment is modelled by a  unitary channel.

Let us see an example for $N=2$.  Suppose that the environments $E$ and $F$ are isomorphic, and their initial state   is the classically correlated state $\sigma_{EF}   = \sum_i \, p_i  \,  |i\>\<i|\otimes |i\>\<i|$.  We take the interactions with the two environments to be identical, and to be described by the control unitary channel $\widetilde{\map V}_{AE}  =  \widetilde{\map W}_{BF}$ with unitary operator
\begin{align}
\widetilde V   =\left. \left.\left( \sum_i    U_i  \otimes  |i\>\<i| \right)  \oplus   \right(  |{\rm vac}\>\<{\rm vac}  | \otimes I   \right) \, .
\end{align} 

With these settings, the effective channel (\ref{correlatedC}) takes the form    
\begin{align}\label{productaction}
\map C  = \map R \otimes \map I_P \, ,
\end{align} 
where $\map R=  \sum_i p_i \,  \map U_i$    is the random-unitary channel that performs the unitary gate $U_i$ with probability $p_i$, and $\map I_P$ is the identity on the path.  

 Note how the correlations between the channels on two paths result into a product action of the effective channel (\ref{productaction}), which acts non-trivially only on the internal degree of freedom of the particle. In contrast, if there are no correlations between the channels on the paths, the effective channel generally creates correlations between the path and the internal degree of freedom.   

The above example shows that  preexisting correlations between the environments on the two paths can  realise every random unitary channel on the internal degree of freedom,  leaving  the path degree of freedom untouched.   It is worth stressing that this is not the case  for all quantum channels: 

\begin{prop}\label{prop:erasurecorrelations}
Let $\map E_0 (\cdot)  = |\psi_0\>\<\psi_0|  \, \Tr [\cdot ]$ be a pure erasure channel on the internal degree of freedom $M$. Then, the channel $\map C  =    \map E_0  \otimes \map I_P$ does not admit a realisation of the form (\ref{correlatedC}). 
\end{prop}
The proof is given in Appendix E.  The intuition behind the proof is that   a complete erasure channel transfers all the information from the message to the environment.  Due to the no-cloning theorem, it is impossible to have a complete transfer of information taking place simultaneously in two spatially separated regions, even if the environments in these regions are correlated.   In general, the product channel   $ \map  E_0  \otimes \map I_P$  represents two overlapping paths,  going through the {\em same}   region and interacting with the same environment.

\subsection{Channels with correlations in time: realising the output of the quantum SWITCH}

In the previous Section, we analysed  situations where the correlations in the noise on two paths are induced by pre-existing  correlations of the two local environments.  Here we consider a  scenario where the local environments in two  different regions are uncorrelated, but environments within the same region exhibit correlations in time. 

 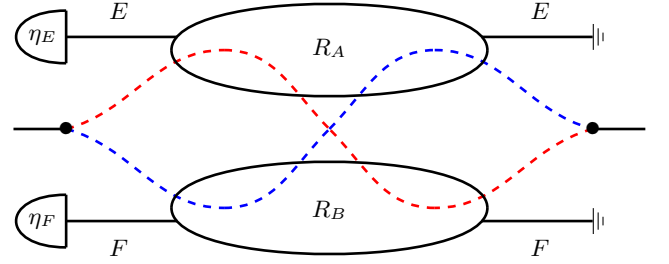
\begin{figure}
 	\centering
	\begin{tikzpicture}[scale=1.4, circuit ee IEC]
	\begin{pgfonlayer}{nodelayer}
		\node [style=none] (0) at (-5, 1.75) {};
		\node [style=none] (1) at (5, 1.75) {};
		\node [style=none] (2) at (5, -1.75) {};
		\node [style=none, xshift=-0.25em] (3) at (3, 1.75) {};
		\node [style=none, xshift=0.25em] (4) at (-3, 1.75) {};
		\node [style=none, xshift=-0.25em] (5) at (3, -1.75) {};
		\node [style=none, xshift=0.25em] (6) at (-3, -1.75) {};
		\node [style=none] (7) at (-5, 2.25) {};
		\node [style=none] (8) at (-5, 1.25) {};
		\node [style=none] (9) at (-5, -1.25) {};
		\node [style=none] (10) at (-5, -2.25) {};
		\node [style=none] (11) at (-5, -1.75) {};
		\node [style=none, font={\large}] (12) at (-5, 0) {$\bullet$};
		\node [style=none] (13) at (-2, 1.5) {};
		\node [style=none, font={\large}] (14) at (5, 0) {$\bullet$};
		\node [style=none] (15) at (0, 0) {};
		\node [style=none] (16) at (-2, -1.5) {};
		\node [style=none] (17) at (2, 1.5) {};
		\node [style=none] (18) at (2, -1.5) {};
		\node [style=none] (19) at (-6, 0) {};
		\node [style=none] (20) at (6, 0) {};
		\node [style=none] (21) at (-5.5, 1.75) {\hspace{0.08cm}$\eta_E$};
		\node [style=none] (22) at (-5.5, -1.75) {\hspace{0.08cm}$\eta_F$};
		\node [style=none] (23) at (-4, 2.25) {$E$};
		\node [style=none] (24) at (4, 2.25) {$E$};
		\node [style=none] (25) at (-4, -2.25) {$F$};
		\node [style=none] (26) at (4, -2.25) {$F$};
		\node [style=none] (27) at (3, 1.5) {};
		\node [style=none] (28) at (-3, 1.5) {};
		\node [style=none] (29) at (3, -1.5) {};
		\node [style=none] (30) at (-3, -1.5) {};
		
		\node [style=none, ground=none, xshift=-0.45em] (35) at (5.25, 1.75) {};
		\node [style=none, ground=none, xshift=-0.45em] (36) at (5.25, -1.75) {};
		\node[style=none, yshift=0.3em] (37) at (0, 1.4) {$R_A$};
		\node[style=none, yshift=-0.3em] (37) at (0, -1.4) {$R_B$};

	\end{pgfonlayer}
	\begin{pgfonlayer}{edgelayer}
		\draw[line width=1pt] (0.center) to (4.center);
		\draw[line width=1pt] (3.center) to (1.center);
		\draw[line width=1pt] (5.center) to (2.center);
		\draw[line width=1pt] [in=180, out=180, looseness=3.25] (7.center) to (8.center);
		\draw[line width=1pt] (7.center) to (8.center);
		\draw [line width=1pt, bend right=90, looseness=3.25] (9.center) to (10.center);
		\draw[line width=1pt] (9.center) to (10.center);
		\draw[line width=1pt] (11.center) to (6.center);
		\draw [line width=1pt, red, style=dashed, in=180, out=15, looseness=1.00] (12.center) to (13.center);
		\draw [line width=1pt, red, style=dashed, in=135, out=0, looseness=1.00] (13.center) to (15.center);
		\draw [line width=1pt, blue, style=dashed, in=180, out=-15, looseness=1.00] (12.center) to (16.center);
		\draw [line width=1pt, blue, style=dashed, in=-135, out=0, looseness=1.00] (16.center) to (15.center);
		\draw [line width=1pt, blue, style=dashed, in=180, out=45, looseness=1.00] (15.center) to (17.center);
		\draw [line width=1pt, blue, style=dashed, in=165, out=0, looseness=1.00] (17.center) to (14.center);
		\draw [line width=1pt, red, style=dashed, in=0, out=-165, looseness=1.00] (14.center) to (18.center);
		\draw [line width=1pt, red, style=dashed, in=-45, out=180, looseness=1.00] (18.center) to (15.center);
		\draw[line width=1pt] (14.center) to (20.center);
		\draw[line width=1pt] (12.center) to (19.center);
		\draw [line width=1pt, bend left=90, looseness=0.50] (28.center) to (27.center);
		\draw [line width=1pt, bend right=90, looseness=0.50] (28.center) to (27.center);
		\draw [line width=1pt, bend left=90, looseness=0.50] (30.center) to (29.center);
		\draw [line width=1pt, bend right=90, looseness=0.50] (30.center) to (29.center);
	\end{pgfonlayer}
\end{tikzpicture}
	\caption{\label{fig:switchregions}  {\em Quantum particle visiting two spatial regions in two alternative orders.} The particle can travel in a superposition of two alternative paths (in red and blue, respectively), visiting either region  $R_A$ before region  $R_B$, or vice-versa.   In each region, the particle interacts with a local environment, initially in the state $\eta_E$ for region $R_A$ and in the state $\eta_F$ for region $R_B$.     The state of each local environment at later times is generally correlated with the state of the environment at earlier times. After the particle emerges from the two regions, the alternative paths are recombined and the environments are discarded.  }
\end{figure}
 
  \begin{figure*}
  	\centering
		\begin{tikzpicture}[scale=1.9, circuit ee IEC]
	\begin{pgfonlayer}{nodelayer}
		\node [style=none] (0) at (-0.5, 0) {};
		\node [style=none] (1) at (-0.5, 1.5) {};
		\node [style=none] (2) at (1, 1.5) {};
		\node [style=none] (3) at (1, 0) {};
		\node [style=none] (4) at (1.5, 1) {};
		\node [style=none] (5) at (1.5, 2.5) {};
		\node [style=none] (6) at (3, 2.5) {};
		\node [style=none] (7) at (3, 1) {};
		\node [style=none] (8) at (1.5, 0.5) {};
		\node [style=none] (9) at (3, 0.5) {};
		\node [style=none] (10) at (1.5, -1) {};
		\node [style=none] (11) at (3, -1) {};
		\node [style=none] (12) at (4, 1.5) {};
		\node [style=none] (13) at (5.5, 1.5) {};
		\node [style=none] (14) at (5.5, 0) {};
		\node [style=none] (15) at (4, 0) {};
		\node [style=none] (16) at (-2.5, 0.5) {};
		\node [style=none] (17) at (-2.5, 2.5) {};
		\node [style=none] (18) at (-1, 2.5) {};
		\node [style=none] (19) at (-1, 0.5) {};
		\node [style=none] (20) at (-1, 1) {};
		\node [style=none] (21) at (-2.5, -1) {};
		\node [style=none] (22) at (-2.5, 1) {};
		\node [style=none] (23) at (-1, -1) {};
		\node [style=none] (24) at (-3.5, 1.5) {};
		\node [style=none] (25) at (-5, 0) {};
		\node [style=none] (26) at (-3.5, 0) {};
		\node [style=none] (27) at (-5, 1.5) {};
		\node [style=none] (28) at (-1, 1.25) {};
		\node [style=none] (29) at (-0.5, 1.25) {};
		\node [style=none] (30) at (-1, 0.25) {};
		\node [style=none] (31) at (-0.5, 0.25) {};
		\node [style=none] (32) at (1, 1.25) {};
		\node [style=none] (33) at (1.5, 1.25) {};
		\node [style=none] (34) at (1, 0.25) {};
		\node [style=none] (35) at (1.5, 0.25) {};
		\node [style=none] (36) at (3, 0.25) {};
		\node [style=none] (37) at (4, 0.25) {};
		\node [style=none] (38) at (3, 1.25) {};
		\node [style=none] (39) at (4, 1.25) {};
		\node [style=none] (40) at (-2.5, 1.25) {};
		\node [style=none] (41) at (-3.5, 1.25) {};
		\node [style=none] (42) at (-3.5, 0.25) {};
		\node [style=none] (43) at (-2.5, 0.25) {};
		\node [style=none] (44) at (-1, 2.25) {};
		\node [style=none] (45) at (1.5, 2.25) {};
		\node [style=none] (46) at (-1, -0.75) {};
		\node [style=none] (47) at (1.5, -0.75) {};
		\node [style=none] (48) at (3, 2.25) {};
		\node [style=none] (49) at (3, -0.75) {};
		\node [style=none] (50) at (5.5, 0.25) {};
		\node [style=none] (51) at (5.5, 1.25) {};
		\node [style=none] (52) at (-5, 0.25) {};
		\node [style=none] (53) at (-5, 1.25) {};
		\node [style=none] (54) at (3.5, 2.25) {};
		\node [style=none] (55) at (3.5, -0.75) {};
		\node [style=none] (56) at (6.5, 0.25) {};
		\node [style=none] (57) at (6.5, 1.25) {};
		\node [style=none] (58) at (-6.5, 1.25) {};
		\node [style=none] (61) at (4.75, 0) {};
		\node [style=none] (62) at (-4.25, 0) {};
		\node [style=none] (65) at (2.25, 1.75) {$\map{\widetilde{V}}_{AE}$};
		\node [style=none] (66) at (2.25, -0.25) {$\map{\widetilde{W}}_{BF}$};
		\node [style=none] (67) at (-1.75, 1.75) {$\map{\widetilde{V}}_{AE}$};
		\node [style=none] (68) at (-1.75, -0.25) {$\map{\widetilde{W}}_{BF}$};
		\node [style=none] (69) at (0.25, 0.75) {$\map{SWAP}$};
		\node [style=none] (70) at (0.25, 0.75) {};
		\node [style=none] (71) at (0.25, 0.75) {};
		\node [style=none] (72) at (4.75, 0.75) {\hspace{0.1cm}$\map{U^{\dagger}}$};
		\node [style=none] (73) at (-4.25, 0.75) {$\map{U}$};
		\node [style=none] (74) at (-5.5, 0.75) {};
		\node [style=none] (75) at (-5.5, -0.25) {};
		\node [style=none] (76) at (-5.5, 0.25) {};
		\node [style=none] (77) at (-5.5, 0.25) {};
		\node [style=none] (78) at (-3, 1.75) {};
		\node [style=none] (79) at (-2.5, 2.25) {};
		\node [style=none] (80) at (-3, 2.25) {};
		\node [style=none] (81) at (-3, 2.75) {};
		\node [style=none] (82) at (-3, -0.75) {};
		\node [style=none] (83) at (-2.5, -0.75) {};
		\node [style=none] (84) at (-3, -0.25) {};
		\node [style=none] (85) at (-3, -0.75) {};
		\node [style=none] (86) at (-3, -1.25) {};
		\node [style=none] (87) at (-3.5, 2.25) {~$\eta_E$};
		\node [style=none] (88) at (-3.5, -0.75) {~$\eta_F$};
		\node [style=none] (89) at (-6, 0.25) {\hspace{0.03cm} $\omega$};
		\node [style=none] (90) at (2.75, 2) {};
		\node [ground=none, style=none, xshift=-0.45em] (91) at (3.75, 2.25) {};
		\node [ground=none, style=none, xshift=-0.45em] (92) at (3.75, -0.75) {};
		\node [style=none] (95) at (-5.5, -1.25) {};
		\node [style=none, font={\small}] (97) at (-5.75, 1.5) {$M$};
		\node [style=none, font={\small}] (98) at (-5.25, 0) {$P$};
		\node [style=none, font={\small}] (99) at (-3, 1.5) {$\widetilde{A}$};
		\node [style=none, font={\small}] (100) at (-3, 0) {$\widetilde{B}$};
		\node [style=none, font={\small}] (101) at (-0.75, 1.5) {$\widetilde{A}$};
		\node [style=none, font={\small}] (102) at (-0.75, 0) {$\widetilde{B}$};
		\node [style=none, font={\small}] (103) at (1.25, 1.5) {$\widetilde{A}$};
		\node [style=none, font={\small}] (104) at (1.25, 0) {$\widetilde{B}$};
		\node [style=none, font={\small}] (105) at (3.5, 1.5) {$\widetilde{A}$};
		\node [style=none, font={\small}] (106) at (3.5, 0) {$\widetilde{B}$};
		\node [style=none, font={\small}] (107) at (6, 1.5) {$M$};
		\node [style=none, font={\small}] (108) at (6, 0) {$P$};
		
		\node [style=none, font={\small}] (109) at (-2.75, 2.5) {$E$};
		\node [style=none, font={\small}] (110) at (-2.75, -1) {$F$};
		\node [style=none, font={\small}] (111) at (0.25, 2.5) {$E$};
		\node [style=none, font={\small}] (112) at (0.25, -1) {$F$};
		\node [style=none, font={\small}] (113) at (3.25, 2.5) {$E$};
		\node [style=none, font={\small}] (114) at (3.25, -1) {$F$};

	\end{pgfonlayer}
	\begin{pgfonlayer}{edgelayer}
		\draw (5.center) to (6.center);
		\draw (6.center) to (7.center);
		\draw (7.center) to (4.center);
		\draw (4.center) to (5.center);
		\draw (8.center) to (10.center);
		\draw (9.center) to (11.center);
		\draw (10.center) to (11.center);
		\draw (8.center) to (9.center);
		\draw (12.center) to (15.center);
		\draw (15.center) to (14.center);
		\draw (14.center) to (13.center);
		\draw (13.center) to (12.center);
		\draw (2.center) to (1.center);
		\draw (1.center) to (0.center);
		\draw (0.center) to (3.center);
		\draw (3.center) to (2.center);
		\draw (17.center) to (18.center);
		\draw (18.center) to (20.center);
		\draw (20.center) to (22.center);
		\draw (22.center) to (17.center);
		\draw (16.center) to (21.center);
		\draw (19.center) to (23.center);
		\draw (21.center) to (23.center);
		\draw (16.center) to (19.center);
		\draw (24.center) to (27.center);
		\draw (27.center) to (25.center);
		\draw (25.center) to (26.center);
		\draw (26.center) to (24.center);
		\draw (38.center) to (39.center);
		\draw (36.center) to (37.center);
		\draw (33.center) to (32.center);
		\draw (34.center) to (35.center);
		\draw (29.center) to (28.center);
		\draw (30.center) to (31.center);
		\draw (41.center) to (40.center);
		\draw (42.center) to (43.center);
		\draw (44.center) to (45.center);
		\draw (46.center) to (47.center);
		\draw (48.center) to (54.center);
		\draw (49.center) to (55.center);
		\draw (50.center) to (56.center);
		\draw (51.center) to (57.center);
		\draw (58.center) to (53.center);
		\draw [bend right=90, looseness=3.25] (74.center) to (75.center);
		\draw (74.center) to (75.center);
		\draw (76.center) to (52.center);
		\draw [bend right=90, looseness=3.25] (81.center) to (78.center);
		\draw (81.center) to (78.center);
		\draw (80.center) to (79.center);
		\draw [bend right=90, looseness=3.25] (84.center) to (86.center);
		\draw (84.center) to (86.center);
		\draw (82.center) to (83.center);
	\end{pgfonlayer}
\end{tikzpicture}
		\caption{\label{fig:switch} {\em  Effective evolution of a particle travelling through regions with time-correlated environments.} The path $P$  is initialised in the state $\omega$. Depending on the state of the path, the message $M$ is routed to one of two spatially separated regions, where it interacts either with the environment  $E$ or with the environment $F$, respectively.  The interaction is modelled by unitary channels $\widetilde{\map V}_{AE}$ and  $\widetilde{\map W}_{BF}$, which act trivially when their input is in the vacuum state.  When the particle emerges from one region   it is routed to the other region by a $\tt SWAP$ operation.  Finally, the message emerges from the two regions, generally with its internal and external degrees of freedom correlated.  The overall evolution reproduces the output of the quantum SWITCH \cite{chiribella2013quantum}. }
\end{figure*}
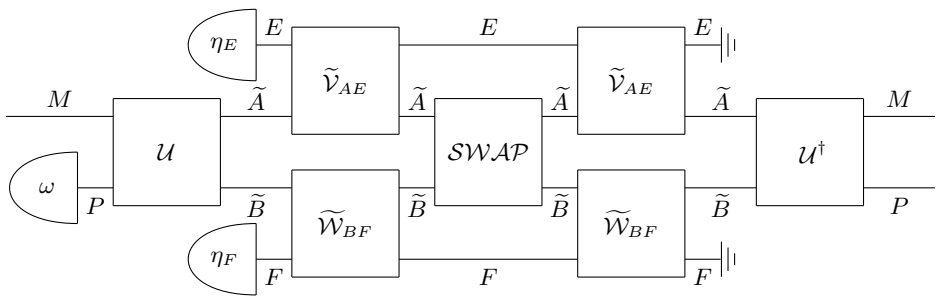

We illustrate the main ideas through an example. Suppose that a quantum particle can visit two regions, $R_A$ and $R_B$, following the two alternative paths shown in Figure \ref{fig:switchregions}. 
 Region $R_A$ contains a quantum system $E$ in some initial state $\eta_E$, while region $R_B$ contains another quantum system $F$ in the state $\eta_F$.    When the information carrier visits one region, it interacts with the corresponding system, thereby experiencing a noisy channel. Let us denote the two channels as $\map  A (\cdot)  =  \Tr_{E}  [   V_{AE}  (\cdot \otimes \eta_E)  V_{AE}^\dag  ]$ and  $\map  B (\cdot)  =  \Tr_{F}  [   V_{BF}  (\cdot \otimes \eta_F)  V_{BF}^\dag  ]$.   In general, the state of a local environment at  later times will be correlated with the state of the same environment at earlier times.   In particular, suppose that environments  $E$ and $F$ behave as ideal quantum memories, whose quantum state does not change in time unless the information carrier interacts with them. 
 Then,  a path visiting region $R_A$ before region $R_B$ will result in the channel $\map B\circ \map A$, while a path visiting region $R_B$ before region $R_A$ will result in the channel $\map A\circ \map B$.

The evolution of a particle sent through the two paths in a superposition is determined by  the local vacuum extensions of the  unitary  channels $\map V_{AE} (\cdot)  =  V_{AE} \cdot  V_{AE}^\dag$ and $\map W_{BF} (\cdot)  =   W_{BF}\cdot  W_{BF}^\dag $,  denoted as  $\widetilde{\map V}_{AE}$ and $\widetilde {\map W}_{BF}$, respectively  (see Equation (\ref{locvacext}) for the definition of local vacuum extension).    The quantum circuit describing the propagation of the particle is illustrated in Figure \ref{fig:switch}, and  
  the corresponding quantum evolution is described  by the effective channel 
\begin{align}\label{outswitch}
\nonumber 
\map S (\map A, &\map B)   (\rho\otimes \omega)   \\
\nonumber
   = &~\map U^\dag \Big(   \Tr_{EF}\Big\{   
 (\widetilde{\map V}_{AE} \otimes \widetilde{\map W}_{BF})  (\map I_E  \otimes \map{SWAP} \otimes \map I_F)  \\
 &(\widetilde{\map V}_{AE} \otimes \widetilde{\map W}_{BF})   [   \eta_E\otimes \map U (\rho\otimes \omega)  \otimes \eta_F ]   \Big\}\Big) \, ,
\end{align}
where the unitary channel $\map U$ implements the isomorphism between $M\otimes P$ and the one-particle subspace $(A\otimes {\rm Vac} ) \oplus ({\rm Vac}  \otimes B)$ [cf.\ Eq. (\ref{QPtoSum})].   The  effective channel    $\map S (\map A, \map B)$   has Kraus operators 
\begin{align}
S_{ij}   =   A_i B_j \otimes |0\>\<0|   +      B_j  A_i  \otimes |1\>\<1| \, ,
\end{align}     
where $\{A_i\}$ and $\{ B_j\}$ are Kraus representations for channels $\map A$ and $\map B$, respectively.  Note
that the channel  $\map S (\map A, \map B)$   depends only on the channels  $\map A$ and $\map B$, and not on their local vacuum extension. 
This situation arises  because of the form of the local vacuum extensions $\widetilde{\map V}_{AE}$ and $\widetilde{\map W}_{BF}$, which act as the identity on $\textrm{Vac} \otimes E$ and  $\textrm{Vac} \otimes F$, respectively, and because of the form of the circuit in Figure \ref{fig:switch},  which forces each of the input states in the two sectors $\widetilde{A}$ and $\widetilde{B}$ to interact with both environments in sequence. These two features of the circuit  erase the dependence of the Kraus operators of $\map S (\map A, \map B)$ on the vacuum amplitudes associated with the channels $\map{A}$ and $\map{B}$.

The channel  (\ref{outswitch}) is equal to the output of  the {\em quantum SWITCH}  \cite{chiribella2013quantum}, {\em i.e.}\  the higher-order quantum operation that takes the two channels $\map A$ and $\map B$  and combines them into the channel  $\map S (\map A, \map B)$. 
Using the direct sum notation 
  \begin{align}\label{switch}
 S_{ij}  =  A_i B_j  \oplus B_j A_i  
 \end{align}
we can see  that  the switched channel $\map S  (\map  A, \map B)$ is  a (correlated) superposition of the channels $\map A\circ  \map B$ and $\map B \circ \map A$, in the  sense of  our general Definition \ref{def:superposition}.

The realisation of the switched channel $\map S (\map A,\map B)$ in Figure \ref{fig:switch}  can be easily  extended from two to $N\ge 2$ independent channels.  In a communication scenario, this realisation  has the following features: 
\begin{enumerate}
\item The encoding operations do not induce signalling from the message to the path.   Specifically,   the encoding operation is of the product form  $\map E  (\rho) =  \rho \otimes \omega$, with the path in the fixed state $\omega$.

\item  Also the intermediate operations  between two subsequent time steps  do not induce signalling from the message to the path.  In Figure \ref{fig:switch},  the intermediate operation is a $\tt SWAP$ gate,  which takes the output of region $R_A$ ($R_B$) and routes it to region $R_B$ ($R_A$).  In terms of the ``message + path''  bipartition, the $\tt SWAP$ gate is just a bit flip  on the path, namely $U^\dag {\tt SWAP }  U  = I_M  \otimes X_P$.  Not only is this operation no-signalling, but in fact it is also a product operation, where $M$ and $P$ evolve independently.   This is important, because it means that  the intermediate operation $\tt SWAP$ respects the separation between internal and external degrees of freedom.
 
\item The realisation of the switched channel $\map S (\map A,\map B)$   is independent of the specific way in which the channels are realised through interactions with the environment,  as long as the unitary channels $\widetilde {\map V}_{AE}$   and $\widetilde {\map W}_{BF}$ are local vacuum extensions of two unitary evolutions that give rise to channels $\map A$ and $\map B$.    Physically, this means that the only assumption in the realisation of the switched channel $\map S (\map A,\map B)$ is that the state of the environment remains unchanged in the lack of interactions with the system. 
 
  \item The realisation of the quantum SWITCH in Figure \ref{fig:switch}  offers  more than just the switched channel $\map S (\map A, \map B)$: it also gives us a vacuum extension. This is important because it makes the superposition of orders {\em composable} with the superposition of paths.  
   \end{enumerate}

It is important to stress the difference between the circuit in Figure \ref{fig:switch} and the quantum SWITCH as an abstract higher-order operation.    The quantum SWITCH is the abstract higher-order map that takes two ordinary channels $\map A$ and $\map B$ as input resources and generates the switched channel $\map S (\map A,\map B)$  as output.  The circuit in Figure \ref{fig:switch} produces the {\em same output} of the quantum SWITCH, using as input resources the local  vacuum extensions of $\map A$ and $\map B$, and two perfect memories in the environments $E$ and $F$. Due to the different input resources, the circuit can therefore not be considered a genuine implementation of the quantum SWITCH, but is rather a simulation of the higher-order map.  In addition,  the  SWITCH map $\map S:   (\map A,\map B) \mapsto \map S (\map A,\map B)$ admits different physical realisations, which are generally different from the circuit in Figure \ref{fig:switch}.  For example, another circuital implementation of the switched channel $\map S (\map A,\map B)$ was proposed by Orehskov in Ref. \cite{oreshkov2018whereabouts}.   The SWITCH also admits realisations based on exotic physics,  such as  superposition of spacetimes \cite{hardy2009quantum,zych2017bell} and closed timelike curves \cite{chiribella2013quantum}.  
 The importance of the circuital realisations, such as that  in Figure \ref{fig:switch}, lies in the fact that they can be implemented with existing photonic technologies, thereby allowing the implementation of communication protocols built from the quantum SWITCH \cite{goswami2018communicating,guo2018experimental}.

In a similar vein, it is important to stress the difference between the communication model proposed in this paper and  the model of quantum communication with superposition of orders  introduced by Ebler, Salek, and Chiribella (ESC) in Ref. \cite{ebler2018enhanced}.  The ESC model describes an  abstract resource theory where an agent (\emph{e.g.}\ a communication company) builds a quantum communication network from a given set of quantum channels, using a subset of allowed higher-order operations, described by quantum supermaps \cite{chiribella2008transforming,chiribella2009theoretical,chiribella2013quantum}.  The purpose of the model is to analyse how the ability to combine quantum channels through the quantum SWITCH  affects their communication capability. The  set of allowed operations includes  composition in parallel, in sequence, and through the quantum SWITCH \cite{ebler2018enhanced}, without assuming a specific implementation of the quantum SWITCH. 
 This makes the theory applicable  not only to standard quantum theory, but also to future extensions of it to new spacetime scenarios, involving \emph{e.g.} superposition of spacetimes \cite{hardy2009quantum,zych2017bell} or  closed timelike curves \cite{chiribella2013quantum}.  In contrast, the second-quantised model proposed in  our paper refers to the known physics of quantum particles propagating in a well-defined background spacetime.

\subsection{Communication model with time-correlated channels} 
 In the previous section we gave a physical model for the realisation of time-correlated channels through interactions with an environment. It is important to stress, however,  that the superposition of time-correlated channels can be realised without access to the environment. 
  
   Generally, correlations between multiple time steps can be described as {\em quantum memory channels}  \cite{kretschmann2005quantum} and can be conveniently represented with the framework of  {\em quantum combs}  \cite{chiribella2008quantum,chiribella2009theoretical} (see also \cite{gutoski2007toward}).  Crucially, the framework of quantum combs does not need the specification of  the internal memories. In a communication scenario, this means that the communication resources  can be described purely in terms of the local input/output systems available to the communicating parties, without the need of specifying the details of the interactions with the environment.

\begin{figure*}
	\centering
	\input{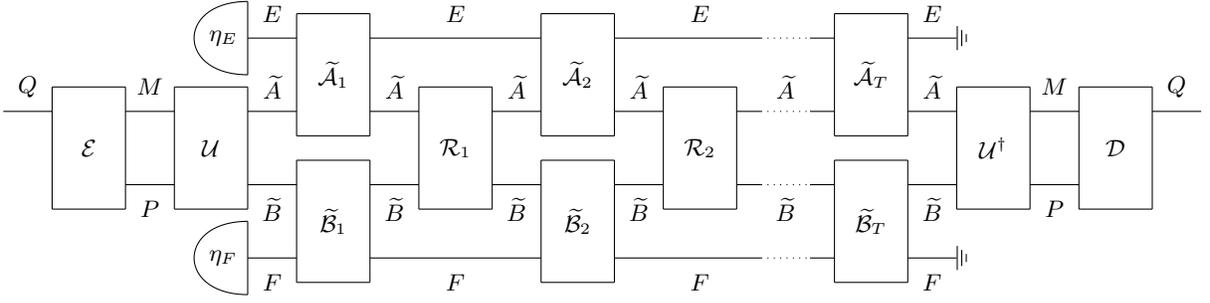}
	\caption{\label{fig:combs} {\em Superposition of two quantum memory channels.} The state of a quantum system $Q$ is encoded into a quantum particle, with internal degree of freedom $M$ (``message'') and external degree of freedom $P$ (``path"). The  composite system  $M\otimes P$ is then mapped onto the one-particle subspace of the composite system $\widetilde{A} \otimes \widetilde{B}$ by the unitary channel $\map U$. Channels $\widetilde{ \map  A}_i, \widetilde{\map{B}}_i, i \in \{1,2,\dots,T\}$ are then applied to the composite systems $E\otimes \widetilde{A}$ and $\widetilde{B} \otimes F$, respectively, where $E$ and $ F$ are internal memories.  Between each successive pair of channels $\widetilde{ \map  A}_i$, $\widetilde{ \map  B}_i$, a repeater $R_i$ acts on the system $\widetilde{A} \otimes \widetilde{B}$, preparing the input for the next step. After $T$ iterations, the  decoding operation $\map{D}$ converts the output back into  system $Q$.
	}
\end{figure*}

    An example of a communication protocol using a superposition of two channels with memory is shown in Figure \ref{fig:combs}. Each channel has $T$ pairs of input/output systems, whose evolution is correlated by an internal memory.  Each input system can either carry a message or be in the vacuum.  The communication protocol works as follows:  
\begin{enumerate}
\item The sender encodes a quantum system $Q$ in the one-particle subspace of the system $\widetilde A^{\rm in}_1\otimes \widetilde B^{\rm in}_1$, representing the inputs of the first time step.  The encoding operation is required to be of the product form $\map E  =  \map M \otimes \omega$, where $\map M$ is a channel from $Q$ to $M$, and $\omega$ is a fixed state of the path. 
\item    The communication channel transfers information from  the (one-particle subspace of the) first input system $\widetilde A^{\rm in}_1\otimes \widetilde B^{\rm in}_1$ to the (one-particle subspace of the)   first output system $\widetilde A^{\rm out}_1\otimes \widetilde B^{\rm out}_1$, which is received by a repeater.  The repeater implements the operation $\map R_1$, which relays the message to the (one-particle subspace of the)  second input system  $\widetilde A^{\rm in}_2\otimes \widetilde B^{\rm in}_2$. The repeater operation is required to be of the product form 
\begin{align}
  \map U^\dag  \map R_1  \map U =     \map M' \otimes \map P  \, , 
\end{align}
where $\map M'$ is a quantum channel acting only on the message  (not necessarily the same channel used in the encoding operation),  and $\map P$ is a quantum channel acting only on the path (for simplicity, we assume here that the input and output systems at all steps are isomorphic to $M$).  
\item The journey  of the message to the receiver  continues through $T$ time steps, alternating transmissions through noisy channels and repeaters. Eventually, the message reaches the receiver, who performs a decoding operation $\map D$. 

 \end{enumerate} 

We finally mention that the correlations in time, represented by quantum memory channels/quantum combs, and the correlations in space, represented by shared states, can be combined together, giving rise to  complex patterns of correlated channels through which information can travel in a superposition of paths.

\subsection{Examples}  
Several examples of Shannon-theoretic advantages of the quantum SWITCH have been recently presented, both for classical \cite{ebler2018enhanced,goswami2018communicating} and for quantum communication \cite{salek2018quantum,chiribella2018indefinite}.   Here we briefly highlight the main features, also providing a  new example of classical communication involving the superposition of two pure erasure channels. 

\subsubsection{Self-switching}
 
Quite counterintuitively, switching a quantum channel with itself gives rise to a number of non-trivial phenomena. 
Suppose that a quantum system $A$ is sent through two independent uses of the same channel $\map A$, with  the control qubit in the $|+\>$ state.    When the message is prepared in the input state $\rho$, the output state is 
\begin{align}
\label{self}
 \map S  (\map A, \map A)     (\rho \otimes |+\>\<+|)  & =    \map C_+ (\rho)  \otimes |+\>\<+|   + \map C_-  (\rho)\otimes |-\>\<-| \\
\nonumber \map C_+    (\rho) &  =     \frac 14 \sum_{i,j}\{ A_i  ,  A_j\}  \rho       \{ A_i  ,  A_j\}^\dag  \\ \nonumber
\map C_-  (\rho)    &=  \frac 14 \sum_{i ,j}   [ A_i  ,  A_j]  \rho       [ A_i  ,  A_j]^\dag   \, , 
\end{align}
where  $\{A_i\}$ is any Kraus representation of channel $\map A$, while  $[A,B]$ and $\{A,B\}$  denote the commutator and anti-commutator, respectively. 

Equation (\ref{self}) gives a number of insights into the scenarios where the superposition of orders offers advantages.  First of all, note that the action of the quantum SWITCH is trivial if the Kraus operators of channel $\map A$ commute with one another. In that case, $\map C_+$ is equal to $\map A^2$, while $\map C_-$ is zero.    Instead, non-trivial effects take place when some of the operators
 do not commute.  For example, suppose that $\map A$ is a pure erasure channel, viz.  $\map A (\rho)  =  |\psi_0\>\<\psi_0|$ for every state $\rho$.   Then, the self-switching formula  (\ref{self})  yields the output state 
\begin{align}
 \map S  (\map A, \map A)     (\rho \otimes |+\>\<+|)       =   & \, |\psi_0\>\<\psi_0| \otimes  \left[ p  |+\>\<+|  +  (1-p)\frac I 2 \right]   \\
\nonumber &  p  =  \<\psi_0  |\rho|\psi_0\> \, . 
 \end{align}

The resulting communication channel is identical to the communication channel in Eq. (\ref{erasureonthepath}), and its classical capacity is $\log_2  (5/4)\approx .32$.  

\subsubsection{Perfect communication through a coherent superposition of orders}

Another consequence of the self-switching formula (\ref{self}), is that one can obtain perfect quantum communication using a  noisy channel $\map A$.  A qubit example was recently discovered in \cite{chiribella2018indefinite} and involves the random unitary channel $\map A  =  1/2  (\map X  +  \map Y)$, with $\map X=  X\cdot X$ and $\map Y =  Y\cdot Y$.  With this choice, the channels $\map C_+ $ and $\map C_-$ in the self-switching formula (\ref{self}) are the identity or the phase flip channel $\map Z  =  Z\cdot Z$, respectively.  Hence, perfect communication can be achieved by measuring the path and conditionally performing a correction operation on the message.   More generally, it is clear that the same effect takes place for a random-unitary channel that performs either the unitary gate $U$ or the unitary gate $V$,   provided that   the conditions $U^2  =  V^2$ and $\{  U,  V\}= 0$ are satisfied.  The possibility of a complete removal of noise through the self-switching effects is in stark contrast with the irreducible amount of noise characterising the superposition of noisy channels on independent paths, cf. the discussion around Equation (\ref{muchcleaner}).

\section{Conclusions}\label{sec:conclusions}
  
We have developed a Shannon-theoretic framework  for  communication protocols  where information propagates along a superposition of multiple paths, experiencing either independent or correlated  processes along them.  Central to our framework is a separation between the internal and external degrees of freedom of the information carrier. Information is encoded only in the internal degrees of freedom, while its propagation is determined by the  state of the external degrees of freedom.    As information propagates through space and time,  the internal and external degrees of freedom become correlated, and such correlations can be exploited by a receiver to enhance their ability to decode the sender's message.  Several examples have been provided, including protocols for classical communication with pure erasure channels and for  quantum communication with entanglement-breaking channels.


Our  basic model  assumed that the external degrees of freedom are not subject to noise. This assumption can be easily relaxed by introducing noisy channels, such as dephasing  or loss of particles along different paths.    An important direction  for future research is to quantify how much noise can be tolerated while still having an advantage over conventional communication protocols where information travels along a single, well-defined path.    

The step from a first to 
a second quantisation  is  in tune with other recent developments in quantum Shannon theory, such as the study  of network scenarios  \cite{leung2010quantum}.    As technology advances towards the realisation of quantum communication networks, we expect that scenarios involving the superpositions of paths  will become  accessible, enabling new communication protocols as well as new fundamental experiments of quantum mechanics in spacetime.

\begin{acknowledgments}
This work is supported by the National
Natural Science Foundation of China through grant~11675136, the Croucher Foundation,  the John Templeton Foundation through grant  60609, Quantum Causal Structures, the Canadian Institute for Advanced Research~(CIFAR), the Hong Research Grant Council through grants~17326616 and 17300918, the Foundational
Questions Institute through grant~FQXi-RFP3-1325, and the HKU Seed Funding for Basic Research.  
HK is supported by funding from the UK EPSRC. 
This publication was made possible through the support of a grant from the John Templeton Foundation. The opinions expressed in this publication are those of the
authors and do not necessarily reflect the views of the John Templeton Foundation. This research was supported in part by the Perimeter Institute for Theoretical 
Physics. Research at the Perimeter Institute is supported by the Government of 
Canada through the Department of Innovation, Science and Economic Development 
Canada and by the Province of Ontario through the Ministry of Research, 
Innovation and Science.

GC is indebted to S Popescu,  B Schumacher, and P Skrzypczyk  for  stimulating  discussions  on  the relation between error filtration \cite{gisin2005error} and communication protocols using  the quantum SWITCH \cite{ebler2018enhanced}.    Discussions with R Renner, P Grangier, A Steinberg, O Oreshkov, R Spekkens, D Schmidt, C Zoufal, D Ebler, S Salek, V Giovannetti, L Rozema, G Rubino, \v{C} Brukner,  P Walther, N Pinzani and WX Mao are also acknowledged. The authors would like to thank C Branciard and another anonymous referee for detailed comments and suggestions on the manuscript.
\end{acknowledgments}

{\small
\bibliographystyle{RS}

}
\newpage
\onecolumngrid
\appendix

\section{Proof of Theorem  \ref{theo:allsup}}
\label{app:allsup}

The starting point of the proof is the characterisation of the No Leakage Condition (\ref{noleak}) in terms of  Kraus operators. 
\begin{lemma}\label{lem:noleak}
	Let $\widetilde  {\map C}   (\rho)   =   \sum_{i=1}^r \widetilde {C}_i  \rho \widetilde {C}_i^\dag$ be a Kraus representation of channel $\widetilde {\map C}$.   Then, channel     $\widetilde {\map C}$ satisfies the No Leakage Condition if and only if  
	\begin{align}\label{noleakkraus_app}
		P_A  \widetilde C_i  P_A  =    \widetilde C_i  P_A \qquad \forall  i\in  \{1,\dots, r\} \, .
	\end{align}   
\end{lemma}
\Proof  The No Leakage Condition can be rewritten as  
\begin{align*}
	\Tr\left [  \left(    \sum_i  P_A     \widetilde  C_i^\dag  P_A   \widetilde C_i     P_A\right) \rho  \right]  =      \Tr [  P_A \rho]  \qquad \forall \rho  \in \St (A) \, ,
\end{align*}
or equivalently
\begin{align}
	\sum_i  P_A     \widetilde  C_i^\dag  P_A   \widetilde C_i     P_A  =   P_A   \, .
\end{align}
On the other hand,  one has the inequality  
\begin{align}
	\nonumber P_A   &=   P_A  \left(\sum_i \widetilde C_i^\dag \widetilde C_i \right)   P_A \\
	\nonumber &    =  \sum_i    P_A   \widetilde C_i^\dag  \left(  P_A  +  P_A^\perp\right)      \widetilde C_i  P_A  \qquad P_A^\perp  :  =  I-  P_A  \\  
	&\le \sum_i  P_A  \widetilde C_i^\dag P_A  \widetilde C_i    P_A  \, ,  
\end{align}
where the equality sign holds if and only if 
\begin{align}
	\sum_i  P_A  \widetilde C_i^\dag P_A^\perp  \widetilde C_i    P_A   =  0   \, ,
\end{align}
or equivalently, 
\begin{align}
	\sum_i  \Big (  P_A^\perp  \widetilde C_i P_A\Big)^\dag   \, \Big(  P_A^\perp  \widetilde C_i    P_A \Big)   =  0   \, ,
\end{align}

Since every term in the sum is a positive semidefinite operator, the equality holds if and only if each term is zero, namely if and only if $P_A^\perp  \widetilde C_i    P_A   =  0$ for every $i$.  In conclusion, we obtained 
\begin{equation}
\nonumber \widetilde C_i    P_A     =      ( P_A +  P_A^\perp)  \widetilde C_i   P_A    
=     P_A  \widetilde C_i  P_A \, ,
\end{equation}
as stated in Equation (\ref{noleakkraus_app}). 
\qed 

\medskip

{\bf Proof of Theorem \ref{theo:allsup}.}     $1  \Longrightarrow 2$.  Let $\map S  \in  \Chan  (A  \oplus B)$ be a superposition of channels $\map A \in  \Chan (A)$ and $\map B  \in  \Chan (B)$, and let $\map S(\rho)  =  \sum_i  S_i \rho S_i^\dag$ be a Kraus decomposition of $\map S$.     Since $\map S$ satisfies the No Leakage Condition for $A$, we must have  
\begin{align}\label{noleakA}
	S_i    P_A   =      P_A  S_i  P_A  \qquad \forall i\in  \{1,\dots  r\} \, .
\end{align}
Similarly, since  $\map S$ satisfies the No Leakage Condition for $B$, we must have  
\begin{align}\label{noleakB}
	S_i    P_B   =      P_B  S_i  P_B  \qquad  \forall i\in  \{1,\dots  r\} \, .
\end{align}
Combining Equations (\ref{noleakA}) and (\ref{noleakB}) we obtain   $S_i  =   S_i  ( P_A  +  P_B )   =     A_i\oplus  B_i$, with  $A_i  :  = P_A  S_i  P_A$ and $B_i: = P_B  S_i P_B$.   Since the restriction of $\map S$ to sector $A$ must be channel $\map A$, we have the condition $  \map S (   P_A  \rho  P_A)   =  \map A (  P_A \rho  P_A)$. Hence, we conclude that $\{A_i\}_{i=1}^r$ is a Kraus representation of $\map A$.  Similarly, since the restriction of $\map S$ to sector $B$ must be channel $\map B$,  we conclude that $\{B_i\}_{i=1}^r$  must be a Kraus representation of $\map B$.  

$2  \Longrightarrow 1$  is immediate.  

$2  \Longrightarrow 3$.   Consider the Stinespring representation of channel $\map S$, obtained  by introducing an environment $E$ of dimension $r$, equal to the number of Kraus operators of $\map S$.  Explicitly, the Stinespring representation is given by the isometry  $V   =   \sum_{i=1}
^r  \,   S_i \otimes |i\>$, where $\{|i\>\}_{i=1}^r$ is the canonical  basis for  $E$.   Since each $S_i$ is of the form   $S_i  =  A_i \oplus  B_i $, the isometry $V$ is of the direct sum form $V  =  V_A \oplus V_B$, where $V_A :  \spc H_A \to  \spc H_A \otimes \spc H_E $ and $V_B  :  \spc H_B\to \spc H_B\otimes \spc H_E$ are the isometries defined as 
\begin{align}
	\label{VA}  
	V_A &:  =  \sum_{i=1}^r  \,  A_i\otimes |i\>   \\
	\label{VB}
	V_B  &:  =  \sum_{i=1}^r \, B_i  \otimes |i\> \, .  
\end{align}   Now, each isometry   $V_A$ and $V_B$ can be extended to a unitary $U_A$ and $U_B$, so that $V_A    =    U_A   (   I_A  \otimes |\eta_A\>)     $ and $V_B   =      U_B   (   I_B  \otimes |\eta_B\>)     $, where $|\eta_A\>$ and $|\eta_B\>$  are  unit vectors in $\spc H_E$.  

Note that  (i)   one can choose $|\eta_A\>  = |\eta_B \>  =  |\eta\>$ without loss of generality, and (ii) 
each unitary $U_A$ and $U_B$ can be realised as a  time evolution for time $T$ with   Hamiltonian $H_{AE}$ and $H_{BE}$, respectively.   Hence,  one can define the unitary  evolutions $U_A   :  =    \exp  [  -i     H_{AE} \,  T/\hbar] $,   $U_B   :  =    \exp  [  -i     H_{BE} \,  T/\hbar]$, and   $U   :=   \exp  [  -i    ( H_{AE}  \oplus H_{BE})  T/\hbar]  =   U_A \oplus U_B$. 

With these definitions, we have  
\begin{align}
	\Tr_E  \left  [   U_{AE} (\rho  \otimes |\eta\>\<\eta|)  U_{AE}^\dag \right]  &  
	=  \sum_i   K_i \rho K_i^\dag \,,
\end{align} 
with 
\begin{align} 
	\nonumber K_i    & :=  \big(  I_A  \otimes \<  i|  \big) \,   U_{AE} \,  \big( I_A \otimes |\eta\>\big)  \\
	\nonumber   & =\big(  I_A \otimes  \<i|  \big)   V_A  \\
	&  =  A_i   \, ,
\end{align}
having used Equation (\ref{VA}) in the last equality. 
Similarly, we have 
\begin{align}
	\Tr_E  \left  [   U_{BE} (\rho  \otimes |\eta\>\<\eta|)  U_{BE}^\dag \right]  &  
	=  \sum_i   L_i \rho L_i^\dag \,,
\end{align} 
with 
\begin{align} 
	\nonumber L_i    & :=  \big(  I_B  \otimes \<  i|  \big) \,   U_{BE} \,  \big( I_B \otimes |\eta\>\big)  \\
	\nonumber   & =\big(  I_B \otimes  \<i|  \big)   V_B  \\
	&  =  B_i   \, ,
\end{align}
having used Equation (\ref{VB}) in the last equality,  and
\begin{align}
	\Tr_E  \left  [   U (\rho  \otimes |\eta\>\<\eta|)  U^\dag \right]  &  
	=  \sum_i   M_i \rho M_i^\dag \,,
\end{align} 
with 
\begin{align} 
	\nonumber M_i    & :=  \big(  I_S  \otimes \<  i|  \big) \,   U \,  \big( I_S \otimes |\eta\>\big)    \qquad \qquad S:  =  A\oplus B \\
	\nonumber   &        = \big(  I_A  \otimes \<  i|  \big) \,   U_{AE} \,  \big( I_A \otimes |\eta\>\big)   ~ \oplus ~ \big(  I_B  \otimes \<  i|  \big) \,   U_{BE} \,  \big( I_B \otimes |\eta\>\big)    \\
	\nonumber  
	&=  L_i  \oplus K_i\\
	&  =  A_i \oplus B_i  \, . 
\end{align}

$3  \Longrightarrow 1$.     
Let $E$ be an environment, let $\ket{\eta} \in \spc H_E$  be a pure state, and let $\spc H_{AE}, \spc H_{BE}$ be Hamiltonians with supports in $\spc H_A\otimes \spc H_E$ and $\spc H_B \otimes \spc H_E$, respectively, such that
\begin{align}
	\nonumber \map A (\rho)   &=   \Tr_E  [  U_{AE}  (\rho\otimes \eta) U_{AE}^\dag ]  \qquad &   U_{AE}  =  \exp[-i  H_{AE}\,  T/\hbar ]\\   
	\nonumber \map B (\rho)   &=   \Tr_E  [  U_{BE}  (\rho\otimes \eta) U_{BE}^\dag ]  \qquad &   U_{BE}  =  \exp[-i  H_{BE}\,  T/\hbar ]\\   
	\map S (\rho)   &=   \Tr_E  [  U  (\rho\otimes \eta) U^\dag ]  \qquad &   U   =   \exp[-i    (H_{AE}  \oplus H_{BE}  )  \,  T/\hbar ]   \equiv  U_{AE}  \oplus  U_{BE} \, . 
\end{align}
By construction, if $\rho$ has support in $\spc H_A$, one has 
\begin{align}
	\nonumber \map S (\rho)     &=      \Tr_E  [  U     P_{AE} \, (\rho\otimes \eta)   \, P_{AE}  U^\dag ]      \qquad P_{AE} :  =  P_A \otimes I_E   \\
	\nonumber &  =   \Tr_E  [  U_{AE} \, (\rho\otimes \eta)   \, U_{AE}^\dag ] \\
	&=  \map A(\rho) \, .
\end{align}
Similarly, if $\rho$ has support in $\spc H_B$, one has 
\begin{align}
	\nonumber \map S (\rho)     &=      \Tr_E  [  U     P_{BE} \, (\rho\otimes \eta)   \, P_{BE}  U^\dag ]      \qquad P_{BE} :  =  P_B \otimes I_E   \\
	\nonumber &  =   \Tr_E  [  U_{BE} \, (\rho\otimes \eta)   \, U_{BE}^\dag ] \\
	&=  \map B(\rho) \, .
\end{align}
Hence, $\map S$ is a superposition of $\map A$ and $\map B$. 
\qed

\section{Vacuum extensions  with non-trivial vacuum dynamics}\label{app:vacuum}

Let $\spc H_{\rm Vac}$ be the vacuum sector, {\em i.e.} the subspace corresponding to the vacuum degrees of freedom.  

\begin{defi}
	Let $\map C  \in  \Chan  (A)$ be a quantum channel.  
	A vacuum extension of channel $\map C$ is any channel $\widetilde {\map C}  \in   \Chan  (A\oplus {\rm Vac})$ such that  {\em (i)}   $\widetilde {\map C}$ satisfies the No Leakage Condition with respect to $A$ and $\rm Vac$, and {\em (ii)}  the restriction of $\widetilde {\map C}$ to sector $A$ is channel $\map C$.  
\end{defi}   
The proof of Theorem \ref{theo:allsup}, provided in Appendix \ref{app:allsup}, shows that every vacuum extension $\widetilde {\map C}$ must have Kraus operators of the form 
\begin{align}\label{generalvacext}
	\widetilde {C}_i   =  C_i  \oplus  C_{{\rm Vac}, i}  \qquad i\in \{1,\dots, r\} 
\end{align}
where $\{  C_i\}_{i=1}^r$ is a Kraus representation of $\map C$ and $\{C_{{\rm Vac},  i}\}_{i=1}^r$ are Kraus operators of a channel $\map C_{\rm Vac}  \in  \Chan ({\rm Vac})$, representing the dynamics of the vacuum sector.  

The simplest case is when the vacuum does not evolve under the action of the device, in which case $\map C_{\rm Vac}  $ is the identity channel.  
In this case, the Kraus operators of the vacuum extension have the simpler form 
\begin{align}
	\widetilde {C}_i   =  C_i  \oplus  \gamma_i  \, I    \qquad i\in \{1,\dots, r\}  \, ,
\end{align}
with $\sum_i |\gamma_i|^2 =1$, which is essentially equivalent to a  vacuum extension with a one-dimensional vacuum subspace.

We now use the vacuum extension to define  an operational superposition of two channels $\map A$ and $\map B$. For simplicity, we assume that the ``vacuum for system $A$'' is the same as the ``vacuum for system $B$'', and we will denote it as $\rm Vac$. 
The operational superposition of channels  is built  in the following way. First,   the direct sum sector  $A \oplus B  $ is embedded into the tensor product $\widetilde A \otimes \widetilde B$, with $\widetilde A   =  A  \oplus {\rm Vac}$ and $\widetilde B  =  B \oplus {\rm Vac}$ using the isometry 
\begin{align}  
	\nonumber V :   &  ~   \spc H_A  \oplus  \spc H_B  \to    \spc H_{\widetilde A} \otimes \spc H_{\widetilde B}  \\
	\label{embedding}& ~  |\alpha\>  \oplus  |\beta\>  \mapsto  |\alpha\>  \otimes  |\upsilon_0\>    \oplus  |\upsilon_0\>\otimes |\beta\>  \, ,
\end{align}
where $|\upsilon_0\> $ is a fixed unit vector in $\spc H_{\rm  Vac}$.  

As an inverse of the isometry $V:  A\oplus B  \to  \widetilde A\otimes \widetilde B$, we  use the following {\em vacuum-discarding map}: 
\begin{defi}   A map $\map T   \in  \Chan  (  \widetilde A\otimes \widetilde B  ,  A\oplus B)$   is a {\em vacuum-discarding map} if it has the form  
	\begin{align}\label{vacdiscard}
		\map T  (\rho)   =     \map T^{\rm succ}    (\rho)       +  \map T^{\rm fail}  (\rho)  \, ,
	\end{align} 
	where $\map T^{({\rm succ})}$ is the quantum operation with  Kraus operators 
	\begin{align}
		T^{\rm succ}_k      &=   P_A  \otimes  \< \upsilon_k|    \oplus    \< \upsilon_k|  \otimes P_B   \, ,
	\end{align}   
	$\{ |\upsilon_k\>\}_{k=1}^{d_{\rm Vac}}$ being an orthonormal basis for the vacuum subspace, and 
	\begin{equation}
	\map T^{({\rm fail})}  (\rho)   =   (1   -   \Tr [  P_{\rm succ}  \, \rho])  \,    |\psi_0\>\<\psi_0|,
	\end{equation}
	where $|\psi_0\> $ is a fixed vector in $\spc H_A  \oplus \spc H_B$ and $P_{\rm succ} :  =   P_A \otimes P_{\rm Vac}   +  P_{\rm Vac} \otimes P_B$.   
\end{defi}  

Physically, the map $\map T$ inverts the isometry $V$ when the total system is in the one-particle sector $(A\otimes {\rm Vac} ) \oplus ({\rm Vac} \otimes B)$, and outputs a ``failure state''  $|\psi_0\>$ when the system is in the two-particle or zero-particle sectors.  Note that the definition of the map $\map T^{\rm succ}$ is independent of the choice of orthonormal basis $\{|\upsilon_i\>\}$ for the vacuum subspace.  

Using the above notions,  one can define  a superposition of two   channels in the following way:  
\begin{defi}\label{sup_def_formal}
	Let $\map A  \in  \Chan (A)$ and $\map B  \in \Chan (B)$ be two quantum channels with vacuum extensions $\widetilde {\map A}$ and $\widetilde {\map B}$, respectively. Let  $V$ be the isometry defined in Equation (\ref{embedding}),  and  $\map T$ be the vacuum-discarding map defined in Equation (\ref{vacdiscard}).  Then, the  {\em superposition} of $\map A$ and $\map B$ specified by $\widetilde {\map A}$,   $\widetilde {\map B}$,  $V$, and $\map T$ is the quantum channel $\map S  \in  \Chan  (A \oplus B)$ defined by  
	\begin{align}\label{canonicalsup'}
		\map S   :  =          \map T \circ (  \widetilde {\map A} \otimes \widetilde {\map B}) \circ  \map V \, ,
	\end{align}
	with $\map V (\cdot): =   V \cdot  V^\dag$. 
\end{defi}

Note that in the case where we restrict the overall system to be in the one-particle sector, $ \Tr [  P_{\rm succ}  \, \rho] = 1$, so $\map T (\rho) = \map T^{({\rm succ})} (\rho)$. If in addition the vacuum is taken to be one-dimensional, then $\map T$ reduces to the unitary channel $\map V^\dagger$ of Equation (\ref{canS}), in which case the above Definition \ref{sup_def_formal} reduces to Definition \ref{sup_def_simple} in the main text.

Using the definition  (\ref{canonicalsup'}),  one can express the superposition as 
\begin{align}
	\nonumber \map S (\rho)    =  &  \map A  (  P_A  \rho  P_A)   \oplus  \map B   (  P_B  \rho  P_B  )       \\
	\nonumber &   +   \sum_{i,j}  \, A_i  \rho  B_j^\dag  ~      \< \upsilon_0  |   A_{{\rm Vac},  i}^\dag   B_{{\rm Vac},  j  }|\upsilon_0\>      \\
	&   +   \sum_{i,j}  \,   B_j \rho  A_i^\dag  ~        \< \upsilon_0  |   B_{{\rm Vac},  j}^\dag   A_{{\rm Vac},  i }|\upsilon_0\>   \, .
\end{align}
The merit of this expression is that it shows how the interference between the two channels $\map A$ and $\map B$ is mediated by the vacuum.  More explicitly, the superposition has a 
Kraus representation of the form   
\begin{align}\label{superpositionijk}
	S_{ijk}   =  A_i  \,  \beta_{jk}  \oplus       B_j  \,   \alpha_{ik}    \, ,     
\end{align} 
where $\{A_i\}$ and $\{ B_j\}$ are the Kraus representations used in the definition  of  the vacuum extensions $\widetilde {\map A}$ and $\widetilde {\map B}$, respectively, and   
\begin{align}
	\alpha_{ik}  :  =  \< \upsilon_k |   A_{{\rm Vac} ,  i}  |\upsilon_0\>   \qquad {\rm and} \qquad \beta_{jk}  :  =  \< \upsilon_k |   B_{{\rm Vac },  j}  |\upsilon_0\>  \, ,
\end{align}
where $\{  A_{{\rm Vac},  i}\}$ and $\{  B_{{\rm Vac},  j}\}$ are the Kraus representations of the vacuum dynamics associated to the extensions  $\widetilde {\map A}$ and $\widetilde {\map B}$, respectively.

We have seen that every vacuum extension leads to a superposition of channels with Kraus operators as in Equation (\ref{superpositionijk}). It is worth noting that the converse also holds:  
\begin{prop}\label{prop:allsuperpositions}
	Let $\{A_i\}_{i=1}^{r_A}$ and $\{  B_j\}_{j=1}^{r_A}$ be Kraus decompositions for $\map A$ and $\map B$, respectively, and let $\{\alpha_{ik}\}_{i\in \{1,\dots, r_a\},   k  \in  \{1, \dots  , v\}}$ and $\{\beta_{jk}\}_{j\in \{1,\dots, r_B\},   k  \in  \{1, \dots  , v\}}$ be complex numbers such that $\sum_{i,k}  |\alpha_{i,k}|^2  =  \sum_{j,k}  |\beta_{j,k}|^2  =1$ (note that $r_A$ and $r_B$ need not be equal here).  Then, there exist two vacuum extensions $\widetilde {\map A}$ and $\widetilde {\map B}$, with $v$-dimensional vacuum sector, such that the Kraus operators $S_{ijk}  : =  A_i  \,  \beta_{jk}  \oplus       B_j  \,   \alpha_{ik}  $ define a superposition of channels $\map A$ and $\map B$. 
\end{prop}
\Proof     Define the probabilities $p_i  :  =  \sum_k   |\alpha_{i,k}|^2$ and $q_j :  =  \sum_k \,   |\beta_{j,k}|^2$ and the unit   vectors 
\begin{align}
	|\alpha^{(1)}_i\> : =  \frac{ \sum_k  \,  \alpha_{i,k}    \,  |\upsilon_k\> }{\sqrt{p_i}}  \qquad {\rm and} \qquad |\beta^{(1)}_j \> : = \frac{\sum_k  \, \beta_{j,k} \,  |\upsilon_k\> }{\sqrt {q_j}}\, . 
\end{align}
Then,  define the Kraus operators 
\begin{align}
	\nonumber A_{{\rm Vac} ,i}   &:  =    \sqrt{p_i}  \,    U_{A,i}   \qquad &&U_{A,i}  :  = \sum_{k=1}^v  \,   |\alpha^{(k)}_i\>\<\upsilon_k  |          \\
	B_{{\rm Vac} ,j}   &:  =    \sqrt{q_j}  \,    U_{B,j}    \qquad &&U_{B, j}   :  = \sum_{k=1}^v  \,   |\beta^{(k)}_j\>\<\upsilon_k  |     \, ,            
\end{align}
where, for every fixed $i$ and $j$,  
$\{|\alpha^{(k)}_i\> \}$ and $\{|\beta^{(k)}_j\> \}$ are two orthonormal bases of $\spc H_{\rm Vac}$, containing the vectors $|\alpha^{(1)}_i\>$ and $|\beta^{(1)}_j\>$, respectively. 
Then,    $U_{A,i}$ and $U_{B,j}$ are unitary operators acting on the vacuum subspace, and $\{A_{{\rm Vac} ,i} \}$ and $\{B_{{\rm Vac} ,j}\}$ are the Kraus representations of two (random-unitary) channels. 
The thesis follows by defining the vacuum extensions  $\widetilde {\map A}$ and $\widetilde {\map B}$ through their Kraus representations  $\widetilde A_i  :  = A_i \oplus  A_{{\rm Vac},  i}$ and $\widetilde B_j  :  = B_j \oplus  B_{{\rm Vac},  j}$, and by setting the initial state of the vacuum to be the first state of the basis $\{  |\upsilon_k\>\}$.   

\qed

\section{Vacuum extensions and unitary dilations
}\label{app:hamiltonian}  

Here we clarify the relation between the superposition of channels defined through their action on the vacuum and the superposition of channels  defined through their unitary implementation.

Oi  \cite{oi2003interference} defined the superposition  of two channels $\map A\in \Chan  (S)$ and $\map B\in \Chan (S)$   in terms of their unitary implementations  
\begin{align}
	\map A  (\rho)    &=  \Tr_{E}  \left[    U_{S E}\,   (  \rho  \otimes  |\eta\>\<\eta|) \,  U_{SE}^\dag \right]\\
	\map B  (\rho)   & =  \Tr_{F}  \left[    V_{SF}\,   (  \rho  \otimes  |\phi\>\<\phi|) \,  V_{SF}^\dag \right] \, ,
\end{align}
where $E$ and $F$ are  quantum systems (the ``environments'' for $\map A$ and $\map B$, respectively),    $U$ and $V $ are unitary operations, representing the joint evolution of  system and environment,  and $|\eta\>$ and $|\phi\>$ are initial pure states of the environments $E$ and $F$, respectively.    
The superposition of channels $\map A$ and $\map B$   is defined as  the channel  $\map S$, taking  system $S $ and a control qubit $C$ as input, and satisfying the relation
\begin{align*}
	\map S_{\rm Oi}   (\rho_{S }\otimes \rho_C)   : &=  \Tr_{EF}   \big[ W (\rho_S\otimes |\eta\>\<\eta|  \otimes |\phi\>\<\phi| \otimes \rho_C) W^\dag  \big]  
\end{align*}
with 
\begin{align}
	W  &  =      \Big(  U_{SE}\otimes I_F    \otimes |0\>\,0| \Big)  +    \Big(   V_{S  F} \otimes  I_E \otimes |1\>\<1|  \Big)\, .
\end{align}

One can extend Oi's definition  to the case of channels $\map A$ and $\map B$ acting on generally different systems $A$ and $B$.  In this case,  the unitary implementations   read
\begin{align}
	\map A  (\rho)    &=  \Tr_{E}  \left[    U_{A E}\,   (  \rho  \otimes  |\eta\>\<\eta|) \,  U_{AE}^\dag \right]\\
	\map B  (\rho)   & =  \Tr_{F}  \left[    V_{BF}\,   (  \rho  \otimes  |\phi\>\<\phi|) \,  V_{BF}^\dag \right] \, ,
\end{align}
and the superposition channel $\map S  \in  \Chan  (A\oplus B)$ is defined as 
\begin{align}\label{generalisedoi1}
	\map S_{\rm Oi}   (\rho)   : &=  \Tr_{E} \Tr_F   \big[ W (\rho\otimes |\eta\>\<\eta|  \otimes |\phi\>\<\phi| ) W^\dag  \big]  
\end{align}
with 
\begin{align}\label{generalisedoi2}
	W  &  =   \left(   U_{AE}\otimes I_F    \right)  \oplus   \left(     V_{B  F} \otimes  I_E\right) \, .
\end{align}
For brevity, we will  denote the unitary extensions of $\map A$ and $\map B$ as $(  U, |\eta\> )$ and $(V,  |\phi\>)$, respectively.

We now show that the superpositions of channels arising from Oi's definition coincide with the superpositions specified by vacuum extensions defined in this paper, provided that the vacuum subspace is one-dimensional.  More generally, we have the following theorem: 
\begin{theo}\label{theo:stinevsvacuum}
	The following are equivalent:  
	\begin{enumerate}
		\item  channel  $\map S$ is a  standard superposition of  independent channels $\map A$ and $\map B$, specified by vacuum extensions $\widetilde {\map A}$ and $\widetilde {\map B}$ with  vacuum subspace of dimension $v$
		\item channel $\map S$ has the unitary realisation of the form  
		\begin{align}\label{vdimunitary}
			\map S  (\rho)   : &=  \Tr_{E} \Tr_F \Tr_G  \big[ W (\rho\otimes |\eta\>\<\eta|  \otimes |\phi\>\<\phi| \otimes |\gamma\>\<\gamma|) W^\dag  \big]  ,
		\end{align}
		where $G$ is a $v$-dimensional system, $|\gamma\>$ is a fixed pure state of $G$,  and
		\begin{align}
			W  &  =   \Big(   U_{AE}\otimes U_{FG}    \Big)  \oplus   \Big(     V_{B  F} \otimes  V_{EG}\Big) \, .
		\end{align}
		Here, $(U_{AE}, | \eta\>)$ is a unitary extension of $\map A$, $(V_{BF}, |\phi\>)$ is a unitary extension of $\map B$, and $U_{FG}$ and $V_{EG}$ are unitary operators.  
	\end{enumerate}
\end{theo}
\Proof  $1  \Longrightarrow 2$.  Suppose that $\map S$ is a standard superposition with $v$-dimensional vacuum.  Then, $\map S$ has Kraus operators of the form $S_{ijk}   =  A_i  \,  \beta_{jk}  \oplus       B_j  \,   \alpha_{ik} $, where $\{A_i\}$ and $\{B_j\}$ are Kraus operators of $\map A$ and $\map B$, respectively, while $\{\alpha_{ik}\}$ and $\{\beta_{jk}\}$ are complex amplitudes, cf. Eq. (\ref{superpositionijk}). 
Then, one can construct a Stinespring isometry for $\map S$  as  
\begin{align}
	\nonumber  V & =   \sum_{ijk} S_{ijk}  \otimes  |i\>_{E}\otimes |j\>_F  \otimes |k\>_{G}  \\
	&=    \Big(  V_{\map A}  \otimes  |\beta\>_{FG} \Big) \oplus  \Big  (  V_{\map B} \otimes |\alpha\>_{EG} \Big)\, ,           
\end{align}
with
\begin{align}
	\nonumber V_{\map A}   & :  =  \sum_i  A_i \otimes |i\>_E   \,  \qquad  && |\beta\>_{FG} :=  \sum_{j,k}  \beta_{jk}  \,  |j\>_F \otimes  |k\>_G  \\
	V_{\map  B}  &: =    \sum_j    B_j   \otimes |j\>_F  \qquad && |\alpha\>_{EG}: = \sum_{i,k}  \alpha_{i,k} \,  |i\>_E \otimes |k\>_G  \, . 
\end{align}
Now, the isometries $V_{\map A}$ and $V_{\map B}$ can be extended to unitary operators $U_{AE}$ and $V_{BF}$ such that 
\begin{align}
	\nonumber V_{\map A}  &=  U_{AE}  \big(I_A \otimes |\eta\>\big)\\
	V_{\map B}   & =  V_{BF}  \big(I_B \otimes |\phi\>\big) \,.
\end{align}  
Likewise, for every fixed pure state   $|\gamma\>  \in  \spc H_G$,  one can find unitary operators $ U_{FG}$ and  $V_{EG}$ such that 
\begin{align}
	\nonumber |\beta\>   &=    U_{FG}  \Big(|\phi\>_F\otimes |\gamma\>_G\Big)\\
	|\alpha\>    &=   V_{EG}  \Big(  |\eta\>_E \otimes |\gamma\>_G\Big) 
\end{align}
Hence, we obtain  
\begin{align}
	\nonumber \map S (\rho )  &   =     \Tr_{EFG}  [   V  \rho   V^\dag]  \\
	\nonumber & = \Tr_{EFG}   \left\{  \left[       \Big(  V_{\map A}  \otimes  |\beta\>_{FG} \Big) \oplus  \Big  (  V_{\map B} \otimes |\alpha\>_{EG} \Big) \right]     \,\rho  \,   \left[   \Big(  V_{\map A}  \otimes  |\beta\>_{FG} \Big) \oplus  \Big  (  V_{\map B} \otimes |\alpha\>_{EG} \Big)\right]^\dag \right\}\\
	\nonumber & = \Tr_{EFG}   \left\{  \left[       \Big(  U_{AE}    \otimes   U_{FG} \Big)   \oplus  \Big  (  V_{BF} \otimes V_{EG} \Big) \right]     \,\Big (\rho \otimes \eta_E \otimes \phi_F\otimes \gamma_G \Big)  \right.  \\   \nonumber &  \qquad \qquad \qquad \qquad \qquad \qquad \qquad\qquad \times  \left. \left[ \Big(  U_{AE}    \otimes   U_{FG} \Big)   \oplus  \Big  (  V_{BF} \otimes V_{EG} \Big)\right]^\dag \right\}\\
	&  =   \Tr_{EFG}  [  W \Big (\rho \otimes \eta_E \otimes \phi_F\otimes \gamma_G \Big)   W^\dag ] \, ,
\end{align}
where $\eta_E = \ket{\eta}\bra{\eta}_E$, $\phi_F = \ket{\phi}\bra{\phi}_F$ and $\gamma_G = \ket{\gamma}\bra{\gamma}_G$.

$2 \Longrightarrow 1.$ Suppose that channel $\map S$ has the unitary extension   (\ref{vdimunitary}).  
Then, it has a Stinespring isometry of the form  
\begin{align}
	V      =    \left(  V_{\map A}  \otimes  |\beta\>_{FG} \right) \oplus  \left(  V_{\map B} \otimes |\alpha\>_{EG}\right) \,, 
\end{align}
where $V_{\map A}:  =   U_{AE}  (  I_A\otimes |\eta\>)$  and $V_{\map B}: =  V_{BF}  (I_B\otimes |\phi\>_F)$ are  Stinespring isometries for $\map A$ and  $\map B$, respectively,  while $|\alpha\>_{EG}:  =    U_{FG}  (  |\phi\>\otimes |\gamma\>)$ and $|\beta\>_{FG}:  =    V_{EG}  (  |\eta\>\otimes |\gamma\>)$ are pure states.

Now, one has $\map S  (\rho)   =  \Tr_{EFG}[  V \rho  V^\dag]   =  \sum_{i,j,k} S_{ijk}  \rho  S_{ijk}^{ \, \dag}$, with 
\begin{align}
	\nonumber S_{ijk}     &:=   (I_S \otimes  \<i|_E \otimes \<j|_F \otimes \<  k|_G)  V \\
	\nonumber   &    =  (I_S \otimes  \<i|_E \otimes \<j|_F \otimes \<  k|_G)   \,     \Big[   \left(  V_{\map A}  \otimes  |\beta\>_{FG} \right) \oplus  \left(  V_{\map B} \otimes |\alpha\>_{EG}\right)  \Big] \\ 
	\nonumber  &  =  \Big[(I_A  \otimes  \<i|_E)  V_{\map A}    ~~   (\<j_F \otimes \<k|_G )  \,  |\beta\>_{FG}  \Big] \,  \oplus  \,\Big[ (I_B  \otimes  \<j|_F)  V_{\map B}     ~ ~  (\<i_E \otimes \<k|_G )  \,  |\alpha\>_{EG} \Big] \\
	&=    A_i  \,  \beta_{jk}   \oplus  B_j    \, \alpha_{ik}   \, ,  
\end{align}
having defined $A_i  :  =  (I_A  \otimes  \<i|_E)  V_A$, $B_j  : =  (I_B\otimes  \<j|_F) V_B$, $\alpha_{ik}   : =  (\<i|_E \otimes  \<k|_G)  |\alpha\>_{EG}$, and $\beta_{jk} : =  (\<j|_F \otimes  \<k|_G )  |\beta\>_{FG}$.  

By construction $\{A_i\}$ and $\{B_j\}$ are Kraus representations of $\map A$ and $\map B$, and the amplitudes $\{\beta_{jk}\}$ and $\{  \alpha_{ik}\}$ satisfy the normalisation conditions $\sum_{j,k} |\beta_{jk}|^2  =1$ and $  \sum_{i,k}|\alpha_{ik}|^2=1 $.    By Proposition \ref{prop:allsuperpositions} we conclude that $\map S$ is a superposition, specified by vacuum extensions, with $v$-dimensional vacuum. \qed

\medskip  

Theorem \ref{theo:stinevsvacuum} shows that Oi's superpositions coincide with our superpositions specified by vacuum extensions in the special case of one-dimensional vacuum: in this case, system $G$ is not present and the unitaries $U_{FG}$ and $V_{EG}$ can be taken to be the identity without loss of generality, {\em e.g.} by redefining $|\eta'\>   =   V_{EG}  |\eta\>$ and $|\phi'\>   =  U_{FG}  |\phi\>$.  In this way, one retrieves Equations (\ref{generalisedoi1}) and (\ref{generalisedoi2}).

\section{Extreme vacuum extensions}\label{app:extreme}  

For a fixed dimension $v$ of the vacuum subspace, the vacuum extensions of a given channel $\map C$ form a convex set, denoted as  ${\sf Vac}  (\map C,  v)$. 
The extreme points of the set are characterised by a straightforward generalisation of Choi's extremality theorem \cite{choi1975completely}: 
\begin{prop}[Extreme vacuum extensions]\label{prop:extreme}
	Let  $\widetilde {\map C} \in  \Chan  (  A  \oplus {\rm Vac})$ be a vacuum extension of $\map C$ with $v$-dimensional vacuum subspace, and let $\{\widetilde C_i  =   C_i  \oplus C_{\rm Vac, i}\}_{i=1}^r$ be a Kraus representation of $\widetilde {\map C}$ consisting of linearly independent operators.   The channel $\widetilde {\map C}$ is an exteme point of   ${\sf Vac}  (\map C,  v)$ if and only if the operators $\left\{ C_j^\dag C_i  \oplus C_{{\rm Vac },j}^\dag C_{{\rm Vac}, i}   \right\}_{i,j  \in  \{1,\dots, r\}}$ are linearly independent. 
\end{prop}

\Proof  Channel $\widetilde {\map C}$ is an extreme point if and only if no pair of channels $\widetilde {\map C}_1  \in  {\sf Vac}  (\map C,  v)$ and $\widetilde {\map C}_2 \in  {\sf Vac}  (\map C,  v)$ exist such that $\widetilde {\map C}  =   (\widetilde {\map C}_1  + \widetilde {\map C}_2)/2$.   Equivalently, channel $\widetilde {\map C}$ is an extreme point if and only if there exists no Hermitian-preserving map $\map P$ such that  $\widetilde {\map C}  \pm  \map P$ is in  $ {\sf Vac}  (\map C,  v)$.  
Now, the same argument of Choi's theorem \cite{choi1975completely} shows that the maps  $\widetilde {\map C}  \pm  \map P$  are completely positive if and only if the map $\map P$ is of the form $\map P(\rho)  =  \sum_{i,j}  \, p_{ij} \,    \widetilde C_i  \rho   \widetilde C_j^\dag $, for some Hermitian matrix $[p_{ij}]$. 
Then, the maps $\widetilde {\map C}  \pm  \map P$ are trace-preserving if and only if    $\sum_{i,j}  p_{ij} \,   \widetilde  C_j^\dag \widetilde C_i =  0$.    This condition implies the condition $p_{ij}  =  0$  for all $i,j$ if and only if  the operators $\{\widetilde C_j^\dag \widetilde C_i\}_{i,j  = 1}^r$ are linearly independent.  The condition in Proposition \ref{prop:extreme} then follows from the block diagonal form  (\ref{generalvacext}).  \qed 

\medskip 

Proposition \ref{prop:extreme} yields several  necessary conditions for a vacuum extension to be extreme.  
\begin{prop}\label{prop:boundv} Let  $\widetilde {\map C} \in  \Chan  (  A  \oplus {\rm Vac})$ be a vacuum extension of $\map C$ with a $v$-dimensional vacuum subspace, let $\{\widetilde C_i  =   C_i  \oplus C_{\rm Vac, i}\}_{i=1}^r$ be a Kraus representation of $\widetilde {\map C}$ consisting of linearly independent operators,  and let $L$ be the number of linearly independent operators in the set $\{  C_j^\dag C_i\}_{i,j=1}^r$.  If $\widetilde {\map C}$ is an extreme vacuum extension, then  the bound  $  r^2  \le L  +  v^2$ holds.  
\end{prop}  
\Proof  Let us use the shorthand notation $(i,j): =  k$,  $  O_k  :  =    C_j^\dag C_i$, and $O_{{\rm Vac},  k}    =  C_{{\rm Vac}, j}^\dag C_{{\rm Vac}, i}$.    Let $\sf S$ be a set of values of $k$ such that the operators $\{  O_k  \, , ~ k\in  {\sf S}\}$ are linearly independent. 
Every operator $O_l$ with $l\not \in \sf S$ can be decomposed as $O_l   =  \sum_{ k\in \sf S}  \, \lambda_{lk} \,  O_k $.  
Now, let $\{c_k\}$ be coefficients such that 
\begin{align}\label{linindipkraus}
	\sum_k  \,   c_k  \,  (  O_k  \oplus   O_{{\rm Vac},  k})  =  0 \,.
\end{align} 
Projecting on the subspace $\spc H_A$, we obtain the condition 
\begin{align}
	\sum_{k\in\sf S}    \left( c_k    +   \sum_{l\not \in  \sf S}   c_l  \, \lambda_{lk}        \right)  \,O_k =  0 \, ,
\end{align}
which implies 
\begin{align}\label{aaa}
	c_k   =   -\sum_{l\not \in  \sf S}   c_l  \, \lambda_{lk}    \qquad \forall k\in\sf S \, .
\end{align} 
Projecting on the subspace $\spc H_{\rm Vac}$ and using relation (\ref{aaa}), we obtain the condition 
\begin{equation}
\label{bbb} \sum_{l\not \in  \sf S}   c_l  \,    A_l   = 0, \quad  
A_l  :  =    O_{{\rm Vac },  l }    -  \sum_{k\in\sf S}     \lambda_{lk}      \,  O_{{\rm Vac}, k }  \, . 
\end{equation}
Now, the number of terms in the  sum  (\ref{bbb}) is  $  r^2  - L$.   If this number exceeds   $v^2$, then some of the operators $\{A_l\}_{l \not \in  \sf S} $ must be  linearly dependent, and therefore there exist non-zero coefficients $\{c_l\}_{l\not \in \sf S}$ such that Equations (\ref{bbb}) and  Equation (\ref{linindipkraus}) hold.   In that case, $\widetilde {\map C}$ would not be extreme.    
Hence, an extreme vacuum extension must satisfy the relation $r^2   -  L\le v^2$.    \qed  

\medskip  

An easy corollary is that the evolution of the system and the evolution of the vacuum must be coherent with one another,  meaning that the Kraus operators  $\widetilde C_i   =  C_i\oplus C_{{\rm Vac}, i}$ cannot be separated into a set with  $C_i  =  0$ and another set with  $C_{{\rm Vac}, i} = 0$.   Quantitatively, we have the following:
\begin{prop} Let  $\widetilde {\map C} \in  \Chan  (  A  \oplus {\rm Vac})$ be a vacuum extension of $\map C$ with $v$-dimensional vacuum subspace, let $\{\widetilde C_i  =   C_i  \oplus C_{\rm Vac, i}\}_{i=1}^r$ be a Kraus representation of $\widetilde {\map C}$ consisting of linearly independent operators,  and let $z$ be the number of values of $i$ such that $C_i = 0$.  If $\widetilde {\map C}$ is an extreme vacuum extension, then  the bound  $  z  \le  \sqrt{ v^2  +1}   -1$ holds.   In particular, for a one-dimensional vacuum $(v=1)$, none of the  Kraus operators $C_i$  can be zero.  
\end{prop}   
\Proof   Since $z$ Kraus operators are null, the number  $L$ of linearly independent operators of the form $\{ C_j^\dag C_i\}$ is at most $(r-z)^2$.  Hence, Proposition \ref{prop:boundv} implies  the bound   
\begin{align}
	r^2  \le   L  +  v^2  \le  (r-z)^2   +  v^2 \, ,
\end{align} which implies 
\begin{align}\label{ccc}
	2rz\le z^2  + v^2 \, .
\end{align}  
Now,   since $\widetilde {\map C}$ is trace-preserving,  there exists at least one value of $i$ such that $C_i \not =0$.  Hence, $r  \ge z+1$ and one has $z(z+2) \le v^2$.   Solving in $z$, one obtains $z  \le  \sqrt{v^2  +1 }  -1$. \qed

\section{Proof of Proposition \ref{prop:erasurecorrelations}}\label{app:erasurecorrelations} 

{\bf Proof of Proposition \ref{prop:erasurecorrelations}.} The proof is by contradiction.  Let $\map E_0 (\cdot)  =  |\psi_0\>\<\psi_0| \Tr [\cdot]$ 
be a pure erasure channel acting on the message system $M$.  Suppose that channel $\map C  =  \map E_0\otimes \map I_P $   has the form 
\begin{align}\label{bis}
	\map C  (\rho_M  \otimes \omega_P)    =\map U^\dag \! \left (  \Tr_{EF} \! \left\{ \! (\widetilde {\map V}_{AE}\otimes \widetilde {\map W}_{BF})   [ \map U (\rho_M \! \otimes  \! \omega_P)   \otimes \sigma_{EF}  ]\right\} \! \right)
\end{align} 
where  $\sigma_{EF}$ is a suitable state of $EF$, and  $\widetilde{\map V}_{AE}(\cdot)  =  \widetilde V_{AE}  \cdot  \widetilde V_{AE}^\dag$ and $\widetilde{\map W}_{BF} (\cdot)  =\widetilde  W_{BF} \cdot \widetilde W_{BF}^\dag$ are local vacuum extensions satisfying the conditions 
\begin{align}
	\widetilde V_{AE}  =  V_{AE}  \oplus  \Big( |{\rm vac}\>\<{\rm vac}|  \otimes I_E \Big)  \qquad  {\rm and}  \qquad    \widetilde W_{AE}  =  W_{AE}  \oplus \Big(  |{\rm vac}\>\<{\rm vac}|  \otimes I_F \Big) 
\end{align}
Without loss of generality, we assume the initial state $\sigma_{EF}$ to be pure, namely  $\sigma_{EF}  =  |\Phi\>\<\Phi|_{EF}$. 

Since $\map C  =  \map E_0 \otimes \map I_P$,  we must have 
\begin{align} 
	\label{cond0} \map C  (\rho_M  \otimes|0\>\<0|_P)  &=   |\psi_0\>\<\psi_0|_M \otimes |0\>\<0|_P \qquad \forall \rho  \in  \St (M)
\end{align}
and
\begin{align}
	\label{cond1} \map C  (\rho_M  \otimes|1\>\<1|_P)  &=   |\psi_0\>\<\psi_0|_M \otimes |1\>\<1|_P\qquad \forall \rho \in \St (M) \, .
\end{align} 
Now, suppose that the input state $\rho$ is pure, say $\rho   =  |\psi\>\<\psi|$. 
Then, condition (\ref{cond0}) yields   
\begin{align}
	\nonumber |\psi_0\>\<\psi_0|_M \otimes |0\>\<0|_P  &  =  \map C   (  |\psi\>\<\psi|_M\otimes |0\>\<0|_P) \\
	\nonumber &  =   \map U^\dag \! \left (  \Tr_{EF} \! \left\{ \! (\widetilde {\map V}_{AE}\otimes \widetilde {\map W}_{BF})   [ \map U (|\psi\>\<\psi|_M \! \otimes  \! |0\>\<0|_P)   \otimes \sigma_{EF}  ]\right\} \! \right)     \\
	&  = \map U^\dag  \Big\{  \Tr_{EF}  [    (\map V_{AE} \otimes \map I_{BF} )( |\psi\>\<\psi|_A  \otimes |{\rm vac}\>\<  {\rm vac}|_B  \otimes |\Phi\>\<\Phi|_{EF}) ]\Big\} \, ,
\end{align}
or equivalently, 
\begin{align}
	|\psi_0\>\<\psi_0|_A   \otimes |{\rm vac} \>\<{\rm vac}|_B  =  \Tr_{EF}  [    (\map V_{AE} \otimes \map I_{BF} )( |\psi\>\<\psi|_A  \otimes |{\rm vac}\>\<  {\rm vac}|_B  \otimes |\Phi\>\<\Phi|_{EF}) ]  \, ,
\end{align}
which in turn is equivalent to
\begin{align}
	|\psi_0\>\<\psi_0|_A    =  \Tr_{EF}  [    (\map V_{AE} \otimes \map I_{F} )( |\psi\>\<\psi|_A  \otimes |\Phi\>\<\Phi|_{EF}) ]  \, .
\end{align}
Since the pure state $|\psi\>$ is generic, this condition implies 
\begin{align}\label{S}
	(V_{AE} \otimes I_F)  (   I_A \otimes   |\Phi\>_{EF})  =    |\psi_0\>_A  \otimes     S \, ,  
\end{align}
where $S  :  \spc H_A \to \spc H_{EF}$ is an isometry. 
Similarly, condition (\ref{cond1})   yields the relation
\begin{align}\label{T}
	(W_{BF} \otimes I_E)  (   I_B  \otimes   |\Phi\>_{EF})  =    |\psi_0\>_B  \otimes     T \, ,  
\end{align}
where $T:  \spc H_B\to \spc H_{EF}$ is an isometry.  

Now,   the condition $\map C   =  \map E_0 \otimes \map{I}_P$ also implies  
\begin{align}
	\nonumber |\psi_0\>\<\psi_0|_M \otimes  |+\>\<+|_P   &=   \map C   (  |\psi\>\<\psi|_M\otimes |+\>\<+|_P)     \\
	\nonumber   &  =      \map U^\dag \! \left (  \Tr_{EF} \! \left\{ \! (\widetilde {\map V}_{AE}\otimes \widetilde {\map W}_{BF})   [ \map U (|\psi\>\<\psi|_M \! \otimes  \! |+\>\<+|_P)   \otimes \sigma_{EF}  ]\right\} \! \right)\\
	\nonumber  &  =       \map U^\dag \! \left \{  \Tr_{EF} \! \left[    (\widetilde {V}_{AE}\otimes \widetilde {W}_{BF})   \left(\frac{    |\psi\>_A \otimes |{\rm vac}\>_B  +  |{\rm vac}\>_A  \otimes |\psi\>_B}{\sqrt 2} \otimes |\Phi\>_{EF}  \right) \right.\right. \\
	\nonumber    & \quad   \times \left. \left. \left( \frac{  \< \psi|_A \otimes \<{\rm vac}|_B  +  \<{\rm vac}|_A  \otimes   \<\psi|_B) }{\sqrt 2}     \otimes   \< \Phi|_{EF} \right) \, (\widetilde {V}_{AE}\otimes \widetilde {W}_{BF})^\dag \right  ] \! \right\}  \, .
\end{align} 
We combine this equality with  the relations 
\begin{align}
	\nonumber     \Big(\widetilde {V}_{AE}\otimes \widetilde {W}_{BF}\Big)  \,\Big(  |\psi\>_A \otimes |{\rm vac}\>_B \otimes |\Phi\>_{EF}\Big)   &=    \Big(V_{AE}\otimes   I_{BF}\Big)  \,\Big(  |\psi\>_A \otimes |{\rm vac}\>_B \otimes |\Phi\>_{EF}\Big )   \\
	&  =  |\psi_0\>_A \otimes  |{\rm vac}\>_B \otimes S  |\psi\>       
\end{align}
and 
\begin{align}
	\nonumber     \Big(\widetilde {V}_{AE}\otimes \widetilde {W}_{BF}\Big)  \,\Big(  |{\rm vac}\>_A \otimes |\psi\>_B \otimes |\Phi\>_{EF}\Big)   &=    \Big(I_{AE}\otimes   W_{BF}\Big)  \,\Big(  |\psi\>_A \otimes |{\rm vac}\>_B \otimes |\Phi\>_{EF}\Big )   \\
	&  =   |{\rm vac}\>_A \otimes |\psi_0\>_B \otimes  T  |\psi\> \, ,      
\end{align}
thus obtaining  
\begin{align}
	\nonumber |\psi_0\>\<\psi_0|_M \otimes  |+\>\<+|_P   &  =       \map U^\dag \! \left \{  \Tr_{EF} \! \left[    \left(\frac{    |\psi_0\>_A \otimes |{\rm vac}\>_B  \otimes  S|\psi\>+  |{\rm vac}\>_A  \otimes |\psi_0\>_B\otimes T|\psi\>}{\sqrt 2}  \right) \right.\right. \\
	\nonumber    & \quad   \times \left. \left.      \left(\frac{ \<\psi_0|_A \otimes \<{\rm vac}|_B  \otimes  \<\psi| S^\dag +  \<{\rm vac}|_A  \otimes \<\psi_0|_B\otimes \<\psi|T^\dag}{\sqrt 2}  \right)   \right] \! \right\} \\
	\nonumber &  =        \Tr_{EF} \! \left[    \left(\frac{    |\psi_0\>_M \otimes |0\>_P  \otimes  S|\psi\>+  |\psi_0\>_M \otimes |1\>_P  \otimes T|\psi\>}{\sqrt 2}  \right) \right.  \\
	\nonumber    & \quad   \times \left.      \left(\frac{ \<\psi_0|_M \otimes \<0|_P  \otimes  \<\psi| S^\dag +  \<\psi_0|_M  \otimes \<1|_P  \otimes \<\psi|T^\dag}{\sqrt 2}  \right)   \right]  \\
	&  =  |\psi_0\>\<\psi_0|_M \otimes \Tr_{EF}\left[ |\Psi\>\<\Psi|_{PEF}\right] \, ,  \label{alloraaa}
\end{align}
with 
\begin{align}
	|\Psi\>_{PEF}   :  =    \frac{  |0\>_P\otimes S |\psi\>  +  |1\>_P  \otimes T|\psi\>  }{\sqrt 2}
\end{align}
Since Equation (\ref{alloraaa}) must be satisfied for every $|\psi\>$, we conclude that $T$ and $S$ must be equal.  

To conclude the proof, consider the channels  $\map M (\rho)   :=  \Tr_{F}  [  S\rho S^\dag]$  and $\map M^{\rm c} (\rho) : =\Tr_{E}  [  S\rho S^\dag] $. Here, we regard both channels as having input $M$, owing to the identification $A  \simeq B\simeq M$. 
Using Equation (\ref{T}), we obtain 
\begin{align}
	\nonumber  \map M  (\rho)  &  =  \Tr_{MF}  [  |\psi_0\>\<\psi_0| \otimes S  \rho  S^\dag] \\
	\nonumber &  =   \Tr_{MF}  \Big[  (\map W_{MF} \otimes \map I_E) (  \rho_M  \otimes  |\Phi\>\<\Phi|_{EF})  \Big]\\
	\label{correctableMc} &  = \Tr_F  [  |\Phi\>\<\Phi|_{EF}] \qquad \forall \rho \in \St (M) \, .
\end{align}
Similarly, Equation (\ref{S}) implies 
\begin{align}
	\nonumber  \map M^{\rm c}  (\rho)  &  =  \Tr_{ME}  [  |\psi_0\>\<\psi_0| \otimes S  \rho  S^\dag] \\
	\nonumber &  =   \Tr_{ME}  \Big[  (\map V_{ME} \otimes \map I_F) (  \rho_M  \otimes  |\Phi\>\<\Phi|_{EF})  \Big]\\
	\label{correctableM} &  = \Tr_E  [  |\Phi\>\<\Phi|_{EF}]  \qquad \forall \rho  \in\St (M) \, .
\end{align}

Now, Equation  (\ref{correctableMc}) implies that $\map M^{\rm c}$ is correctable (\emph{i.e.}\ can be inverted to recover the state $\rho$), while Equation (\ref{correctableM}) implies that $\map M$ is correctable \cite{wilde2013quantum}.  Since $\map M$ and $\map M^{\rm c}$ are complementary to each other, this is in contradiction with the no-cloning theorem: by correcting  $\map M$ one would retrieve one copy of $\rho$, and by correcting $\map M^{\rm c}$ one would retrieve another copy.     In conclusion, the channel $\map E_0\otimes \map I_P$ does not admit a realisation of the form (\ref{bis}). \qed   

\medskip

\end{document}